\documentclass[a4paper,fleqn,usenatbib]{mnras}
\usepackage{mathptmx}
\usepackage[T1]{fontenc}
\usepackage{ae,aecompl}
\usepackage{times}
\usepackage{amssymb}
\usepackage{amsmath}
\usepackage{upgreek}
\usepackage{color}
\usepackage{graphicx}
\usepackage{pdflscape}

\usepackage{pifont}
\newcommand{\cmark}{\ding{51}}%
\newcommand{\xmark}{\ding{55}}%

\usepackage{newtxtext,newtxmath}

\DeclareRobustCommand{\VAN}[3]{#2}
\let\VANthebibliography\thebibliography
\def\thebibliography{\DeclareRobustCommand{\VAN}[3]{##3}\VANthebibliography}

\title[Outflows in protoclusters]{An ALMA study of outflow parameters of protoclusters: outflow feedback to maintain the turbulence}

\author[T. Baug et al.]{T. Baug,$^{1,2,3}$\thanks{E-mail: tapas.polo@gmail.com (TB)}
Ke Wang,$^{1,3}$\thanks{Corresponding Author: kwang.astro@pku.edu.cn (KW)}
Tie Liu,$^{4}$
Yue-Fang Wu,$^{5}$
Di Li, $^{6,7,8}$
Qizhou Zhang,$^{9}$
Mengyao Tang,$^{10}$
\newauthor
Paul F. Goldsmith,$^{11}$
Hong-Li Liu,$^{12}$
Anandmayee Tej,$^{13}$
Leonardo Bronfman,$^{14}$
L. Viktor Toth,$^{15}$
\newauthor
Kee-Tae Kim,$^{16}$
Shang-Huo Li,$^{4}$
Chang Won Lee,$^{16,17}$
Ken'ichi Tatematsu,$^{18}$
and
Tomoya Hirota$^{18}$
\\
$^{1}$Kavli Institute for Astronomy and Astrophysics, Peking University, 5 Yiheyuan Road, Haidian District, Beijing 100871, China\\
$^{2}$S. N. Bose National Centre for Basic Sciences, Block-JD, Sector-III, Salt Lake, Kolkata 700106, India\\
$^{3}$PKU - SHAO Joint Research Center for Astrophysics, Peking University, 5 Yiheyuan Road, Haidian District, Beijing 100871, China\\
$^{4}$Shanghai Astronomical Observatory, Chinese Academy of Sciences, 80 Nandan Road, Shanghai 200030, China\\
$^{5}$Department of Astronomy, Peking University, 100871, Beijing, People's Republic of China\\
$^{6}$CAS Key Laboratory of FAST, National Astronomical Observatories, Chinese Academy of Sciences, Beijing 100101, China\\
$^{7}$University of Chinese Academy of Sciences, Beijing 100049, China\\
$^{8}$NAOC-UKZN Computational Astrophysics Centre, University of KwaZulu-Natal, Durban 4000, South Africa\\
$^{9}$Harvard-Smithsonian Center for Astrophysics, 60 Garden Street, Cambridge, MA 02138, USA\\
$^{10}$Department of Astronomy, Yunnan University, Kunming, 650091, China\\
$^{11}$Jet Propulsion Laboratory, National Aeronautics and Space Administration, United States\\
$^{12}$Departamento de Astronom\'{\i}a, Universidad de Concepci\'{o}n, Av. Esteban Iturra s/n, Distrito Universitario, 160-C, Chile\\
$^{13}$Indian Institute of Space Science and Technology, Thiruvananthapuram 695 547, Kerala, India\\
$^{14}$Departamento de Astronom\'{i}a, Universidad de Chile, Casilla 36-D, Santiago, Chile\\
$^{15}$E\"{o}v\"{o}s Lor\'{a}nd University, Department of Astronomy, P\'{a}zm\'{a}ny P\'{e}ter s\'{e}t\'{a}ny 1/A, H-1117, Budapest, Hungary\\
$^{16}$Korea Astronomy and Space Science Institute, 776 Daedeokdae-ro, Yuseong-gu, Daejeon 34055, Republic of Korea\\
$^{17}$University of Science and Technology, Korea (UST), 217 Gajeong-ro, Yuseong-gu, Daejeon 34113, Republic of Korea\\
$^{18}$National Astronomical Observatory of Japan, National Institutes of Natural Sciences, 2-21-1 Osawa, Mitaka, Tokyo 181-8588, Japan\\
}

\date{Accepted XXX. Received YYY; in original form ZZZ}

\pubyear{2020}

\begin{document}
\label{firstpage}
\pagerange{\pageref{firstpage}--\pageref{lastpage}}
\maketitle


\begin{abstract}
With the aim of understanding the role of outflows in star formation, we performed a statistical study of the physical parameters of outflows in eleven massive protoclusters associated with ultra-compact H{\sc ii} regions. A total of 106 outflow lobes are identified in these protoclusters using the ALMA CO (3–2), HCN (4--3) and HCO$^{+}$ (4--3) line observations. Although the position angles of outflow lobes do not differ in these three tracers, HCN and HCO$^{+}$ tend to detect lower terminal velocity of the identified outflows compared to CO. The majority of the outflows in our targets are young with typical dynamical time-scales of 10$^{2}-$10$^{4}$ years, and are mostly composed of low-mass outflows along with at least one high-mass outflow in each target. An anti-correlation of outflow rate with dynamical time-scale indicates that the outflow rate possibly decreases with time. Also, a rising trend of dynamical time-scale with the mass of the associated core hints that the massive cores might have longer accretion histories than the low mass cores. Estimation of different energies in these protoclusters shows that outflows studied here cannot account for the generation of the observed turbulence, but can sustain the turbulence at the current epoch as the energy injection rate from the outflows is similar to the estimated dissipation rate.
\end{abstract}




\section{Introduction}
\label{sec:intro}
Star formation is a complex process that involves the collapse and accretion of gas onto protostars \citep{lada85}. Molecular outflows are ubiquitous in the early stages of star formation. It is expected that at the protostellar phase a fraction of the accreted material from the envelope/disk is expelled as a result of angular momentum conservation \citep[see reviews by][]{frank14, bally16}. The gas is typically ejected in the form of high-velocity collimated jets that sweep up the surrounding material and form molecular outflows around the jet axis \citep[][and references therein]{lee00, arce07, hartmann16}. Compared to accretion disks, these outflows are more easily detectable in star-forming clouds \citep[][and references therein]{bally16}. Since the first-ever observational detection of a molecular outflow in Orion A \citep{kwan76, zuckerman76}, many outflows have been found in Galactic star-forming regions. Over the past few decades, studies of the outflows have increased significantly \citep[see the review by][]{bally16}, and the jets/outflows were detected in all mass regimes, starting from brown dwarfs \citep[e.g.,][]{whelan05, riaz17} to intermediate-mass protostars \citep[e.g.,][]{zapata10, reiter17, takahashi19} to high-mass protostars \citep[e.g.,][and references therein]{caratti15, li19b}.

One of the main debates in the star formation community is whether the massive young stellar objects are a scaled-up version of low-mass young stellar objects (YSOs) where disk-accretion plays the dominating role for gaining the stellar mass. For the formation of low-mass stars, bipolar outflows driven by the accretion disks are proposed to be the basic formation mechanism theoretically \citep{shu87}, and are also verified observationally \citep[e.g.,][and references therein]{bontemps96,richer00,arce07}. However, on the other hand understanding of the formation mechanism of massive stars is still elusive \citep{tan14}. Two major competing models for massive star formation are (i) core accretion via disk \citep{mckee03} and (ii) competitive accretion \citep{bonnell01}. The most obvious way to distinguish between these two models might be the detection of the accretion disk around massive protostars. But a direct detection of accreting disk is difficult because the accretion disk is small and short-lived, and also because of complicated gas dynamics at that scale \citep{kim06}. Here, the study of the properties of molecular outflows which are the manifestation of disk-accretion in young sources, could help us to improve our understanding of the underlying formation process \citep{shepherd96, beuther02, molinari02, arce07}. If it is assumed that massive stars do form via an accretion disk similar to the low-mass stars, they should generate massive and powerful outflows \citep[see][and references therein]{devilliers14}. A few recent studies \citep[e.g.,][]{devilliers14, li18} indeed found the applicability of the same scaling between outflow activity and clump masses for both low-mass and massive objects, suggesting a similar formation mechanism. However, an extensive study of molecular outflows toward massive star-forming regions is still lacking owing to their large distances and high level of clustering. Recent interferometric observations with the ALMA enable us to target such regions thanks to its high spatial resolution and sensitivity. Studies of outflows associated with Galactic massive star-forming regions may provide us clues to better understand the launching mechanism of molecular outflows, and hence, the underlying star formation mechanism.

The outflows also inject a large amount of mechanical energy into the parent molecular cloud \citep{solomon81, lada85, bachiller96}. Such energetic feedback from young stars may significantly influence the self-regulation of star formation \citep{franco83}. Feedback may provide the required turbulence to the parent molecular cloud for stabilizing it against the gravitational collapse \citep{shu87, nakamura07, carroll09, matzner15}. The impact of outflows on surrounding gas has been studied in several massive and low-mass star-forming regions \citep[][and references therein]{arce10, nakamura11, narayanan12, li15, feddersen20}. Majority of these studies reported that the outflows do not have sufficient energy to sustain the observed turbulence in their parent molecular clouds \citep[see e.g.,][and references therein]{li20}. In a survey on Perseus molecular cloud, \citet{arce10} found that even though outflows have a large impact on the local clouds near the active star-forming area, the energy from outflows is not sufficient to produce the observed turbulence in the entire Perseus complex. A similar result was also found by \citet{narayanan12} and \citet{li15} for Taurus region. However, \citet{nakamura11} found that the protostellar outflows play a crucial role in replenishing the supersonic turbulence in $\rho$ Ophiuchi Main Cloud. \citet{plunkett13} also found the outflow energy comparable to the turbulent energy in NGC 1333 star-forming region. A similar result was also obtained by \citet{yang18} in hundreds of star-forming clumps.

Protostellar outflows have been extensively studied using several different atomic and molecular tracers at different wavelength regime, for example, H$_2$, [Fe{\sc ii}], CO, SiO, HCN, HCO+, HNCO, CS, and so on \citep[see][for a detailed discussion]{bally16}. The emission lines of CO, particularly the low-$J$ pure rotational transitions ($J \leq$4) in millimeter and sub-millimeter bands are widely used as tracers of outflow activity \citep{shepherd96, beuther02, yang18}. This is because of the high abundance of CO and also because observations of these CO lines are relatively straightforward using ground-based telescopes. In this paper, we report a study of outflow parameters derived from the CO ($J$ = 3--2) line toward 11 Galactic massive protoclusters \citep[1--24$\times$10$^3$ M$_\odot$;][]{liu16} using data from ALMA. The details of these 11 target regions have been presented in Table 1 of our previous paper \citep[][hereafter Paper I]{baug20}. In total, we identified 106 lobes in these 11 target regions. A study of the orientation of the outflow lobes with respect to their host filaments is already reported in Paper I. The present paper concentrates on the derived outflow parameters (e.g., outflow mass, outflow rate, mechanical luminosity, dynamical timescale of outflow, etc.), their relation with the mass of the host cores, and the energy injection by these outflows to their parent clouds.

The paper is organized in the following manner. In Section~\ref{sec:data}, we present a brief introduction to the data. Identification of outflow lobes, comparison of the outflow parameters obtained with different tracers (i.e., CO(3--2), HCN(4--3), HCO$^{+}$ (4--3); hereafter CO, HCN and HCO$^{+}$ throughout the paper), and derivation of physical parameters of the CO outflows are presented in Section~\ref{sec:results}. A comparison of the CO outflow parameters with the parameters of the host cores and analysis on the energy budget of the host clouds are presented in Section~\ref{sec:discussion}. A conclusion of this study is presented in Section~\ref{sec:conclusions}.

\section{Data}
\label{sec:data}
\subsection{ALMA observations}
Details of the ALMA observations (project 2017.1.00545.S; PI: Tie Liu) have been presented in Paper I. These data were observed as a pilot project for the ALMA Three-millimeter Observations of Massive Star-forming regions (ATOMS) survey \citep{liu20a, liu20b}.

\section{Results}
\label{sec:results}
\subsection{Identification of Outflows}
\label{sec:IndetifyOutflows}
The procedure to identify outflows adopted in this work was already reported in Paper I. For clarity, we briefly mention the procedure here. First, we cropped the data for the whole spectral window into smaller cubes that only cover $\pm$200 km s$^{-1}$ centering on the systemic velocity of each target. We carefully examined these small position-position-velocity (PPV) data cubes in SAOImageDS9\footnote{https://sites.google.com/cfa.harvard.edu/saoimageds9} looking for the redshifted and blueshifted lobes around continuum sources starting from the highest velocity channels, as these channels are expected to be least contaminated by emission from the central clouds. The outflow lobes in a couple of regions are substantially crowded. There were a few confusing CO outflow lobes that overlap with each other. For those, we compared the outflow lobes in different tracers (i.e., CO, HCN, HCO$^{+}$, and SiO) to confirm their credibility. However, in this paper we do not include any analysis from SiO observations. Analysis using this particular line will be reported for the full ATOMS survey sample in a future paper. Here, we considered as separate outflow lobes only those for which the emission between the two lobes at terminal velocity is separated by more than 5 pixels in the PPV space. 

The terminal velocity and extent of each outflow lobe were considered up to a 5$\sigma$ level where $\sigma$ is the rms measured from a few line-free channels. In addition to the bipolar outflows, we identified unipolar outflows that are associated with continuum sources, and also a few outflow lobes without having any association with detected continuum sources. Here, we report a total of 106 outflow lobes. Among them 32 are bipolar and 42 are unipolar. Note that in this paper, we report one additional lobe which was confused and thought to be associated with another lobe in Paper I.

\begin{figure}
\includegraphics[width=\columnwidth]{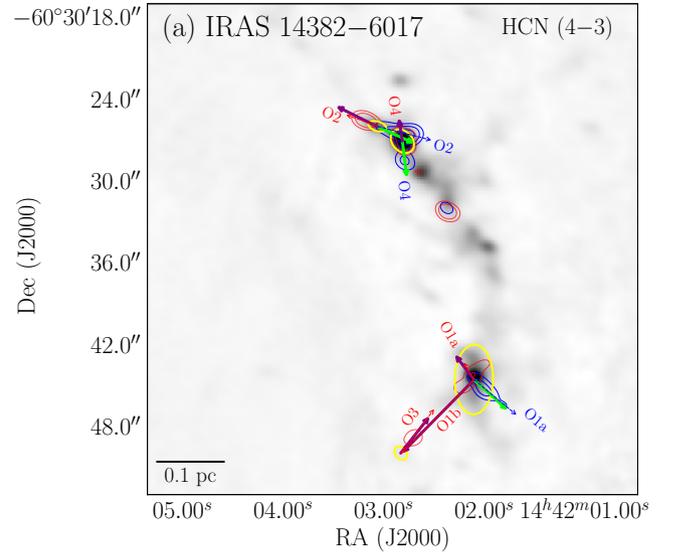}
\caption{ HCN outflow lobes toward the IRAS 14382-6017 region overlaid on the ALMA 0.9 mm continuum map of the region. The red and blue contours correspond to redshifted and blueshifted HCN emission integrated over carefully selected velocity ranges to depict the outflow lobes. The extents of blueshifted and redshifted outflow lobes in HCN are also marked by solid green and purple arrows, while the extent of outflow lobes identified using CO are shown by blue and red dashed arrows, respectively. The driving sources detected in the continuum map are marked in yellow ellipses. }
\label{fig1}
\end{figure}

A total of 99 and 80 outflow lobes are detected in HCN and HCO$^{+}$, respectively. Among the 106 lobes identified in CO, 80 are identified in HCN and 62 are identified in HCO$^{+}$. We also named the outflow lobes for convenience (see Tables~\ref{table1}~and~\ref{table2}). The naming of an outflow is done as follows. It contains the name of the region, followed by an alphanumeric character (that starts with the letter `O') to mark the source number, and a letter if multiple outflow lobes are identified with a single core. Details of all the identified outflow lobes, such as coordinates of the continuum source, orientations of outflow lobes on the plane of the sky, terminal velocity, and extent of each outflow are presented in Paper I and are thus not presented here. The outflow lobes identified with HCN overlaid on the 0.9 mm ALMA continuum maps for one region are presented in Figure~\ref{fig1}. Figures for the identified outflows (in both HCN and HCO$^{+}$) in the remaining regions are presented in Appendix~\ref{appendix}. The outflow lobes identified in CO are also marked by arrows in Figure~\ref{fig1}. For reference, similar figures corresponding to the CO outflows are available in Paper I.

\subsection{Comparison of outflow lobes}
\label{CompareLobes}
\begin{figure}
\includegraphics[width=\columnwidth]{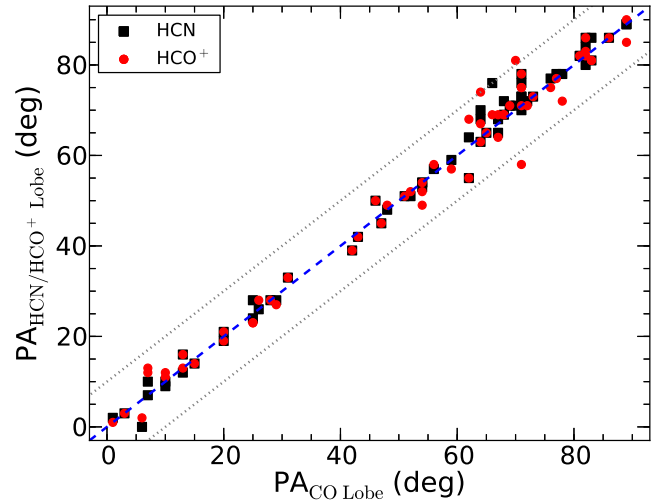}
\caption{Comparison of the position angles of the lobes in HCN (black squares) and HCO$^{+}$ (red circles) with respect to the position angles of the lobes in CO. The blue dashed line shows the ratio PA$_{\rm HCN/HCO^{+}}$/PA$_{\rm CO}$ = 1. Grey dotted lines show the track for PA variation of $\pm$10$^\circ$.}
\label{fig2}
\end{figure}

\begin{figure}
\includegraphics[width=\columnwidth]{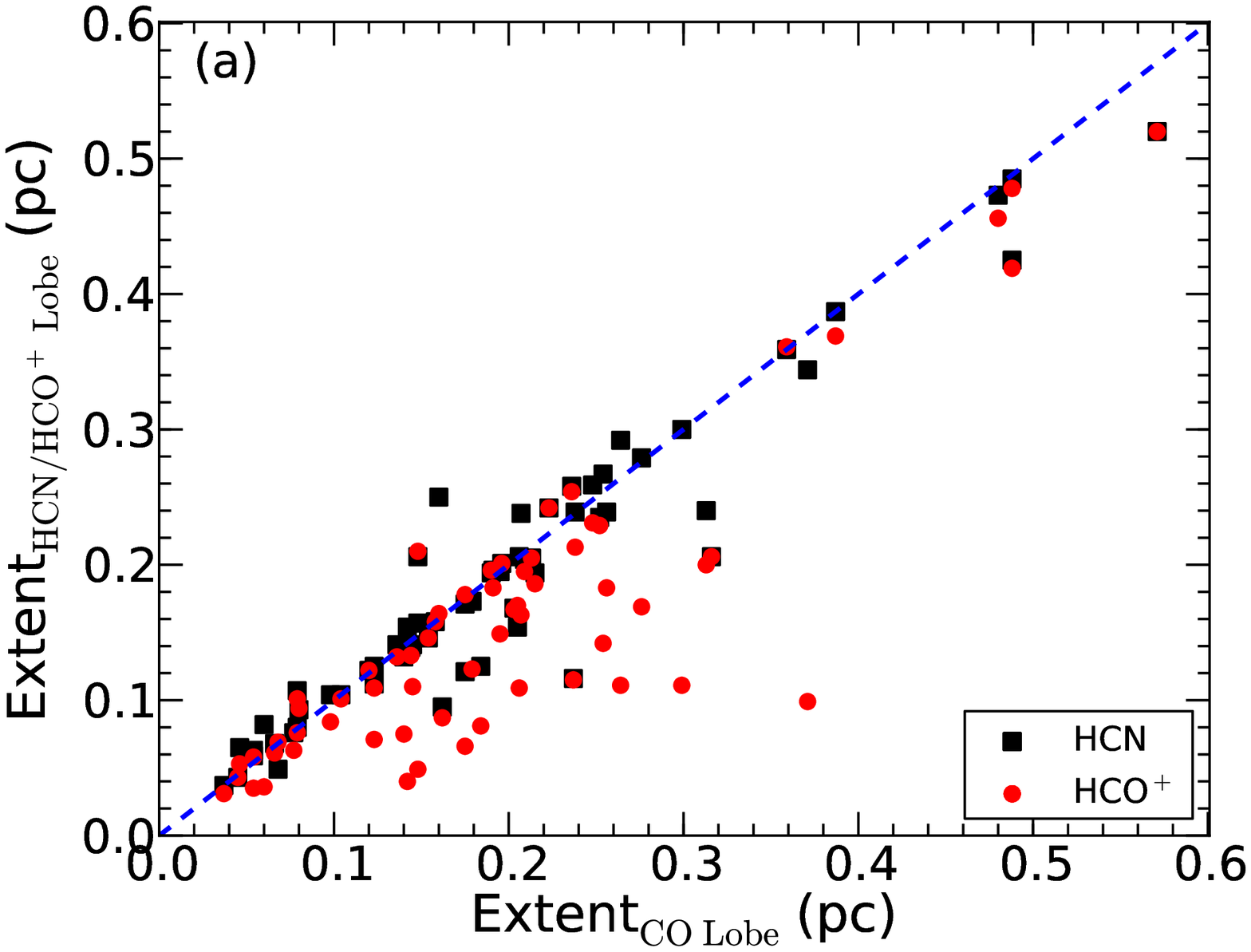}
\includegraphics[width=\columnwidth]{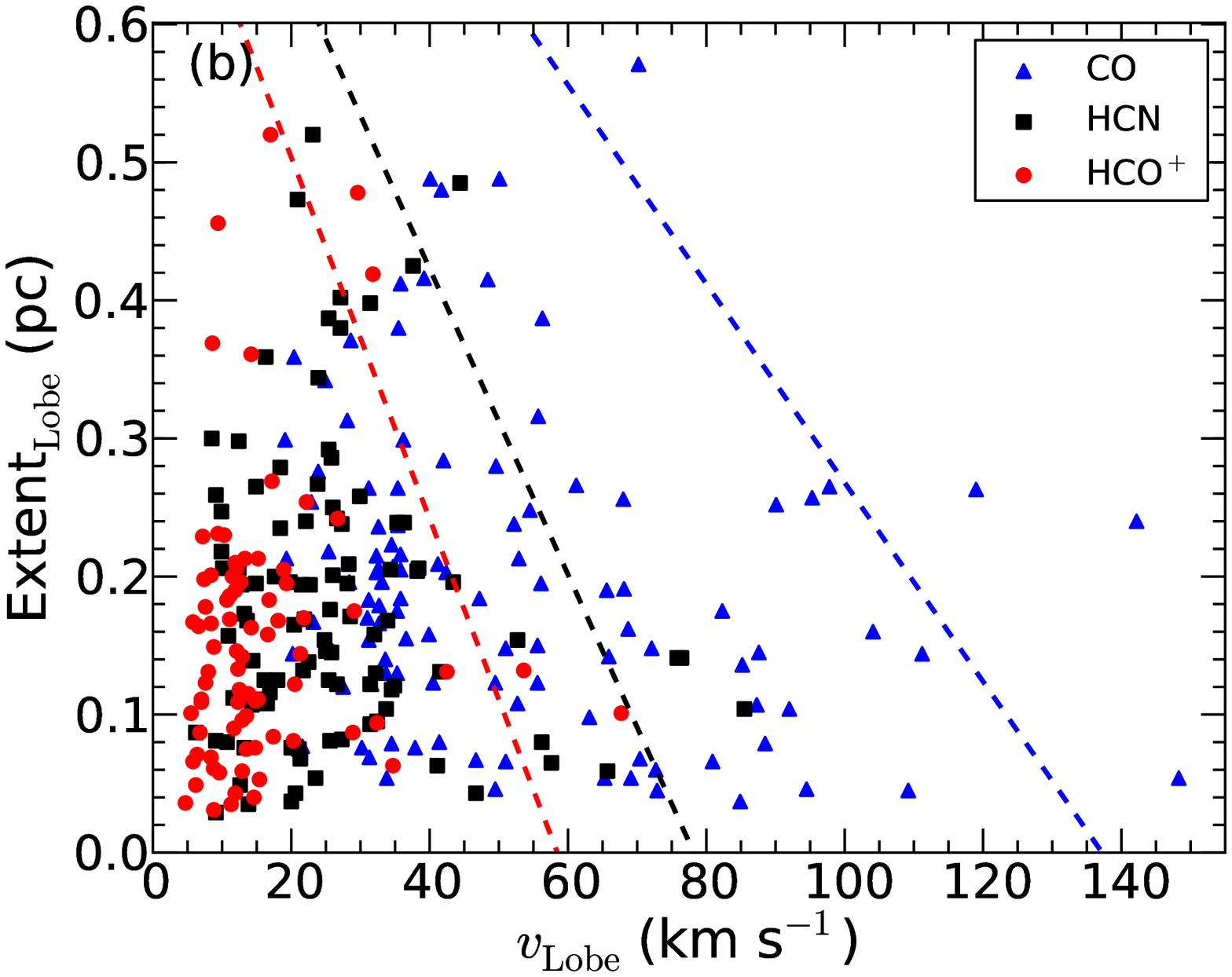}
	\caption{(a) Comparison between the extent of the outflow lobes in HCN and HCO$^{+}$ with respect to the extent of the outflow lobes in CO. (b) Comparison of the extent of the outflow lobes with the terminal velocity of the outflow. In both the panels, black square and red circles represent parameters for HCN and HCO$^{+}$, respectively, while the blue triangles in the bottom panel represent the values for CO. The blue, black and red dashed lines are drawn `by-eye' for visualization (no fit involved) to portray typical limits and trends of terminal velocities and the extent of the lobes in CO, HCN and HCO$^{+}$, respectively. }
\label{fig3}
\end{figure}

In order to examine how the detected outflow lobes conform in three different tracers, we plot the position angles, velocities, and extents of the outflow lobes in Figures~\ref{fig2}~and~\ref{fig3}. 

As seen in Figure~\ref{fig2}, the position angles of outflow lobes do not differ much in all three tracers and have typical variations $<\pm$ 10$^\circ$. This particular variation is of the order of uncertainties in the measurements of the position angles. Figures~\ref{fig3}a~and~\ref{fig3}b show the projected extent of the outflow lobes in HCN and HCO$^{+}$ against CO emission, and the extent of the lobes against the terminal velocities of the outflow lobes in different tracers, respectively. Even though the extents of the lobes are generally found to be similar in CO and HCN (Figure~\ref{fig3}a), a few HCO$^{+}$ lobes are found to be 10-20\% shorter compared to the other two tracers. Also, HCN and HCO$^{+}$ typically have lower terminal velocities compared to those traced in CO. A similar result was noted earlier by \citet{lopez10}. These authors found lower values of outflow momenta and kinetic energies estimated using HCO$^{+}$ (1--0) emission compared to the commonly used tracers like CO. Typically, HCN and HCO$^{+}$ trace the outflow in the vicinity of the driving source, while CO is sensitive enough to detect the low density and high-velocity outflow material. This is possibly because CO is more abundant than HCN and HCO$^{+}$, and also because the excitation of HCN (4--3) and HCO$^{+}$ (4--3) requires relatively higher temperatures compared to CO (3--2). Note that different tracers trace various components of the outflows, and that explains the variation seen in the observed outflow parameters in three different tracers \citep[see][for a detailed discussion]{bally16}. 
 
\subsection{Dynamical properties of outflows}
\label{SecDynProp}
In this paper, we only calculated the parameters for the outflows that are identified in CO. The calculation of outflows parameters using HCN and HCO$^{+}$ are not performed in this paper.
The physical properties of all the identified CO outflow lobes, such as mass ($M_{\rm out}$), momentum ($P_{\rm out}$), energy ($E_{\rm out}$), dynamical ages, outflow rate ($\dot{M}_{\rm out}$), mechanical luminosity ($L_{\rm mech}$), and outflow force ($F_{\rm out}$) were derived following the similar procedure described in \citet{wang11}. We first simplified the equation given in \citet{mangum15} to calculate the total CO column density of the outflows assuming that they are in local thermodynamic equilibrium and the emission is optically thin. Note that the calculated mass considering the emission to be optically thin might be underestimated by a factor of a few \citep[see][]{offner11, arce13, dunham14}. However, opacity correction for the estimated mass of the outflows is primarily crucial if the lower bound velocity of the lobes are below 3 km s$^{-1}$ \citep[see][for a detailed discussion]{dunham14}. We do not apply any opacity correction as the lower bound velocity of our idenntifed lobes are typically $\sim$5 km s$^{-1}$.
 The equation for the total CO column density is given by :
\begin{equation}
	N^{\rm{thin}}_{\rm{CO}} ({\rm cm^{-2}}) = 4.81 \times 10^{12} (T_{\rm ex} + 0.92) {\rm exp}\left(\frac{33.12}{T_{\rm ex}}\right)\int T_{\rm B}dv,
\end{equation}
where $T_{\rm ex}$ and $T_{\rm B}$ are excitation temperature and brightness temperature in K, respectively, and $dv$ is in km s$^{-1}$. Estimated outflow parameters may vary depending on the adopted $T_{\rm ex}$. \citet{dunham14} performed a detailed calculation to estimate the correction factor for a  $T_{\rm ex}$ range from 10--200 K. They pointed out that low-$J$ transitions of CO (as in this study) are generally insensitive for $T_{\rm ex} \gtrsim$ 50 K. Note that the targets in this study are massive star-forming regions associated with H{\sc ii} regions. Thus, a higher gas temperature is expected in these regions compared to the gas temperature typically assumed for cold clumps (10--30~K). Hence, we derived the outflow parameters in our target regions assuming a $T_{\rm ex}$ of 50~K. Also, this is a crude assumption to consider a single $T_{\rm ex}$ for all the outflows. In fact, excitation temperatures may even vary over different parts of the lobes, and there may be very warm molecular gas in shocks \citep[see e.g.,][and references therein]{green11}. Thus, consideration of a single $T_{\rm ex}$ can lead to significant underestimates (by up to factors of 3--4) in the outflow kinetic energy and mechanical luminosity which are the most sensitive to the highest-velocity gas \citep[see][for a detail discussion]{downes07, dunham14}. As discussed by \citet{dunham14}, a different value of $T_{\rm ex}$ from 10--50 K might lead to a maximum variation in the outflow parameter by a factor of 1--3 which is of the order of uncertainty of these derived parameters.
Here, we derived the outflow parameters using the equations given in \citet{wang11}:
\begin{align}
	M_{\rm out} & = d^{2}\left[\frac{{\rm H_2}}{{\rm CO}}\right] \overline{m}_{\rm H_2}\int_\Omega N^{\rm{thin}}_{\rm{CO}} (\Omega') d\Omega', \\
	P_{\rm out} & = M_{\rm out}v, \\
	E_{\rm out} & = \frac{1}{2} M_{\rm out} v^{2}, \\
	t_{\rm dyn} & = \frac{L_{\rm flow}}{v_{\rm Lobe}}, \\
	\dot{M}_{\rm out} & = \frac{M_{\rm out}}{t_{\rm dyn}}, \\
	L_{\rm mech} & = \frac{E_{\rm out}}{t_{\rm dyn}}, \\
	F_{\rm out} & = \frac{P_{\rm out}}{t_{\rm dyn}}, 
\end{align}

where $d$ is the distance to the source. $\left[\frac{{\rm H_2}}{{\rm CO}}\right]$ is the abundance ratio of $H_2$ to CO which is taken to be $10^4$ \citep{blake87}. The mean molecular weight, $\overline{m}_{\rm H_2}$, is assumed to be 2.33 $m_{\rm H}$. The parameter $\Omega$ is the solid angle subtended by the outflow lobe, and $v$ is the velocity of the outflow relative to the systemic velocity. The parameters $L_{\rm flow}$ and $v_{\rm Lobe}$ are the extent of the outflow lobe from the driving source and the outflow terminal velocity, respectively.

Since we only measure the radial component of the outflow velocity and the projected size of the lobe on the plane of the sky, it is important to include the corrections for the inclination. 
We have corrected the parameters, $L_{\rm flow}$, $P_{\rm out}$, $E_{\rm out}$, $t_{\rm dyn}$, $\dot{M}_{\rm out}$, $L_{\rm mech}$, and $F_{\rm out}$ multiplying by 1/sin~$i$, 1/cos~$i$, 1/cos$^2~i$, cos~$i$/sin~$i$, sin~$i$/cos~$i$, sin~$i$/cos$^3~i$, and sin~$i$/cos$^2~i$, respectively, for a mean inclination angle ($i$) of 57$\fdg$3 assuming all orientations are equally favorable \citep[see][for a detailed description]{dunham14}. However, the outflow parameters estimated here should be treated only as lower limits because it might include several observational biases. The assumption for the CO line to be optically thin might not always be true, and unlike other low-density gas tracers (e.g., C{\sc ii} 157 $\mu$m line), CO is unable to trace the outflows in low-density regions. Also, a lower limit may be due to the limited sensitivity of our observations. Moreover, the lower bound velocity of our identified outflows is at about 5 km s$^{-1}$ from the systemic velocity of the cloud because we could not separate out the outflow gas component from the host cloud below this velocity limit. However, this low velocity part of the outflow might have a significant contribution to the observed outflow mass \citep{arce01, downes07, offner11}. Beside, it was not always possible to disentangle the overlapping emission from two nearby lobes, particularly when they approached the cloud velocity (within 2-6 km s$^{-1}$). Although, the minimum detectable velocity of our identified outflows is at about 5 km s$^{-1}$ from the cloud velocity, this effect might lead to 15-20\% uncertainties to the estimated outflow mass, and hence, to other calculated parameters.
Together all these biases may lead to underestimate the estimated parameter up to a factor of $\sim$10 even after correcting the estimated parameters for the mean inclination \citep{dunham14}. 
Table~\ref{table1} lists the velocity range used to derive the calculations, $M_{\rm out}$, and several other calculated dynamical parameters (i.e., $P_{\rm out}$, $E_{\rm out}$, $\dot{M}_{\rm out}$, $F_{\rm out}$, and $L_{\rm mech}$); the values are listed for all the outflows after correcting for the inclination. As mentioned before, in this paper we estimated the outflow parameters based on the CO emission. So, in Table~\ref{table1} we include only the outflows detected in CO. The lobes identified in HCN or HCO$^+$ but not detected in CO are not listed in the table. Additionally, in Table~\ref{table1} we indicate whether the corresponding outflow is detected in either of the other two tracers. For reference, all the outflow lobes detected in all three tracers are marked in Figures~\ref{fig1}~and~\ref{figA1}--\ref{figA4}.

\begin{figure}
\includegraphics[width=\columnwidth]{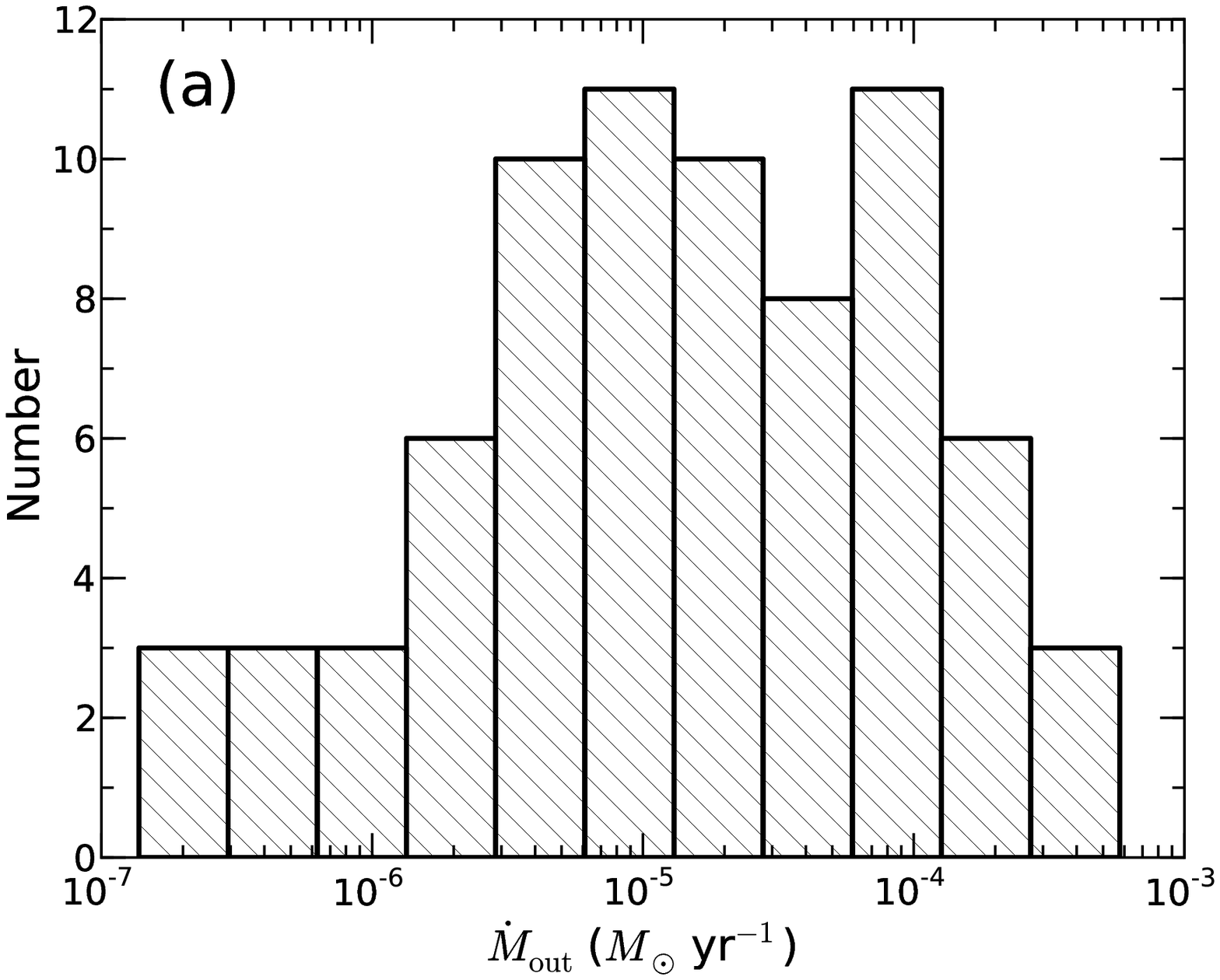}
\includegraphics[width=\columnwidth]{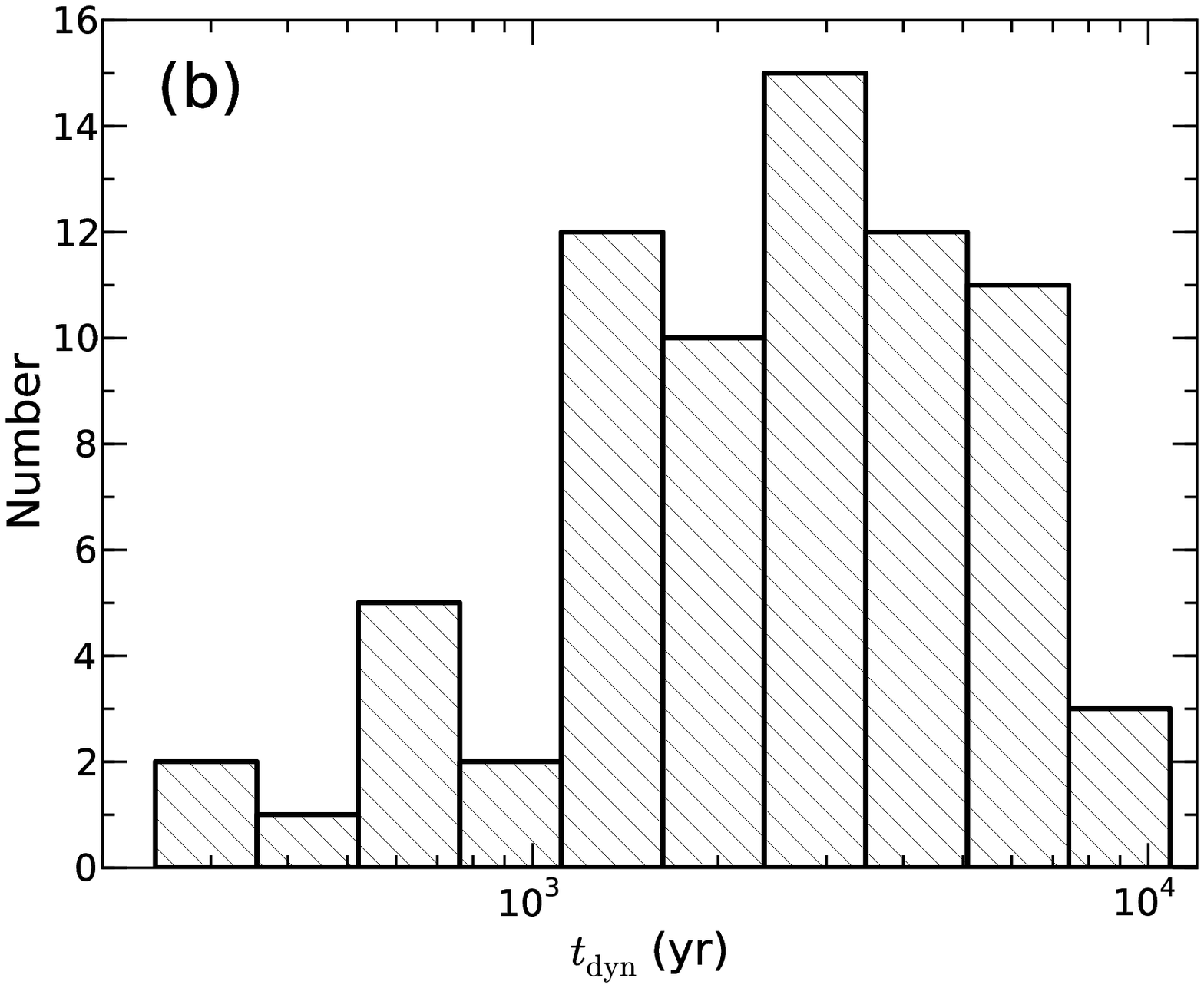}
\includegraphics[width=\columnwidth]{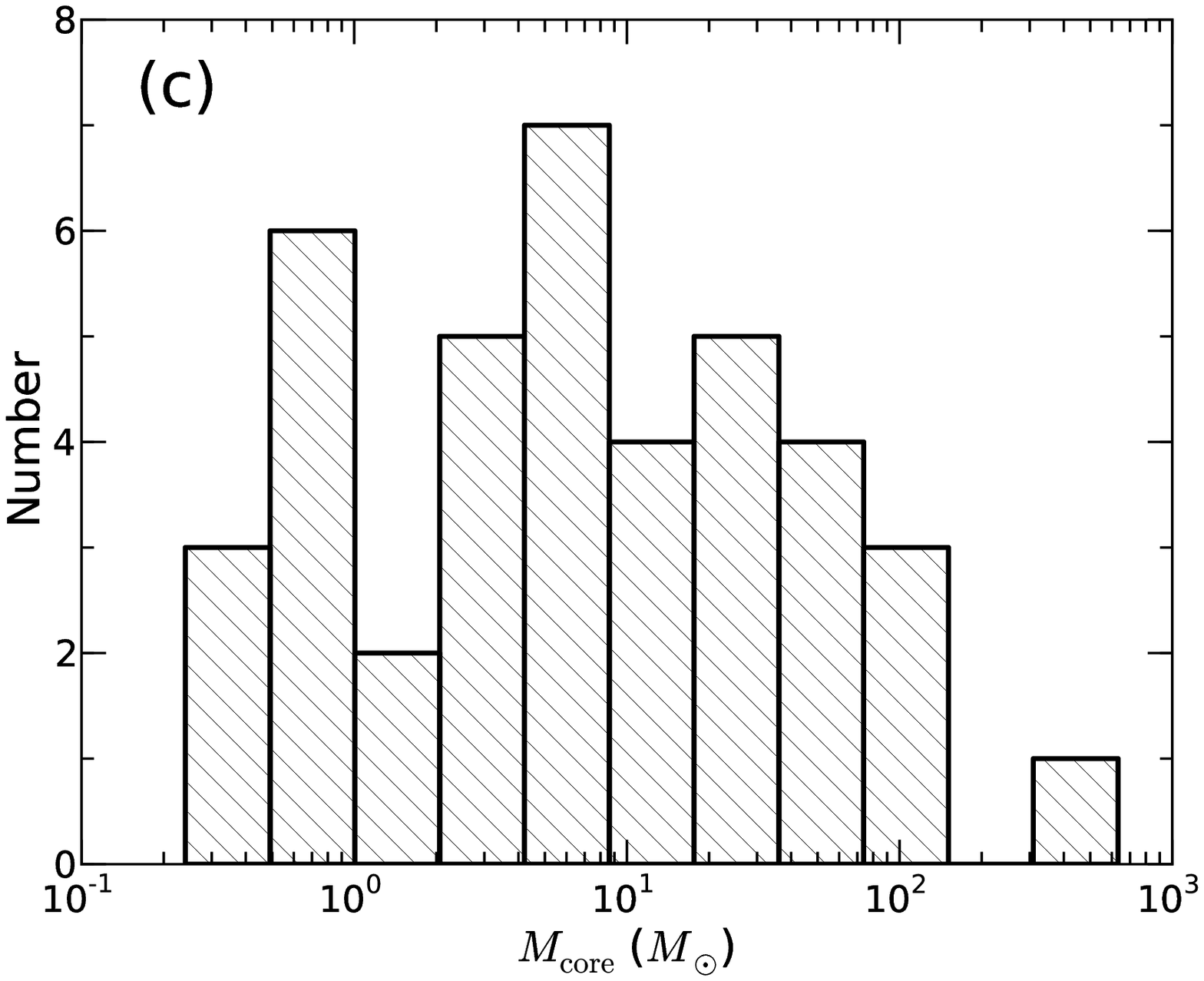}
\caption{(a) Histogram of the outflow rate. (b) Histogram of the estimated dynamical ages of all the outflows, and (c) Histogram of the calculated mass of the host cores considering a dust temperature of 50 K and spectral index, $\beta$, of 2.}
\label{fig4}
\end{figure}

A histogram of $\dot{M}_{\rm out}$ is shown in Figure~\ref{fig4}a. It can be seen that the values $\dot{M}_\mathrm{out}$ have a wide range starting from $\sim$10$^{-7}$ to $\sim$10$^{-3}$ $M_\odot$~yr$^{-1}$. This provides us a unique opportunity to study the behavior of young outflows in a wide range of mass in Galactic massive protoclusters. Several authors reported that outflows associated with massive star formation are comparatively more powerful than those associated with low-mass star formation \citep{shepherd96, beuther02, wu04, hatchell07, yang18, li18}. 
While outflows with $M_{\rm out}$ of 0.1--1 $M_\odot$ and $\dot{M}_{\rm out}$ of 10$^{-7}$--10$^{-6}$ $M_\odot$~yr$^{-1}$ are known as low-mass outflows, the outflows with $M_{\rm out}$ of 10--1000 $M_\odot$ and $\dot{M}_{\rm out}$ of 10$^{-5}$--10$^{-3}$ $M_\odot$~yr$^{-1}$ are considered as high-mass outflows \citep{yang18}. Thus, a wide spread in the values of physical parameters for our observed outflows indicates that these correspond to low-mass to high-mass outflows.
Note that the targets in this study are massive star-forming regions. We indeed found at least one massive outflow with $\dot{M}_\mathrm{out}$ more than about 10$^{-4}$ $M_\odot$~yr$^{-1}$ in almost every region (except for I14382 and I17204). 

A histogram for the dynamical age of the outflows are also shown in Figure~\ref{fig4}b. The dynamical ages of the outflows in our sample are ranging from 0.1--2.8$\times$10$^4$ years. These results confirm their driving sources to be young and still might be in an early phase when the matter is accreted onto the protostellar condensation. 

\begin{table*}
\scriptsize
\centering
\caption{Outflow parameters (listed only those detected in CO) derived from CO}
\label{table1}
\begin{tabular}{lccrrrrrrrrrrrrrrccc} 
\hline
	Lobe Name & \multicolumn{2}{c}{$v$ (km s$^{-1}$)}  & \multicolumn{2}{c}{$M_{\rm out}$}          &          \multicolumn{2}{c}{$P_{\rm out}$ (10$^{-1}$}  &      \multicolumn{2}{c}{$E_{\rm out}$}          & \multicolumn{2}{c}{$t_\mathrm{dyn}$} & \multicolumn{2}{c}{$\dot{M}_\mathrm{out}$ (10$^{-5}$} & \multicolumn{2}{c}{$F_\mathrm{out}$ (10$^{-4}$ M$_\odot$} & \multicolumn{2}{c}{$L_\mathrm{mech}$} & \multicolumn{2}{c}{Detected in} & WS$^a$ \\
		    &                   &                    & \multicolumn{2}{c}{(10$^{-2}$ $M_\odot$)} & \multicolumn{2}{c}{ M$_\odot$ km s$^{-1}$)} & \multicolumn{2}{c}{($M_\odot$ km$^2$ s$^{-2}$)} & \multicolumn{2}{c}{(10$^3$ yr)}      & \multicolumn{2}{c}{ M$_\odot$ yr$^{-1}$) } & \multicolumn{2}{c}{km s$^{-1}$ yr$^{-1}$) } & \multicolumn{2}{c}{(10$^{-1} L_\odot$)} & HCN & HCO$^{+}$ & \\
	& Blue  &   Red  &  Blue  &  Red  &  Blue  &  Red   &  Blue  &   Red  &  Blue  &  Red  &  Blue  & Red  &  Blue  & Red  &  Blue  & Red & B~~~R & B~~~R & \\
\hline
 I14382\_O1a &  [-103,  -65] & [ -55,  -14] &      0.5 &      0.2 &       1.8 &       0.6 &      4.2 &      1.5 &      0.9 &      0.5 &      0.5 &      0.3 &      2.0 &      1.2 &      3.9 &      2.3  & \cmark~~\cmark & \xmark~~~\xmark & Y\\ 
 I14382\_O1b &            -- & [ -55,   -8] &       -- &      0.2 &        -- &       1.1 &       -- &      3.5 &       -- &      1.7 &       -- &      0.1 &       -- &      0.6 &       -- &      1.7 & --~~\cmark & --~~~\xmark & N\\ 
 I14382\_O2 &  [-129,  -65] & [ -55,    6] &      1.0 &      0.5 &       3.7 &       2.2 &     10.3 &      6.3 &      0.6 &      0.6 &      1.7 &      0.8 &      6.1 &      3.7 &     14.1 &      8.7 & \cmark~~\cmark & \cmark~~\cmark & Y\\ 
 I14382\_O3 &            -- & [ -55,  -32] &       -- &      0.1 &        -- &       0.2 &       -- &      0.4 &       -- &      1.7 &       -- &      0.0 &       -- &      0.1 &       -- &      0.2 & --~~\cmark & --~~~\xmark & N\\
 I14498\_O1a &            -- & [ -45,   65] &       -- &      8.1 &        -- &      23.0 &       -- &     73.3 &       -- &      1.4 &       -- &      5.9 &       -- &     16.6 &       -- &     43.7 & --~~\cmark & --~~~\xmark & y\\ 
 I14498\_O1b &  [-112,  -55] &           -- &     16.8 &       -- &      43.9 &        -- &     77.2 &       -- &      1.8 &       -- &      9.2 &       -- &     24.1 &       -- &     35.1 &       -- & \cmark~~-- & \cmark~~-- & y\\ 
 I14498\_O2 &  [-116,  -55] & [ -45,   11] &      5.7 &      0.8 &      15.5 &       2.0 &     31.8 &      4.3 &      0.5 &      0.5 &     11.5 &      1.5 &     31.7 &      3.8 &     53.6 &      6.9 & \xmark~~~\xmark & \xmark~~~\xmark & N\\ 
 I14498\_O3 &            -- & [ -45,  -34] &       -- &      2.4 &        -- &       3.5 &       -- &      2.8 &       -- &      6.9 &       -- &      0.3 &       -- &      0.5 &       -- &      0.3 & --~~\cmark & --~~~\xmark & y\\ 
 I14498\_O4 &            -- & [ -45,  -28] &       -- &      1.5 &        -- &       2.6 &       -- &      2.6 &       -- &      3.7 &       -- &      0.4 &       -- &      0.7 &       -- &      0.6 & --~~~\xmark & --~~~\xmark & y\\ 
 I15520\_O1a &  [ -70,  -46] & [ -36,  -16] &      0.9 &      5.8 &       2.6 &      11.4 &      4.3 &     13.0 &      4.0 &      7.0 &      0.2 &      0.8 &      0.7 &      1.6 &      0.9 &      1.5 & \cmark~~\cmark & \cmark~~\cmark & Y\\ 
 I15520\_O2a &            -- & [ -36,   -2] &       -- &      0.1 &        -- &       0.4 &       -- &      0.9 &       -- &      4.2 &       -- &      0.0 &       -- &      0.1 &       -- &      0.2 & --~~~\xmark & --~~~\xmark & y\\ 
I15520\_O2b &  [ -70,  -46] &           -- &      8.0 &       -- &      21.8 &        -- &     33.2 &       -- &      4.5 &       -- &      1.8 &       -- &      4.8 &       -- &      6.0 &       -- & \cmark~~-- & \cmark~~-- & y\\ 
I15520\_O3 &  [ -61,  -46] &           -- &      2.0 &       -- &       4.3 &        -- &      4.9 &       -- &      7.3 &       -- &      0.3 &       -- &      0.6 &       -- &      0.6 &       -- & \cmark~~-- & \cmark~~-- & N\\ 
I15520\_O4 &  [ -99,  -46] &           -- &      1.2 &       -- &       7.9 &        -- &     30.3 &       -- &      2.7 &       -- &      0.4 &       -- &      2.9 &       -- &      9.2 &       -- & \xmark~~~-- & \xmark~~~-- & N\\ 
I15520\_O5 &  [-106,  -46] &           -- &      3.0 &       -- &      11.6 &        -- &     32.4 &       -- &      1.8 &       -- &      1.7 &       -- &      6.6 &       -- &     15.2 &       -- & \cmark~~-- & \cmark~~-- & N\\ 
I15596\_O1a &  [-119,  -79] & [ -69,  -36] &      5.2 &      7.0 &      15.3 &      19.4 &     30.1 &     35.4 &      5.4 &      7.2 &      1.0 &      1.0 &      2.8 &      2.7 &      4.6 &      4.1 & \cmark~~\cmark & \xmark~~~\cmark & N \\ 
I15596\_O1b &  [-125,  -79] & [ -69,  -44] &      6.7 &      2.2 &      18.6 &       4.8 &     36.0 &      6.5 &      2.9 &      3.7 &      2.3 &      0.6 &      6.5 &      1.3 &     10.4 &      1.4 & \cmark~~\cmark & \xmark~~~\cmark & N \\ 
I15596\_O2a &  [-122,  -79] & [ -69,   10] &      3.6 &      5.1 &      10.5 &      18.4 &     20.6 &     59.4 &      0.8 &      0.6 &      4.4 &      9.1 &     13.0 &     32.8 &     21.0 &     87.8 & \cmark~~\cmark & \cmark~~\cmark & N\\ 
I15596\_O2b &  [-155,  -79] & [ -69,   -5] &      1.6 &      3.0 &       7.0 &       9.5 &     24.9 &     26.1 &      0.3 &      0.4 &      5.9 &      7.6 &     25.6 &     24.4 &     75.3 &     55.7 & \cmark~~\cmark & \cmark~~\cmark & N\\ 
I15596\_O3 &            -- & [ -69,  -22] &       -- &      1.7 &        -- &       5.5 &       -- &     14.5 &       -- &      1.4 &       -- &      1.2 &       -- &      4.0 &       -- &      8.6 & --~~~\xmark & --~~~\xmark & y\\ 
I15596\_O4 &  [-113,  -79] & [ -69,  -28] &      3.4 &      1.1 &       9.3 &       2.7 &     16.6 &      5.5 &      3.0 &      1.6 &      1.1 &      0.7 &      3.1 &      1.8 &      4.6 &      2.9 & \cmark~~\cmark & \cmark~~\cmark & Y\\ 
I15596\_O5 &  [-166,  -79] & [ -69,   26] &      7.0 &      6.8 &      45.8 &      51.0 &    225.1 &    266.6 &      1.7 &      1.0 &      4.1 &      7.0 &     26.6 &     52.9 &    108.4 &    228.8 & \xmark~~~\cmark & \xmark~~~\cmark & Y\\ 
I15596\_O6 &  [-122,  -79] & [ -69,   -5] &      3.8 &      0.5 &       9.6 &       3.2 &     16.6 &     14.4 &      1.8 &      1.3 &      2.1 &      0.4 &      5.3 &      2.5 &      7.6 &      9.2 & \cmark~~\cmark & \cmark~~\cmark & Y\\ 
I15596\_O7 &  [-104,  -79] &           -- &      0.0 &       -- &       0.1 &        -- &      0.3 &       -- &      2.4 &       -- &      0.0 &       -- &      0.1 &       -- &      0.1 &       -- & \xmark~~~-- & \xmark~~~-- & y\\ 
I15596\_O8 &  [-153,  -79] &           -- &      5.6 &       -- &      20.4 &        -- &     55.6 &       -- &      1.3 &       -- &      4.2 &       -- &     15.3 &       -- &     34.5 &       -- & \cmark~~-- & \cmark~~-- & y\\ 
I16060\_O1 &  [-120,  -97] & [ -81,  -52] &     63.3 &     35.4 &     208.2 &     120.8 &    373.7 &    227.1 &      4.1 &      7.6 &     15.6 &      4.6 &     51.3 &     15.8 &     76.3 &     24.6 & \cmark~~\cmark & \cmark~~\cmark & N\\ 
I16060\_O2 &  [-136,  -97] &           -- &     69.7 &       -- &     274.3 &        -- &    606.8 &       -- &      6.1 &       -- &     11.4 &       -- &     44.8 &       -- &     82.2 &       -- & \cmark~~-- & \cmark~~-- & N\\ 
I16060\_O3 &            -- & [ -81,  -53] &       -- &      3.3 &        -- &       9.3 &       -- &     15.2 &       -- &      6.7 &       -- &      0.5 &       -- &      1.4 &       -- &      1.9 & --~~~\xmark & --~~~\xmark & y\\ 
I16060\_O4 &            -- & [ -81,  -66] &       -- &      0.8 &        -- &       2.1 &       -- &      2.9 &       -- &     15.9 &       -- &      0.1 &       -- &      0.1 &       -- &      0.2 & --~~~\xmark & --~~~\xmark & y\\ 
I16071\_O1a &  [-139,  -94] & [ -78,  -19] &      3.9 &     18.1 &      19.3 &      63.8 &     56.4 &    155.7 &      4.3 &      5.1 &      0.9 &      3.5 &      4.5 &     12.5 &     10.8 &     25.2 & \cmark~~\cmark & \cmark~~\cmark & N\\ 
I16071\_O1b &  [-118,  -94] & [ -78,  -58] &     11.1 &      8.3 &      35.2 &      22.6 &     64.6 &     35.1 &      6.7 &      5.3 &      1.6 &      1.6 &      5.2 &      4.2 &      8.0 &      5.5 & \cmark~~\cmark & \xmark~~~\cmark & N\\ 
I16071\_O1c &  [-122,  -94] & [ -78,  -37] &      4.1 &     12.4 &      11.3 &      36.2 &     18.8 &     64.1 &      2.5 &      2.5 &      1.6 &      4.9 &      4.5 &     14.3 &      6.2 &     21.0 & \cmark~~\cmark & \cmark~~\cmark & Y\\ 
I16071\_O1d &  [-225,  -94] &           -- &     40.0 &       -- &     313.4 &        -- &   2017.0 &       -- &      1.1 &       -- &     37.8 &       -- &    295.7 &       -- &   1576.3 &       -- & \xmark~~~-- & \xmark~~~-- & y\\ 
I16071\_O1e &  [-117,  -94] &           -- &     12.0 &       -- &      33.5 &        -- &     53.1 &       -- &      3.8 &       -- &      3.2 &       -- &      8.8 &       -- &     11.6 &       -- & \cmark~~-- & \cmark~~-- & y \\ 
I16071\_O2 &  [-192,  -94] & [ -78,   58] &      4.8 &      3.8 &      41.2 &      32.5 &    241.3 &    245.1 &      0.3 &      0.2 &     18.6 &     16.7 &    159.3 &    142.1 &    772.5 &    887.9 & \cmark~~\cmark & \xmark~~~\cmark & Y \\ 
I16071\_O3 &  [-111,  -94] & [ -78,  -58] &      6.5 &      1.8 &      19.4 &       4.5 &     31.7 &      6.2 &      3.4 &      7.5 &      1.9 &      0.2 &      5.8 &      0.6 &      7.8 &      0.7 & \cmark~~\cmark & \cmark~~\cmark & Y \\ 
I16071\_O4 &  [-102,  -94] &           -- &      2.3 &       -- &       5.0 &        -- &      5.8 &       -- &      9.8 &       -- &      0.2 &       -- &      0.5 &       -- &      0.5 &       -- & \cmark~~-- & \cmark~~-- & y \\ 
I16071\_O5 &            -- & [ -78,  -58] &       -- &      1.2 &        -- &       3.3 &       -- &      5.3 &       -- &      3.7 &       -- &      0.3 &       -- &      0.9 &       -- &      1.2 & --~~~\xmark  & --~~~\xmark & y \\ 
I16071\_O6a &  [-177,  -94] & [ -78,  -56] &      4.4 &      0.8 &      25.6 &       2.5 &     96.3 &      4.2 &      0.3 &      1.0 &     14.0 &      0.8 &     82.0 &      2.5 &    255.5 &      3.5 & \cmark~~\cmark & \cmark~~\cmark & N \\ 
I16071\_O6b &  [-148,  -94] & [ -78,   -2] &     10.5 &     13.7 &      59.8 &      71.8 &    202.9 &    275.0 &      1.4 &      1.0 &      7.8 &     13.2 &     44.2 &     69.1 &    124.2 &    219.2 & \cmark~~\cmark & \cmark~~\cmark & N \\ 
I16071\_O7 &  [-173,  -94] &           -- &     11.1 &       -- &      61.9 &        -- &    262.4 &       -- &      1.6 &       -- &      7.0 &       -- &     38.8 &       -- &    136.2 &       -- & \cmark~~-- & \cmark~~-- & y \\ 
I16076\_O1a &  [-125,  -97] & [ -77,   -5] &      4.3 &      6.3 &      16.1 &      37.9 &     33.2 &    155.3 &      1.2 &      1.0 &      3.5 &      6.2 &     13.3 &     37.8 &     22.7 &    128.3 & \cmark~~\cmark & \cmark~~\cmark & N \\ 
I16076\_O1b &  [-111,  -97] & [ -77,   20] &      7.5 &      4.3 &      23.3 &      21.4 &     37.9 &     78.0 &      2.7 &      0.8 &      2.7 &      5.3 &      8.5 &     26.3 &     11.4 &     79.5 & \cmark~~\cmark & \xmark~~~\cmark & N \\ 
I16076\_O1c &            -- & [ -77,  -22] &       -- &      5.7 &        -- &      25.7 &       -- &     75.7 &       -- &      1.5 &       -- &      3.9 &       -- &     17.4 &       -- &     42.4 & --~~\cmark & --~~\cmark & N \\ 
I16076\_O1d &  [-171,  -97] & [ -77,  -10] &      0.8 &      1.9 &       4.1 &      10.0 &     15.2 &     32.5 &      0.8 &      0.5 &      1.1 &      3.8 &      5.3 &     19.6 &     16.3 &     52.5 & \xmark~~~\xmark & \xmark~~~\xmark & N \\ 
I16076\_O1e &            -- & [ -77,  -55] &       -- &      0.5 &        -- &       1.4 &       -- &      2.3 &       -- &      3.2 &       -- &      0.1 &       -- &      0.4 &       -- &      0.6 & --~~\cmark & --~~~\xmark & y \\ 
I16076\_O1f &            -- & [ -77,  -55] &       -- &      0.1 &        -- &       0.3 &       -- &      0.5 &       -- &      3.8 &       -- &      0.0 &       -- &      0.1 &       -- &      0.1 & --~~~\xmark & --~~~\xmark & y \\ 
I16076\_O1g &  [-147,  -97] & [ -77,  -55] &      1.8 &      1.5 &       7.8 &       4.3 &     21.9 &      6.5 &      1.0 &      3.6 &      1.8 &      0.4 &      8.0 &      1.2 &     18.6 &      1.5 & \cmark~~\cmark & \cmark~~\cmark & N \\ 
I16076\_O1h &  [-115,  -97] & [ -77,  -58] &      1.2 &      0.5 &       3.6 &       1.4 &      5.6 &      2.1 &      4.1 &      3.3 &      0.3 &      0.1 &      0.9 &      0.4 &      1.1 &      0.5 & \cmark~~\xmark & \xmark~~~\xmark & Y \\ 
I16076\_O1i &  [-140,  -97] &           -- &      2.1 &       -- &      10.2 &        -- &     28.7 &       -- &      2.2 &       -- &      1.0 &       -- &      4.7 &       -- &     10.9 &       -- & \cmark~~-- & \cmark~~-- & N \\ 
I16076\_O1j &  [-116,  -97] &           -- &      2.6 &       -- &       9.1 &        -- &     17.1 &       -- &      3.2 &       -- &      0.8 &       -- &      2.8 &       -- &      4.4 &       -- & \xmark~~~-- & \xmark~~~-- & N \\ 
I16076\_O1k &  [-136,  -97] &           -- &      0.9 &       -- &       2.9 &        -- &      5.6 &       -- &      1.3 &       -- &      0.7 &       -- &      2.2 &       -- &      3.6 &       -- & \xmark~~~-- & \xmark~~~-- & N \\ 
I16076\_O1l &  [-119,  -97] &           -- &      0.2 &       -- &       0.8 &        -- &      1.5 &       -- &      2.3 &       -- &      0.1 &       -- &      0.3 &       -- &      0.6 &       -- & \cmark~~-- & \xmark~~~-- & N \\ 
I16076\_O2 &            -- & [ -77,  -55] &       -- &      3.4 &        -- &      10.8 &       -- &     18.6 &       -- &      7.2 &       -- &      0.5 &       -- &      1.5 &       -- &      2.1 & ~~~~~~\cmark & --~~~\xmark & y \\ 
I16076\_O3a &  [-120,  -97] &           -- &      3.2 &       -- &       9.3 &        -- &     14.9 &       -- &      5.2 &       -- &      0.6 &       -- &      1.8 &       -- &      2.4 &       -- & \xmark~~~--  & \xmark~~~-- & y \\ 
I16076\_O4a &            -- & [ -77,  -54] &       -- &      3.2 &        -- &      12.2 &       -- &     25.3 &       -- &      2.7 &       -- &      1.2 &       -- &      4.6 &       -- &      7.9 & --~~~\xmark & --~~~\xmark & N \\ 
I16076\_O4b &  [-121,  -97] &           -- &      0.4 &       -- &       1.4 &        -- &      2.8 &       -- &      1.3 &       -- &      0.3 &       -- &      1.1 &       -- &      1.8 &       -- & \xmark~~~-- & \xmark~~~-- & N \\ 
I16076\_O5 &  [-107,  -97] &           -- &      0.2 &       -- &       0.7 &        -- &      1.0 &       -- &      4.5 &       -- &      0.1 &       -- &      0.1 &       -- &      0.2 &       -- & \xmark~~~-- & \xmark~~~-- & y \\ 
 I16272\_O1a &  [ -84,  -54] & [ -38,   18] &     13.1 &     15.6 &      40.3 &      62.8 &     72.5 &    173.7 &      3.2 &      2.4 &      4.1 &      6.6 &     12.7 &     26.6 &     18.9 &     60.9 & \cmark~~\cmark & \cmark~~\cmark & Y \\ 
 I16272\_O1b &  [ -63,  -54] & [ -38,  -27] &      0.8 &      0.5 &       1.7 &       1.1 &      1.9 &      1.3 &     11.1 &      7.0 &      0.1 &      0.1 &      0.2 &      0.2 &      0.1 &      0.2 & \cmark~~\cmark & \cmark~~\cmark & Y \\ 
 I16272\_O1c &            -- & [ -38,  -18] &       -- &      0.2 &        -- &       0.7 &       -- &      1.2 &       -- &      2.4 &       -- &      0.1 &       -- &      0.3 &       -- &      0.4 & ~~~~~~\cmark & --~~~\xmark & y \\ 
  I16272\_O2 &            -- & [ -38,   -2] &       -- &      3.2 &        -- &       9.3 &       -- &     15.9 &       -- &      2.4 &       -- &      1.3 &       -- &      3.8 &       -- &      5.4 & --~~~\xmark & --~~~\xmark & y \\ 
 I16351\_O1a &  [ -58,  -45] & [ -35,   48] &     13.7 &     32.3 &      30.7 &     202.6 &     36.8 &    833.8 &      2.7 &      0.6 &      5.1 &     52.6 &     11.5 &    329.7 &     11.4 &   1124.3 & \cmark~~\cmark & \cmark~~\cmark & Y\\ 
 I16351\_O1b &  [ -72,  -45] &           -- &      1.8 &       -- &       4.5 &        -- &      6.8 &       -- &      4.2 &       -- &      0.4 &       -- &      1.1 &       -- &      1.3 &       -- & \xmark~~~-- & \xmark~~~-- & y \\ 
 I17204\_O1a &  [ -54,  -25] & [  -9,    0] &      2.4 &      0.5 &       7.4 &       1.1 &     12.7 &      1.2 &      1.9 &      4.5 &      1.3 &      0.1 &      3.9 &      0.2 &      5.5 &      0.2 & \cmark~~\cmark & \cmark~~\cmark & N \\ 
 I17204\_O1b &  [ -46,  -25] &           -- &      0.9 &       -- &       2.7 &        -- &      4.8 &       -- &      3.4 &       -- &      0.2 &       -- &      0.8 &       -- &      1.2 &       -- & \cmark~~-- & \cmark~~-- & N \\ 
 I17204\_O1c &  [ -49,  -25] &           -- &      2.0 &       -- &       7.3 &        -- &     14.6 &       -- &      3.1 &       -- &      0.6 &       -- &      2.3 &       -- &      3.9 &       -- & \cmark~~-- & \cmark~~-- & N \\ 
 I17204\_O1d &  [ -38,  -25] &           -- &      5.7 &       -- &      14.9 &        -- &     20.1 &       -- &      8.6 &       -- &      0.7 &       -- &      1.7 &       -- &      1.9 &       -- & \xmark~~~-- & \cmark~~-- & y \\ 
  I17220\_O1 &  [-123, -102] & [ -86,  -69] &      2.4 &      3.0 &       7.2 &       8.8 &     12.1 &     14.0 &      4.2 &      4.3 &      0.6 &      0.7 &      1.7 &      2.0 &      2.4 &      2.7 & \cmark~~\cmark & \cmark~~\cmark & Y \\ 
  I17220\_O2 &  [-140, -102] & [ -86,    0] &     11.6 &     10.2 &      38.5 &      60.7 &     80.1 &    286.0 &      3.5 &      1.7 &      3.3 &      6.0 &     10.8 &     35.7 &     18.7 &    139.2 & \cmark~~\cmark & \xmark~~~\xmark & Y\\ 
  I17220\_O3 &  [-146, -102] & [ -86,  -62] &     32.8 &      1.4 &     114.8 &       4.0 &    249.7 &      6.8 &      3.0 &      5.6 &     11.0 &      0.3 &     38.6 &      0.7 &     69.5 &      1.0 & \cmark~~\cmark & \cmark~~\xmark & Y \\ 
  I17220\_O4 &  [-143, -102] & [ -86,  -64] &     15.9 &      2.6 &      58.6 &       6.6 &    135.2 &      9.6 &      2.9 &      2.6 &      5.5 &      1.0 &     20.5 &      2.5 &     39.1 &      3.0 & \cmark~~\cmark & \cmark~~\cmark & Y \\ 
  I17220\_O5 &            -- & [ -86,  -67] &       -- &      0.4 &        -- &       1.2 &       -- &      1.9 &       -- &      1.6 &       -- &      0.3 &       -- &      0.7 &       -- &      1.0 & --~~~\xmark & --~~~\xmark & y \\ 
  I17220\_O6 &            -- & [ -86,  -63] &       -- &      0.6 &        -- &       1.9 &       -- &      3.5 &       -- &      1.5 &       -- &      0.4 &       -- &      1.3 &       -- &      2.0 & --~~\cmark & --~~\cmark & y \\
  \hline
\multicolumn{20}{l}{\footnotesize $^a$ WS imples {\it well-separated} outflows without contamination from nearby lobes; Y and y indicates the bipolar and unipolar outflows, respectively.}\\
\end{tabular}
\end{table*}

\subsection{Mass of the host cores}
\label{SecCoreMass}
Smaller structures within molecular clouds are defined as clumps, cores, and substructures of a core based on their spatial extent in dust continuum and line emission maps. In this paper, we adopted the terminology recommended by \citet{williams00} and \citet{wang14}: clumps have sizes $\sim$1 pc, dense cores have $\sim$0.1 pc, and substructures within cores have sizes $\sim$0.01 pc. Small scale structures in 0.9 mm continuum ALMA data were identified using the Python-based {\sc astrodendro}-package\footnote{https://dendrograms.readthedocs.io/en/stable/index.html} which uses the {\sc dendrogram} algorithm to identify the hierarchical structures (i.e., cores) in the continuum maps. A detailed discussion of the algorithm can be found in \citet{rosolowsky08}. The {\sc astrodendro}-package needs the inputs -- the minimum flux level and the minimum number of pixels -- to be considered. A minimum flux level of 3$\sigma$ was adopted as a good value where $\sigma$ is the background flux estimated from the emission-free areas in the continuum map. We considered structures having an area of more than the beam size of 25 pixels (corresponding to a spatial scale of $\sim$0.007--0.02 pc at distances of 2.6--7.6 kpc) which is sufficient to identify dense cores. A detailed analysis of the cores will be presented in a future paper. In this paper, we only present the parameters (i.e., outflow name, coordinates of the continuum source, size, position angle, and integrated flux) of those 41 cores that are associated with outflows (see Table~\ref{table2}). These cores have typical sizes in the range from 0.01--0.06 pc which confirms them to be cores or smaller substructures within cores. However, from now onward we shall refer to them as cores. 

We further estimated the dust mass of each core assuming the dust emission to be optically thin using the following equation :
\begin{equation}
	M_{\rm{dust}} = \frac{F_\nu d^2}{B_\nu(T_{\rm dust}) \kappa_\nu},
\end{equation}
where $F_\nu$ is the total dust continuum flux of the identified core at frequency $\nu$, $d$ is the distance to the source, $B_\nu(T_{\rm dust})$ is the Planck function at the temperature $T_{\rm dust}$, and $\kappa_\nu$ = 10($\frac{\nu}{1.2 THz}$)$^\beta$ cm$^2$ g$^{-1}$ is the dust opacity \citep{hildebrand83} where $\beta$ is the spectral index. As mentioned before the targets here are massive star-forming regions and are associated with H{\sc ii} regions. In such an environment, the cores in the vicinity of H{\sc ii} regions could be externally heated to a high temperature of more than 100~K \citep[see e.g.,][and references therein]{osorio99, mookerjea07}. Conversely, in the case of a cold star-forming core, $T_{\rm dust}$ is typically assumed to be $\lesssim$30~K \citep[e.g.,][]{plunkett13, wang14}. Hence, in the absence of a line-temperature measurement, we safely considered a dust temperature ($T_{\rm dust}$) of 50 K to calculate masses of the cores in these regions. The estimated mass of the cores for two dust emissivity spectral indices (i.e., $\beta =$ 1.5 and 2.0) are listed in Table~\ref{table2} along with the other details of the identified cores. The number of outflow lobes associated with each core is also listed in the last column of Table~\ref{table2}. The mass estimated using $\beta =$ 2.0 is approximately a factor of 2 higher compared to the mass estimated using $\beta =$ 1.5.
 
A histogram for the core mass considering the $T_{\rm dust}$ of 50 K and $\beta$ of 2, is shown in Figure~\ref{fig4}c. The mass of the cores is typically up to a few tens of solar masses with a couple of sources more than one hundred solar masses. Note that the mass is estimated here assuming the emission to be optically thin which might not always hold. There is also a possibility that multiple cores are located along the line of sight, and together they may appear as a single core with considerably high mass. Such a possibility cannot be ignored given the fact that several cores in our sample drive multiple outflows (see Table~\ref{table2}), indicating the presence of multiple outflow driving sources within 0.1 pc scale. Hence, estimated core mass with such a large uncertainty may show a spurious relation when compared with the outflow parameters. 
To reduce such possibility, we defined a cleaned set of {\it well-defined} outflows (discussed in Section~\ref{SecComOutAndCores}). This definition of {\it well-defined} outflows rejects the most massive cores in our sample which hold the highest possibility of having a companion. Although this precaution helps us to reduce contamination, we still cannot rule out the possibility of multiplicity in the identified cores as the regions are located considerably higher distances. There still might exist unresolved multiples. In an unresolved multiple, a single core may drive outflows while its companion(s) might remain inactive. Thus, the estimated core mass should be considered as an upper limit, and any derived relation or trend of outflow parameters with core masses should be treated with caution.

\begin{table*}
\centering
\caption{Details of the driving cores}
\label{table2}
\begin{tabular}{rcclrrrccc} 
\hline
Outflow & $\alpha$ (J2000) & $\delta$ (J2000)      &     Source Size             & $F_\mathrm{int}$ & \multicolumn{2}{c}{$M_\mathrm{Core, 50 K}$ ($M_\odot$)}             & No. of & Well-Defined \\ 
Name	&  (h m s)         &   ($^\circ ~ ' ~''$ ) & pc$\times$pc, PA ($^\circ$) &  (mJy)           & \multicolumn{1}{c}{$\beta$ = 1.5}&\multicolumn{1}{c}{$\beta$ = 2.0} & lobes  & Outflow?   \\
 \hline
	I14382\_O1 & 14 42 02.106 & -60 30 44.590 &  0.05$\times$0.03,   89 &  416.7 &   14.0 &   26.2 & 3 & N \\ 
	I14382\_O2 & 14 42 03.066 & -60 30 26.020 &  0.01$\times$0.01,  163 &   34.4 &    1.2 &    2.2 & 2 & Y \\ 
	I14382\_O3 & 14 42 02.833 & -60 30 49.990 &  0.01$\times$0.01,  132 &    3.9 &    0.1 &    0.2 & 1 & N \\ 
	I14498\_O1 & 14 53 42.681 & -59 08 52.880 &  0.02$\times$0.01,  177 & 1034.7 &   21.2 &   39.6 & 2 & Y \\ 
	I14498\_O2 &    **        &      **       &         --              &    --  &    --  &    --  & 2 & N \\ 
	I14498\_O3 & 14 53 43.579 & -59 08 43.780 &  0.04$\times$0.02, -135 &   49.0 &    1.0 &    1.9 & 1 & Y \\ 
	I14498\_O4 & 14 53 42.941 & -59 09 00.870 &  0.03$\times$0.02,   87 &  164.6 &    3.4 &    6.3 & 1 & Y \\ 
	I15520\_O1 & 15 55 48.398 & -52 43 06.530 &  0.01$\times$0.00, -158 &  133.1 &    1.8 &    3.4 & 2 & Y \\ 
	I15520\_O2 & 15 55 48.654 & -52 43 08.660 &  0.01$\times$0.01,  172 &  433.4 &    5.9 &   10.9 & 2 & N \\ 
	I15520\_O3 & 15 55 48.393 & -52 43 04.380 &  0.02$\times$0.01,   76 & 1102.8 &   14.9 &   27.8 & 1 & N \\ 
	I15520\_O4 & 15 55 48.848 & -52 43 01.610 &  0.01$\times$0.00,   98 &   22.9 &    0.3 &    0.6 & 1 & N \\ 
	I15520\_O5 & 15 55 49.265 & -52 43 03.020 &  0.01$\times$0.01, -146 &   21.2 &    0.3 &    0.5 & 1 & N \\ 
	I15596\_O1 & 16 03 31.921 & -53 09 22.960 &  0.01$\times$0.01,   47 &   63.3 &    2.5 &    4.6 & 4 & N \\ 
	I15596\_O2 & 16 03 32.646 & -53 09 26.820 &  0.02$\times$0.01,  156 &  108.5 &    4.2 &    7.8 & 4 & N \\ 
	I15596\_O3 & 16 03 32.656 & -53 09 45.760 &  0.03$\times$0.01,   70 &   12.4 &    0.5 &    0.9 & 1 & Y \\ 
	I15596\_O4 & 16 03 32.705 & -53 09 29.570 &  0.02$\times$0.02,   63 &   86.1 &    3.3 &    6.2 & 2 & Y \\ 
	I15596\_O5 & 16 03 31.697 & -53 09 32.090 &  0.03$\times$0.02, -176 &  253.7 &    9.8 &   18.3 & 2 & Y \\ 
	I15596\_O6 &    **        &      **       &         --              &    --  &    --  &    --  & 2 & Y \\ 
	I15596\_O7 & 16 03 32.927 & -53 09 27.850 &  0.01$\times$0.01, -155 &    3.8 &    0.1 &    0.3 & 1 & Y \\ 
	I15596\_O8 & 16 03 30.635 & -53 09 33.990 &  0.02$\times$0.01,  159 &   28.9 &    1.1 &    2.1 & 1 & Y \\ 
	I16060\_O1 & 16 09 52.650 & -51 54 54.860 &  0.01$\times$0.01,  165 & 1205.3 &   65.3 &  121.7 & 2 & N \\ 
	I16060\_O2 & 16 09 52.450 & -51 54 55.790 &  0.02$\times$0.02,   72 & 3157.7 &  171.0 &  318.9 & 1 & N \\ 
	I16060\_O3 &    **        &      **       &         --              &    --  &    --  &    --  & 1 & Y \\ 
	I16060\_O4 & 16 09 52.803 & -51 54 57.900 &  0.01$\times$0.01, -170 &  143.1 &    7.8 &   14.4 & 1 & Y \\ 
	I16071\_O1 & 16 10 59.750 & -51 50 23.540 &  0.06$\times$0.04,  106 & 7058.5 &  339.3 &  632.9 & 8 & N \\ 
	I16071\_O2 & 16 10 59.553 & -51 50 27.510 &  0.02$\times$0.01,   53 &   53.7 &    2.6 &    4.8 & 2 & Y \\ 
	I16071\_O3 & 16 10 59.400 & -51 50 16.520 &  0.03$\times$0.02,  160 &   98.2 &    4.7 &    8.8 & 2 & Y \\ 
	I16071\_O4 & 16 11 00.242 & -51 50 26.220 &  0.01$\times$0.01, -170 &    5.2 &    0.2 &    0.5 & 1 & Y \\ 
	I16071\_O5 & 16 10 58.732 & -51 50 36.370 &  0.04$\times$0.01,  175 &   29.2 &    1.4 &    2.6 & 1 & Y \\ 
	I16071\_O6 & 16 10 59.286 & -51 50 11.780 &  0.01$\times$0.01,   71 &    5.6 &    0.3 &    0.5 & 4 & N \\ 
	I16071\_O7 & 16 10 58.742 & -51 50 17.330 &  0.02$\times$0.01,  151 &   10.0 &    0.5 &    0.9 & 1 & Y \\ 
	I16076\_O1 & 16 11 26.540 & -51 41 57.320 &  0.04$\times$0.02,  101 &  858.8 &   43.0 &   80.2 & 17& N \\ 
	I16076\_O2 & 16 11 27.384 & -51 41 50.210 &  0.02$\times$0.01,   50 &   10.2 &    0.5 &    0.9 & 1 & Y \\ 
	I16076\_O3 & 16 11 27.697 & -51 41 55.360 &  0.05$\times$0.02,  153 &  255.7 &   12.8 &   23.9 & 1 & Y \\ 
	I16076\_O4 & 16 11 26.876 & -51 41 55.920 &  0.01$\times$0.01,  122 &   16.4 &    0.8 &    1.5 & 1 & N \\ 
	I16076\_O5 &    **        &      **       &         --              &    --  &    --  &    --  & 1 & Y \\ 
	I16272\_O1 & 16 30 58.770 & -48 43 53.890 &  0.01$\times$0.01, -168 & 2288.4 &   46.9 &   87.5 & 5 & N \\ 
	I16272\_O2 &    **        &      **       &         --              &    --  &    --  &    --  & 1 & Y \\ 
	I16351\_O1 & 16 38 50.501 & -47 28 00.910 &  0.02$\times$0.01,  109 & 2000.8 &   33.7 &   62.8 & 3 & N \\ 
	I17204\_O1 & 17 23 50.249 & -36 38 59.660 &  0.03$\times$0.02,  148 &  698.2 &   11.8 &   21.9 & 5 & N \\ 
	I17220\_O1 & 17 25 25.635 & -36 12 35.120 &  0.04$\times$0.03, -144 &  175.0 &   20.2 &   37.7 & 2 & Y \\ 
	I17220\_O2 & 17 25 24.796 & -36 12 36.850 &  0.03$\times$0.01,   64 &   12.7 &    1.5 &    2.7 & 2 & Y \\ 
	I17220\_O3 & 17 25 24.357 & -36 12 47.890 &  0.03$\times$0.01,  160 &   20.3 &    2.4 &    4.4 & 2 & Y \\ 
	I17220\_O4 & 17 25 24.453 & -36 12 39.360 &  0.03$\times$0.02, -147 &   47.8 &    5.5 &   10.3 & 2 & Y \\ 
	I17220\_O5 & 17 25 24.926 & -36 12 43.440 &  0.02$\times$0.01,  160 &   29.8 &    3.4 &    6.4 & 1 & Y \\ 
	I17220\_O6 & 17 25 25.697 & -36 12 39.480 &  0.06$\times$0.03,  178 &  170.9 &   19.8 &   36.9 & 1 & Y \\
\hline
\multicolumn{7}{l}{$**$ The outflows without any detected continuum source}\\
\end{tabular}
\end{table*}

\section{Discussion}
\label{sec:discussion}
\subsection{Comparison among outflow parameters}
\label{ComOutflowPars}
\begin{figure*}
\includegraphics[width=\columnwidth]{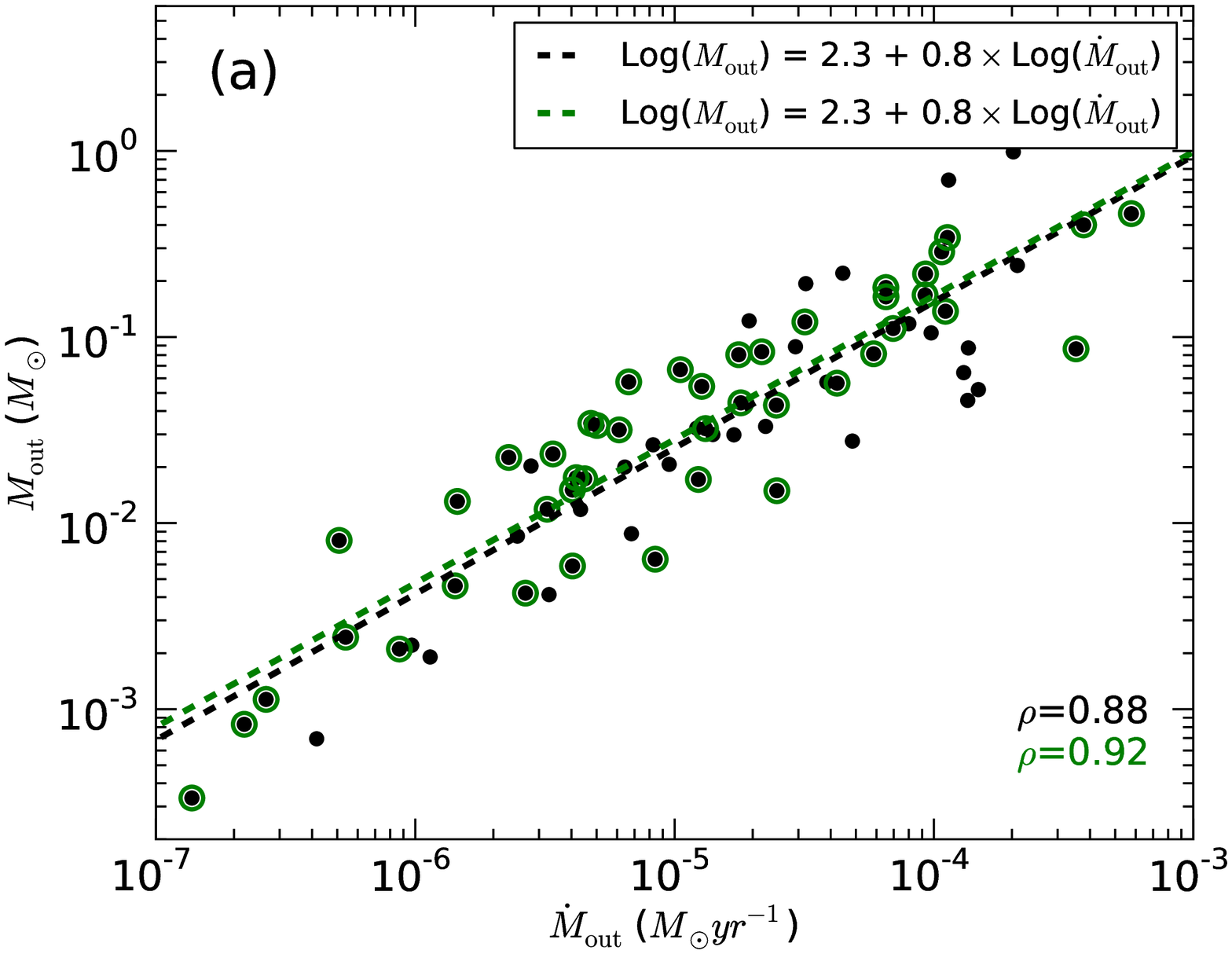}
\includegraphics[width=\columnwidth]{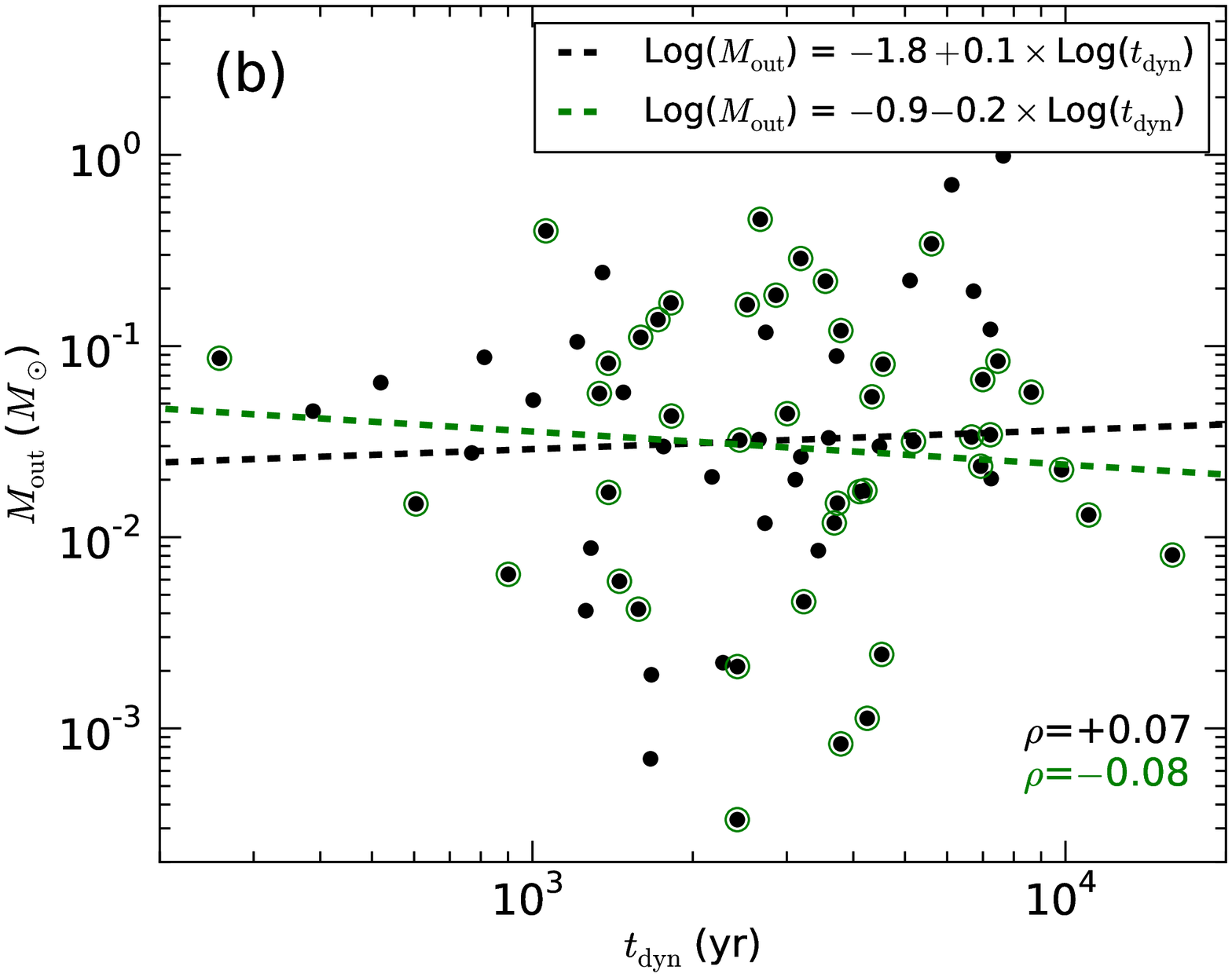}
\includegraphics[width=\columnwidth]{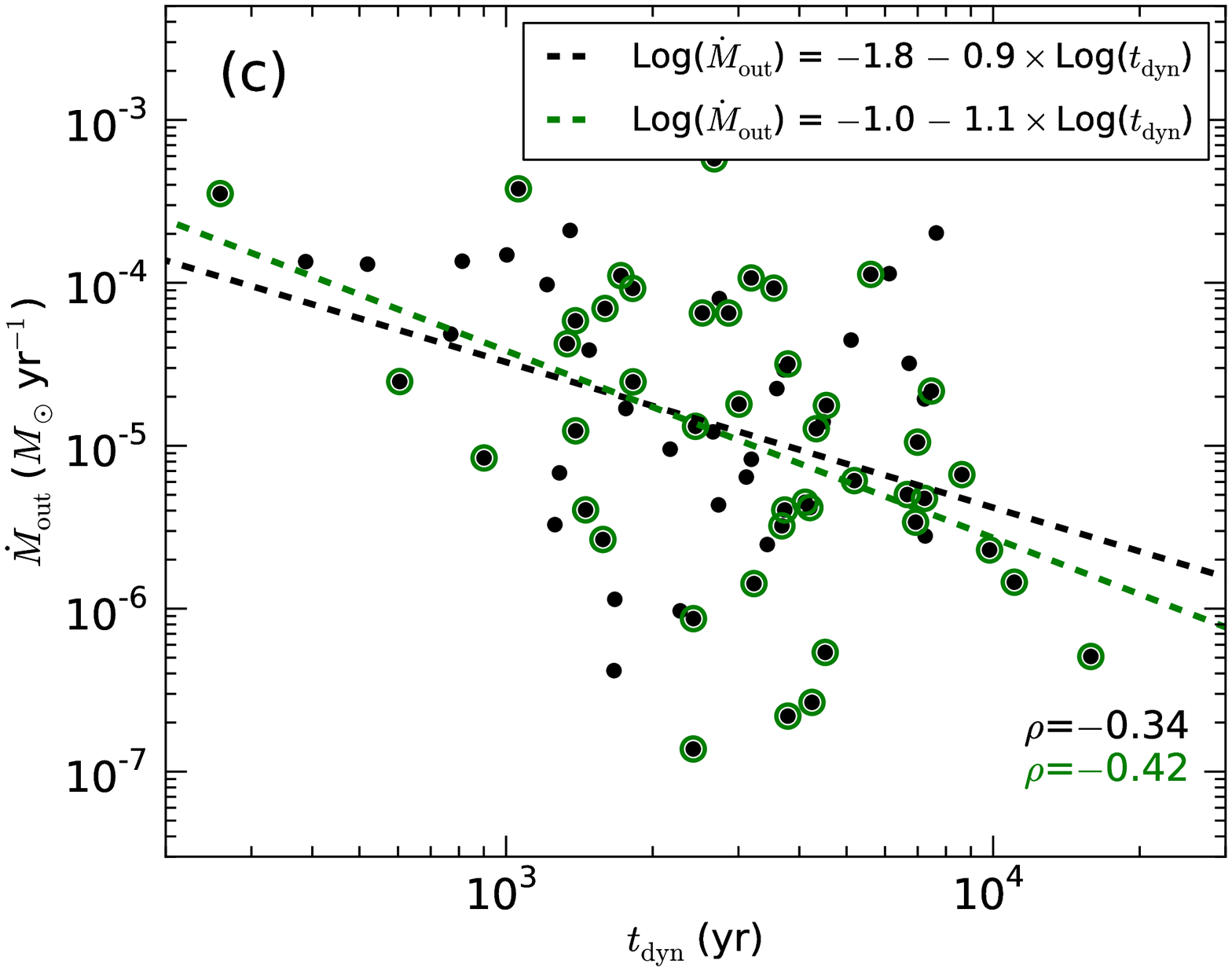}
\includegraphics[width=\columnwidth]{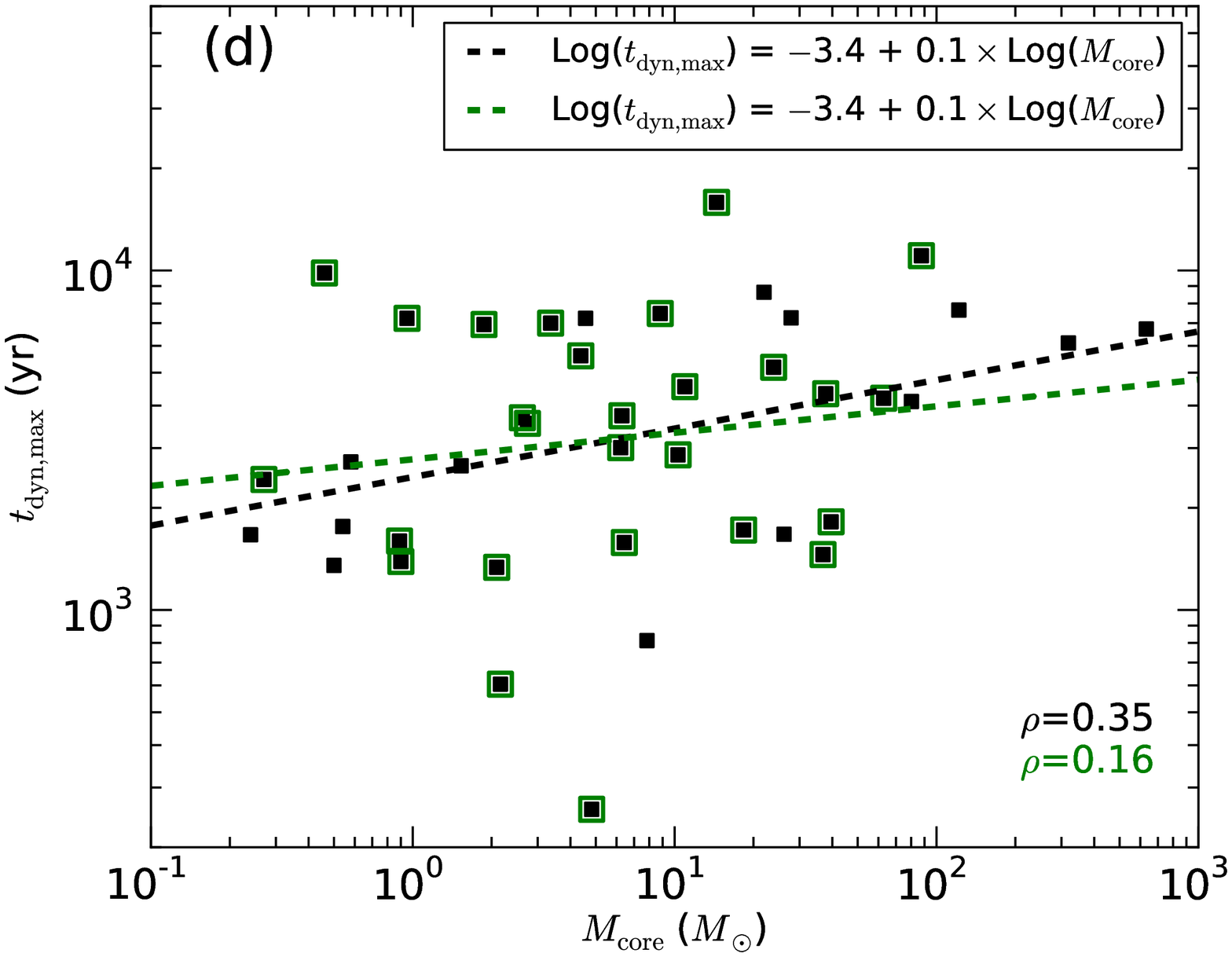}
	\caption{(a) Outflow mass versus the outflow rate for each lobe, (b) outflow mass against the dynamical time-scale of the outflows, (c) outflow rate against the dynamical time-scale of the outflows, and (d) the dynamical time-scale versus the mass of their host cores. The data points marked with green circles/square are for {well-separated} lobes. The black dashed lines in all the panels are the linear least-square fits on a logarithmic scale to all data points while the green dashed lines show the linear least-square fit to the {\it well-separated} data points in logarithminc scale. In each panel, the corresponing Spearman's Rank correlation coefficients ($\rho$) are also written in black and green, respectively.}
\label{fig5}
\end{figure*}

The regions studied here have shown a complex outflow morphology. Thus, estimated outflow parameters might be contamineted by the pixels of the nearby outflow lobes particularly toward the lower velocity ends. Thus, to separate out the contaminated lobes, we examined each region by overlaying the areas used to calculate the outflow parameters for all the identified lobes. Accordingly, we defined a cleaned set of {\it well-separated} outflow lobes which do not have any overlapping pixel (in the PP space) with a neighbouring lobe. The final column of Table~\ref{table1} denotes whether the outflow lobe is {\it well-separated} or not. In Figure~\ref{fig5}a, we have plotted $M_{\rm out}$ versus $\dot{M}_{\rm out}$ for all the lobes. The {\it well-separated} lobes are also marked in the figure.
A nice correlation is noted between $\dot{M}_{\rm out}$ and $M_{\rm out}$ (Figure~\ref{fig5}a) which is possibly expected as a higher gas outflow rate should inject more matter into the surrounding environments. Spearman's rank correlation coefficients ($\rho$) are also marked in the figure for both the sets of data. We further plotted the $M_{\rm out}$ with respect to the $t_{\rm dyn}$ of all the lobes (Figure~\ref{fig5}b). 
A non-correlation ($\rho$ = 0.07 for all and $\rho$ = -0.08 for all {\it well-separated} lobes) can be noted among these two parameters.
A similar non-correlation trend was noted previously by \citet{wu04} for the high-mass clumps. However, they noted an increasing trend of the outflow mass with time for the group of low-mass clumps. This is particularly puzzling for our sample because most of the outflows in this study are low-mass outflows. For further confirmation of this non-correlation, we performed a Student's $t$-test for the significance level of the null hypothesis following the similar method described in \citet{li03}. We found that the test produces $p$-value greater than 0.9 which rejects the null hypothesis. This particular test indicates that the observed non-correlation might be a random result. A possible reason for this could be the comparatively smaller range of dynamical time-scales ($\sim$0.3--11$\times$10$^{3}$ yr) in our sample compared to that reported by \citet[][$\sim$1--550$\times$10$^{3}$ yr]{wu04}.

We further examined the outflow rate against the dynamical time-scale. An anti-correlation trend ($\rho$ = -0.34 and -0.42 for all and {\it well-separated} lobes, respectively) is noted for $\dot{M}_{\rm out}$ with $t_{\rm dyn}$ (Figure~\ref{fig5}c). This observed trend could be due to a decrease in the outflow rate with time as also previously suggested by several authors \citep[][and references therein]{motte07, lopez11, watson16}.

Figure~\ref{fig5}d shows the relation between the maximum dynamical time-scale ($t_{\rm dyn, max}$) of the outflows associated with a single core and the core mass. A hint of positive correlation is found for $t_{\rm dyn, max}$ with the core mass (Spearman's Rank correlation coefficient $\rho$ = 0.35 and 0.16 for all and {\it well-separated} lobes, respectively). A positive trend was o observed in young infrared dark clouds by \citet{li20}. This positive trend might indicate for the massive cores to have comparatively longer outflow dynamical timescale than their low-mass counterparts. If it is generally assumed that the dynamical time-scale of outflows reflects the accretion time-scale of the cores, then this positive trend might also hint at the possibility for the massive cores to have longer accretion history than the low mass cores in these protoclusters as also suggested by \citet{li20}. However, as already discussed before a few cores in our sample might not be well-resolved. The presence of an unresolved companion would overestimate the core mass. Thus, this observed increasing trend might vary depending on an improved core mass estimation. For a more robust explanation, it needs a comparison of $t_{\rm dyn}$ with the bolometric luminosity of the associated cores. However, the determination of bolometric luminosity requires multi-band observations of the source at a similar spatial resolution. No other observations are available for our targets at the angular resolution of ALMA, and it is thus difficult to determine the bolometric luminosity of the cores in our sample.

\subsection{Comparison between outflow parameters and core parameters}
\label{SecComOutAndCores}
\begin{figure*}
\includegraphics[width=\columnwidth]{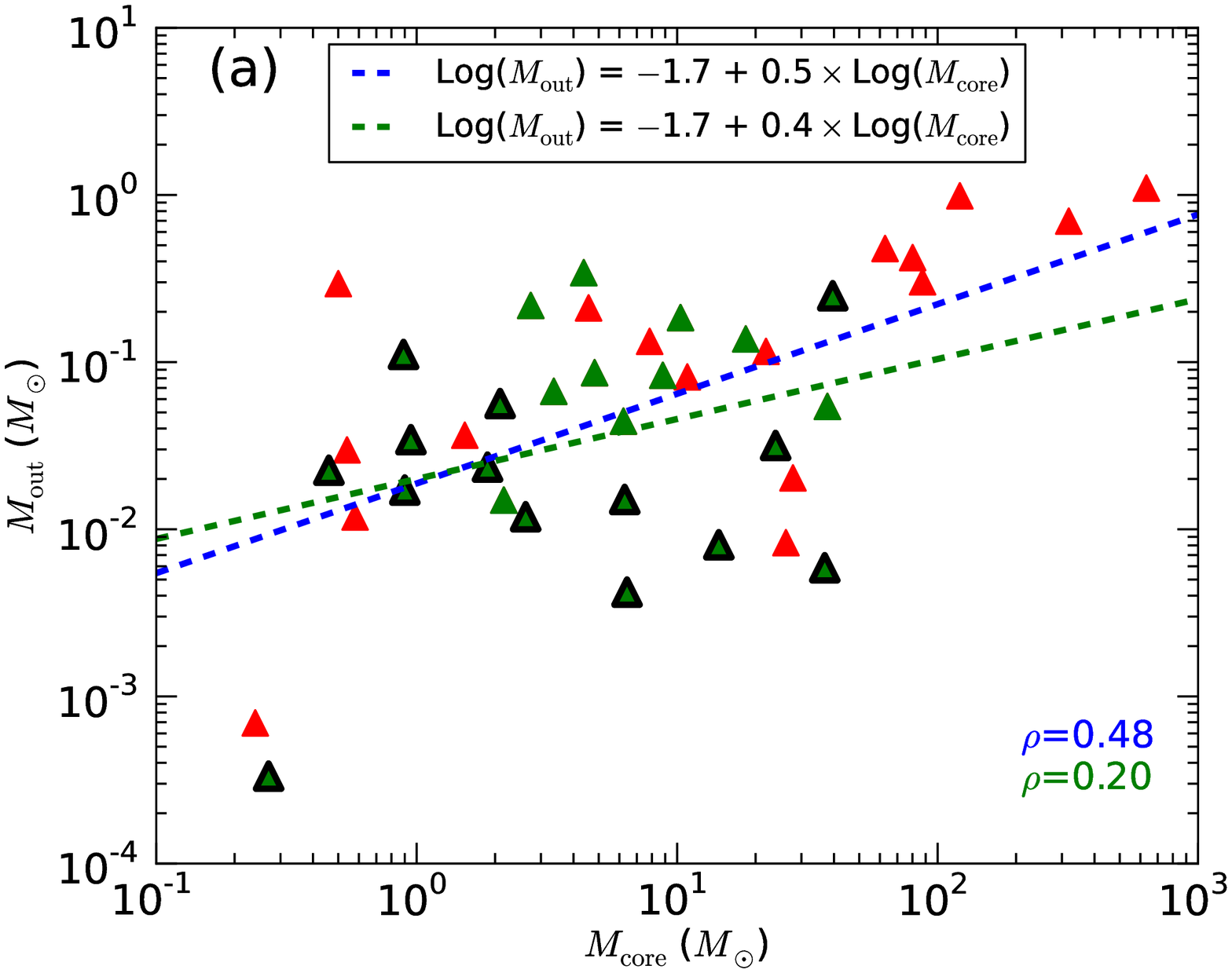}
\includegraphics[width=\columnwidth]{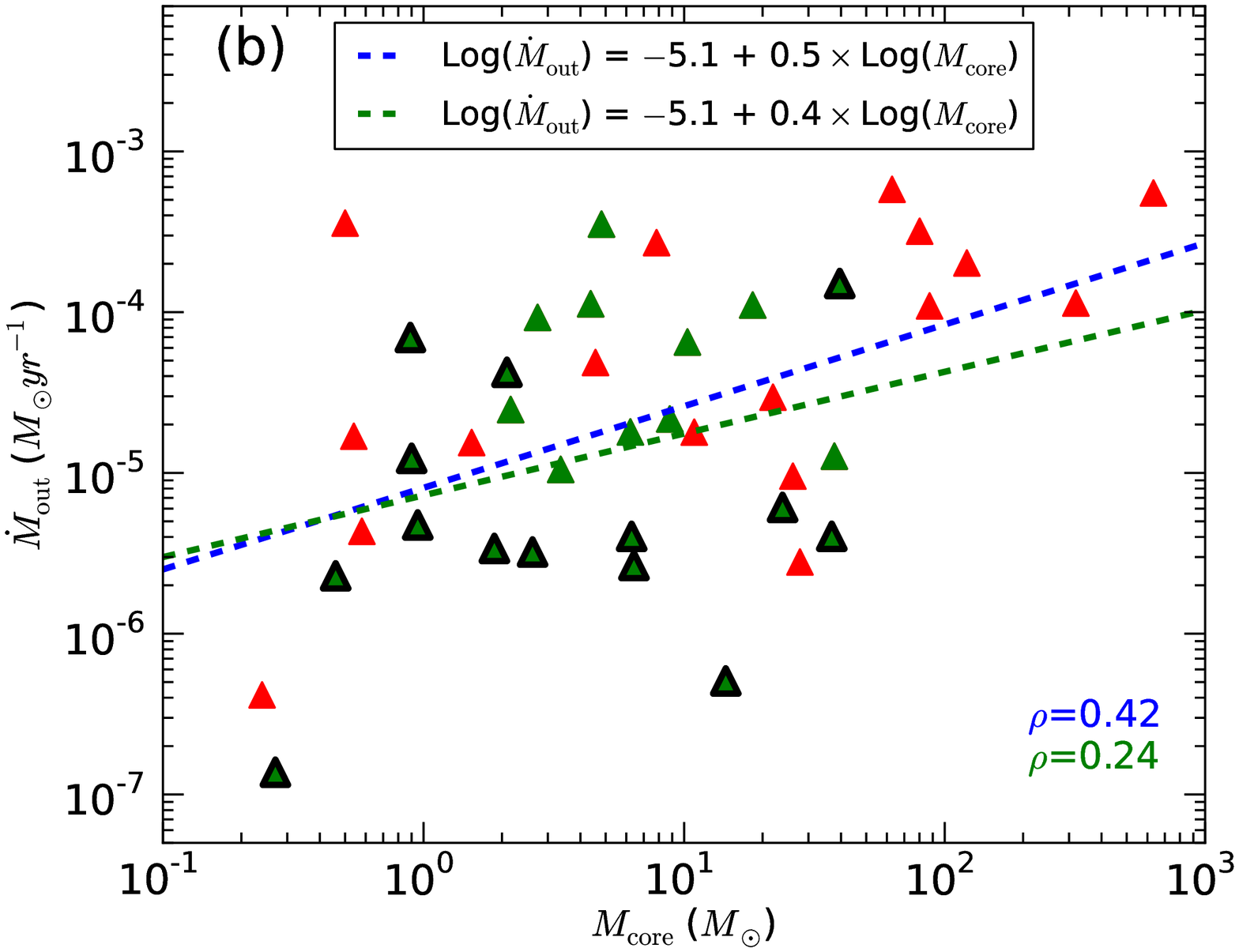}
\includegraphics[width=\columnwidth]{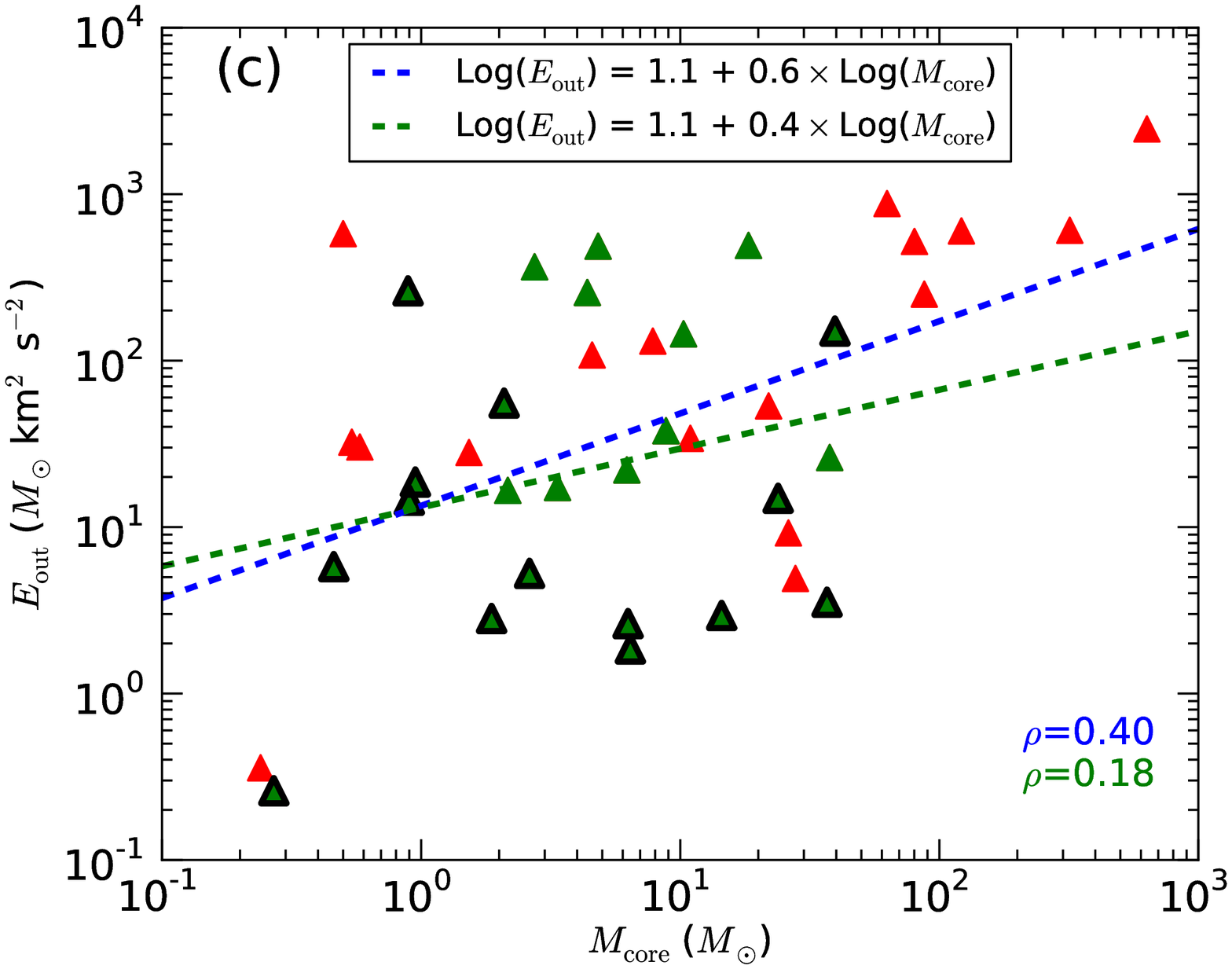}
\includegraphics[width=\columnwidth]{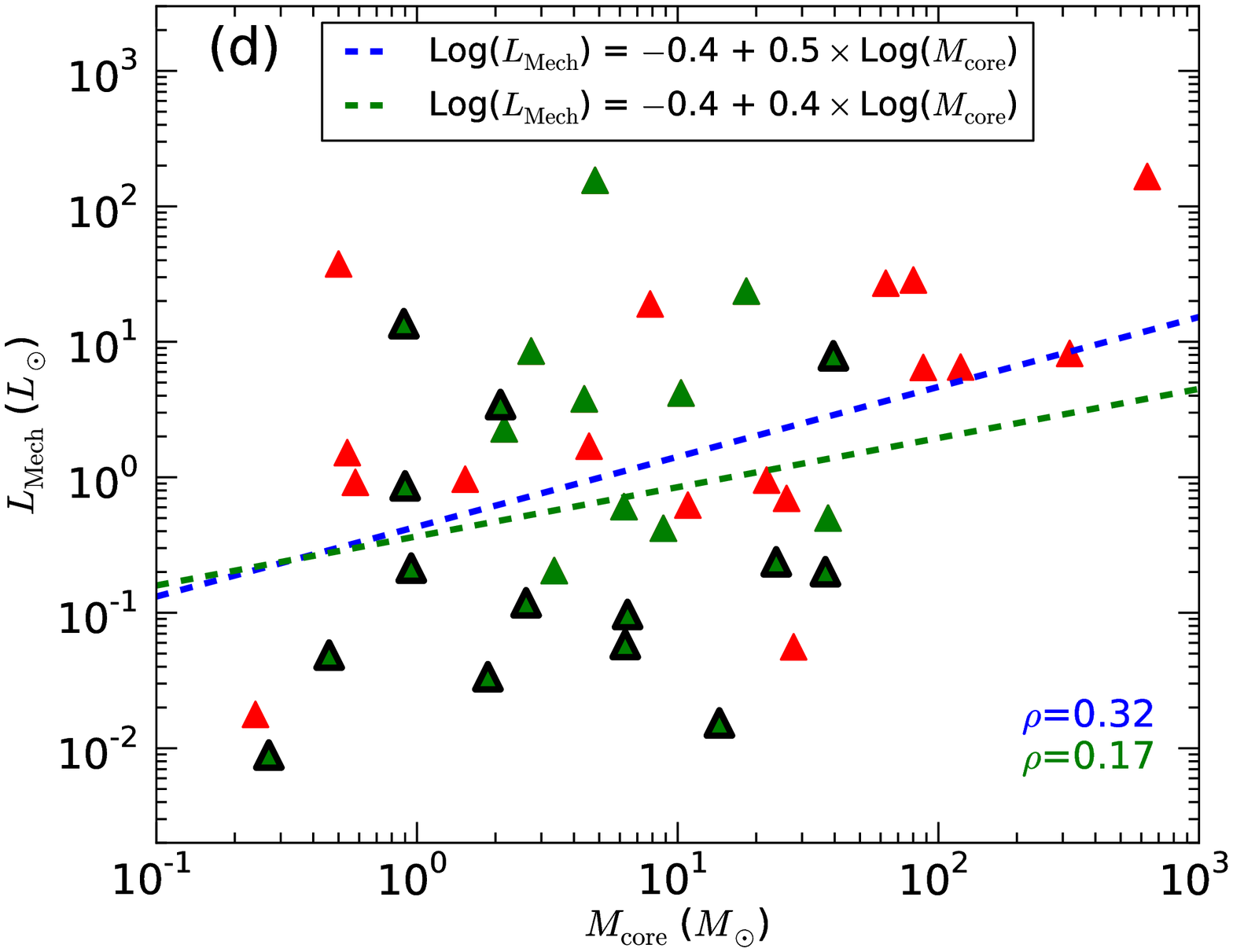}
	\caption{(a) The outflow mass versus the core mass, (b) the outflow rate versus the core mass, (c) the outflow energy versus the core mass, and  (d) the mechanical luminosity of the outflow versus the core mass. The triangles represent the total outflow parameters associated with each core. The green triangles represent the {\it well-defined} outflows (see text). The green triangles with black edges correspond to the unipolar outflows while the remaining green triangles are for bipolar outflows. The red triangles are the sources that are contaminated either by driving multiple outflow lobes or the lobe has overlapping emission with a nearby lobe. The blue and green dashed lines are the linear least-square fit to all and green data points in the logarithmic scale, respectively. The Spearman's rank correlation coefficients ($\rho$ in green for {\it well-defined} and in blue for all outflows) are also marked in all the panels.}
\label{fig6}
\end{figure*}

Figure~\ref{fig6}a shows the outflow mass plotted as a function of core mass, $M_{\rm core}$. Note that several massive cores are found associated with multiple outflows (both bipolar and unipolar). These outflows may be driven by further small-scale cores or multiple cores along the line of sight. However, such individual driving cores/sub-cores are not resolved even at the resolution of our ALMA data, possibly because our targets are densely clustered environment located at comparatively large distances (2.6--7.6 kpc). Note that the presence of several outflows has been noted before in other massive star-forming regions \citep[see e.g.,][]{beuther02,zhang07,fernandez13,nony20}. In such a dense clustered environment, it is difficult to always identify bipolar signatures of outflows. For uniformity, we considered plotting the total outflow parameters (see triangles in Figure~\ref{fig6}) driven by each identified core as that helps us to have a better comparison between the outflow and core properties. 
Additionally, for clarity we defined a cleaned {\it well-defined} set of outflow and their host cores. In Section~\ref{ComOutflowPars}, we already defined a set of {\it well-separated} lobes. Here, we further examine the {\it well-separated} outflows with respect to their host cores, and rejected those for which a host core is associated with more than one outflow (could be bipolar or unipolar). Finally, we named this cleaned set of data as {\it well-defined} outflows. We also marked these {\it well-defined} outflows in Figure~\ref{fig6} (see green triangles). Note that our calculated parameters are based on the emission detected in interferometric data which might lose the extended emission flux from the target. There are other additional components (i.e., assumption of optically thin emission, projection effect of outflow lobes, overlapping outflows) which underestimate the calculated parameters (see Section~\ref{SecDynProp}). Similarly, values of the cores might have a large uncertainty depending on the adopted dust temperature and the spectral index in the calculation, missing extended emission, and presence of unresolved companions (see Section~\ref{SecCoreMass}).
 
A correlation trend ($\rho$ = 0.48) is found between the outflow mass and the core mass for all the outflows (Figure~\ref{fig6}a). Even though an increasing trend is noted, a low correlation coefficient ($\rho$ = 0.20) is obtained when only {\it well-defined} outflows are considered. The general increasing trend fits to a power-law of $M_{\rm out} \propto M^{0.5}_{\rm core} ~and~ M^{0.4}_{\rm core}$ for all and {\it well-defined} cases, respectively. A similar proportionality is seen at clump-scale by \citet[][$M_{\rm out} \propto M^{0.6}_{\rm clump}$]{yang18} for massive outflows associated with hundreds of ATLASGAL clumps. However, a few other studies found a slightly steeper relation between outflow mass and the core mass \citep[i.e., $M_{\rm out} \propto M^{0.8}_{\rm core};$][]{beuther02, lopez09, devilliers14}. Most of the previous studies were based on a few tens of outflows. Thus, the slightly shallower slope in our sample could be a result of a large range of core mass as also pointed out by \citet{yang18}. The ratio $M_{\rm out}/M_{\rm core}$ has an average value of 0.05$\pm$0.13 which implies that on average 5\% of the core gas is entrained in the molecular outflow. This entrainment ratio is comparable to the value of 4--6\% reported by \citet{beuther02}, \citet{yang18}, and \citet{li18} seen at the clump-scale analysis.

Figure~\ref{fig6}b shows $\dot{M}_{\rm out}$ plotted as a function of $M_{\rm core}$. Despite a large scatter in the data points, a positive trend ($\rho$ = 0.42 for all outflows and $\rho$ = 0.24 for {\it well-defined} outflows) can be noted for the wide-range of $M_{\rm core}$ values. A similar positive trend with a similar slope was noted in previous studies toward massive clump-scale outflows \citep[see][and references therein]{lopez10, yang18, li18}. In addition, we also plotted outflow energy ($E_{\rm out}$) and mechanical luminosity of the outflow ($L_{\rm mech}$) with respect to the core mass, $M_{\rm core}$ (see Figure~\ref{fig6}c,d). A positive trend is noted for $E_{\rm out}$ with $M_{\rm core}$ for all outflows. However, the trend is not significant when only {\it well-defined} set of outflows are considered. Similarly, the positive trend for $L_{\rm mech}$ with $M_{\rm core}$ is not significant, specifically for the {\it well-defined} outflows. The correlation here is not statistically significant for both {\it well-defined} and all outflows (i.e. weak or non-correlation). Overall, these results suggest that the energy of an outflow generally increases with increasing mass of the driving core, but without a tight correlation.

\begin{figure}
\includegraphics[width=\columnwidth]{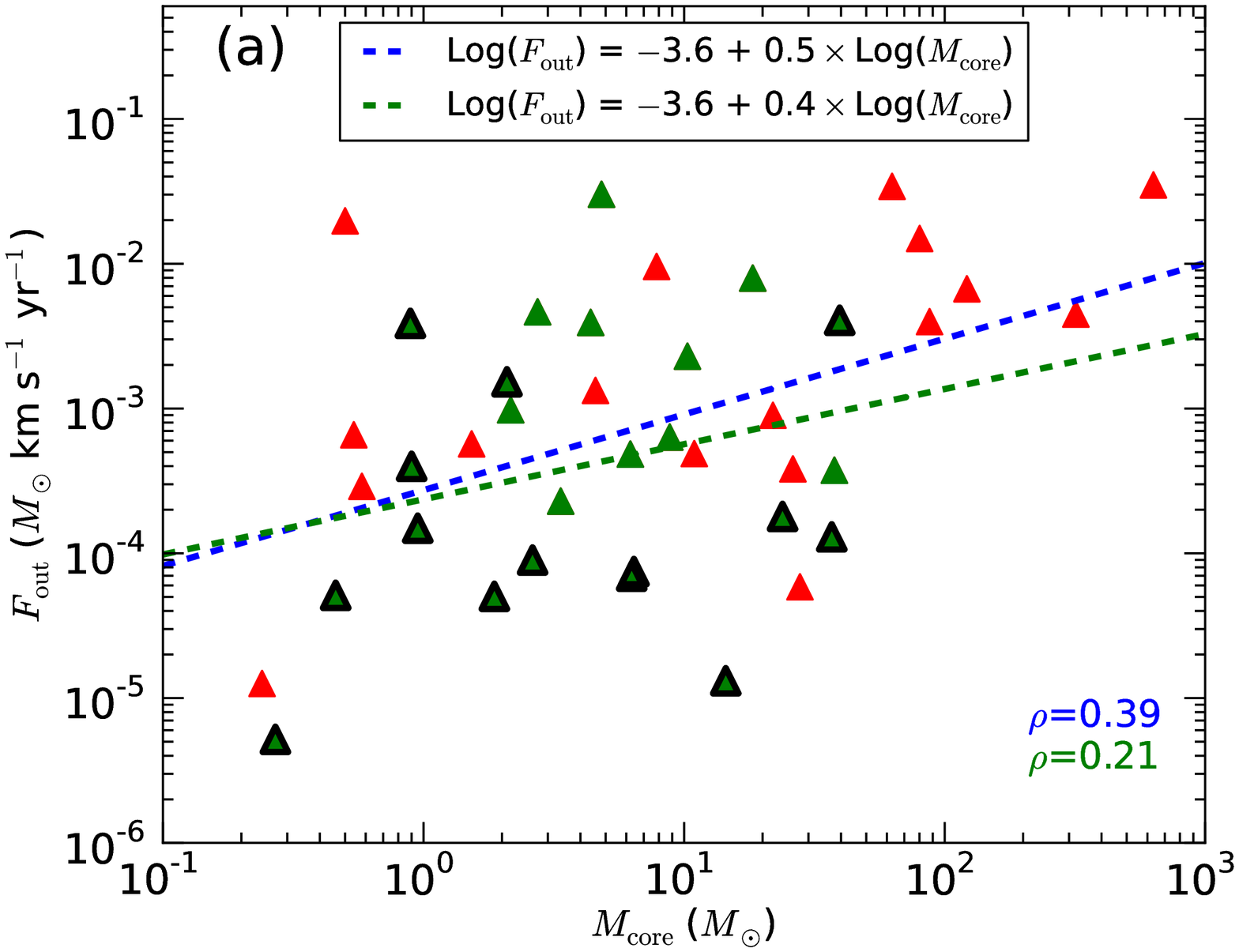}
\includegraphics[width=\columnwidth]{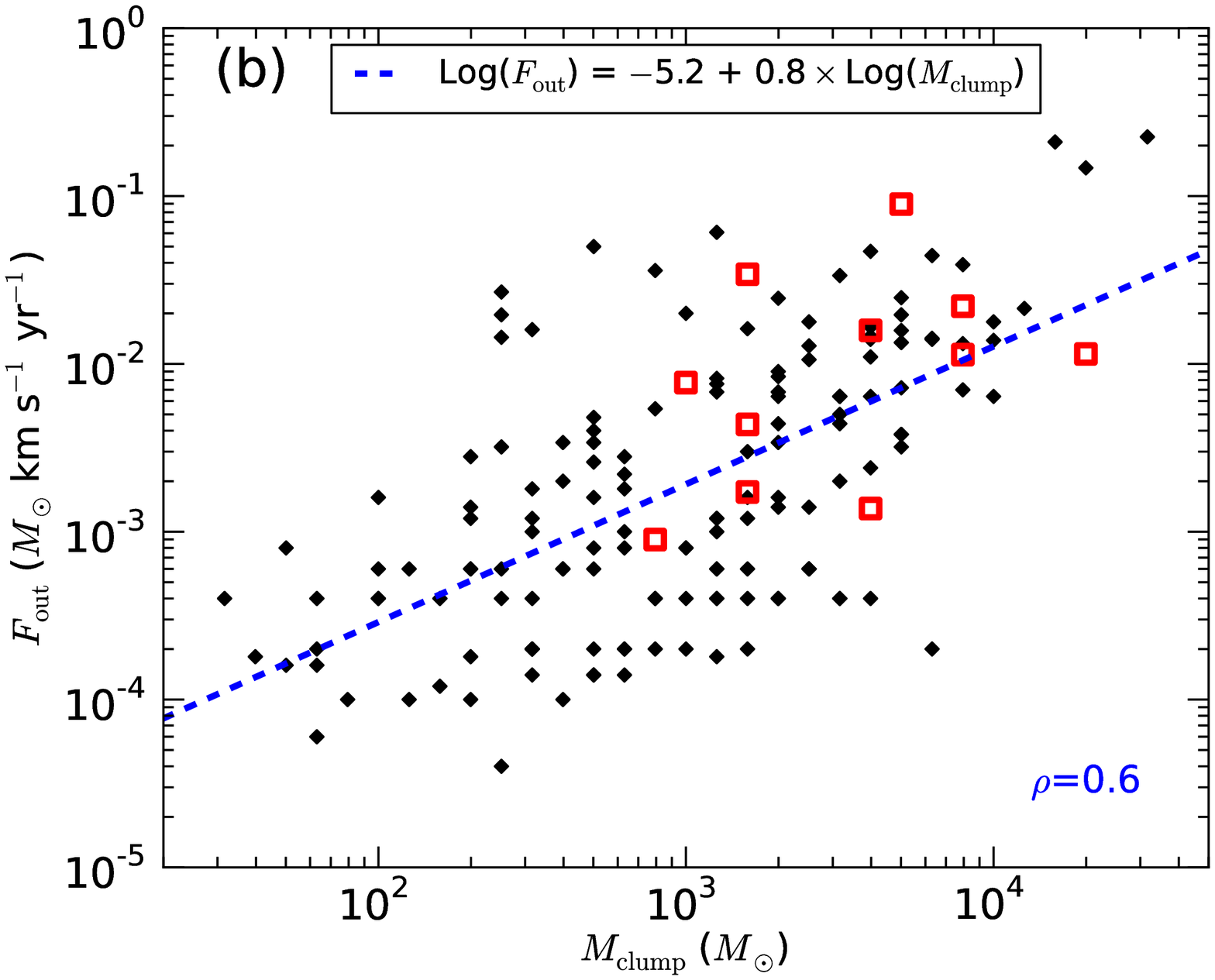}
\includegraphics[width=\columnwidth]{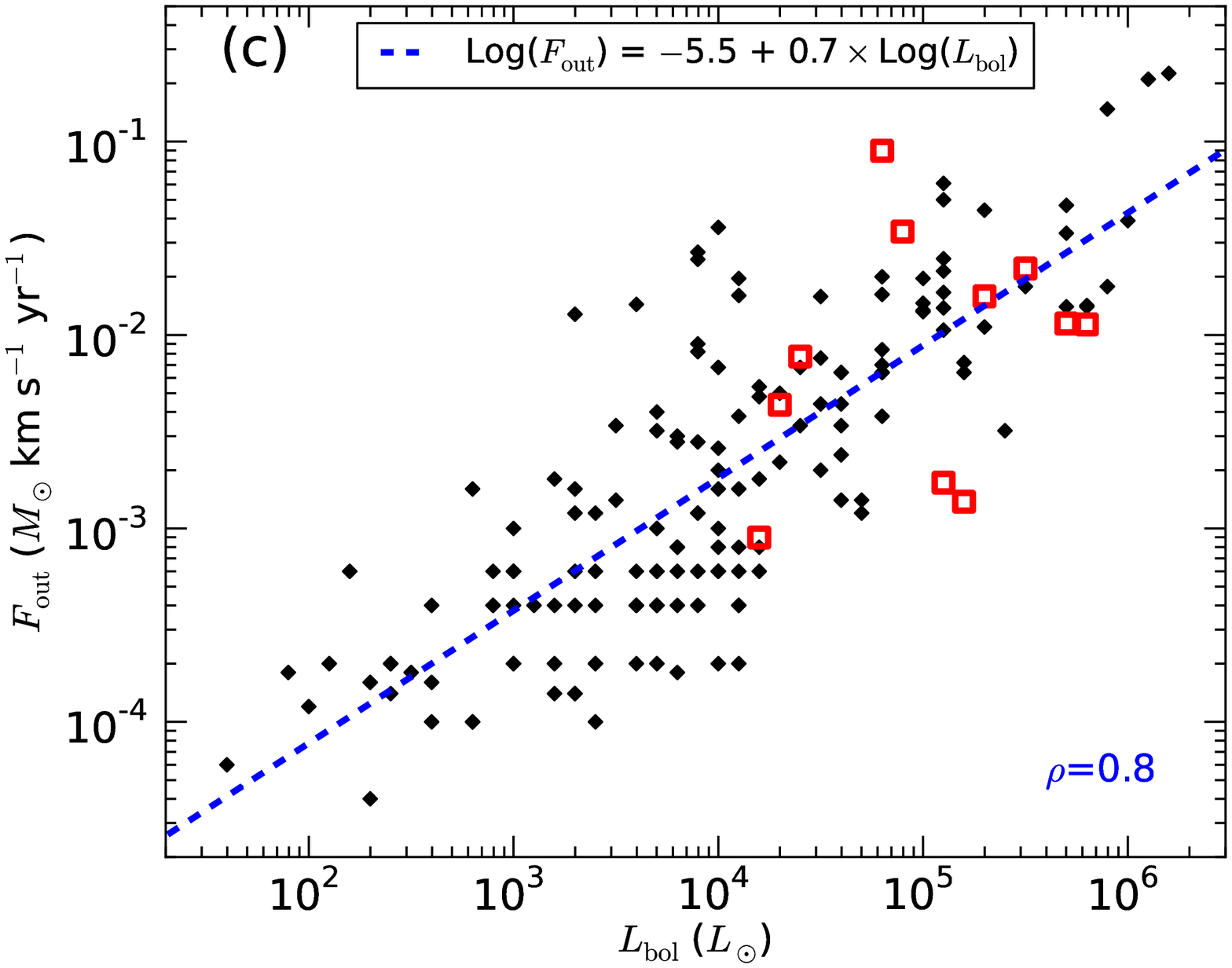}
\caption{(a) Outflow force plotted against the mass of the associated cores. The triangles represent the total outflow parameters associated with each core. The green triangles represent the {\it well-defined} outflows whereas the red triangles are contaminated lobes (see text and caption of Figure~\ref{fig6}). The blue and green dashed lines are the linear least-square fit to all and green data points in the logarithmic scale, respectively. (b) Total outflow force plotted against the mass of the whole clump. (c) Total outflow force plotted against the bolometric luminosity of the clump. Red squares in panels (b) and (c) show the data points for our targets while the black diamonds are obtained from \citet{yang18}. Blue dashed lines in both the panels represent the linear least-square fits on a logarithmic scale. The equation of the best fit lines are also labelled and correlation coefficients are also indicated in each panel.}
\label{fig7}
\end{figure}

The mechanical force of an outflow, $F_{\rm out}$, can be considered as a measure of the strength of the outflow, and the rate at which the outflow injects momentum (at the current epoch) into the surrounding gas \citep{bachiller92, downes07}. Several studies have already reported a positive correlation of $F_{\rm out}$ with core/clump mass and also with the bolometric luminosities \citep[see][for example]{shepherd96, wu04, zhang05, yang18}. Our calculated $F_{\rm out}$ (mainly composed of low-mass outflows) is plotted against $M_{\rm core}$ in Figure \ref{fig7}a. An increasing trend ($\rho$ = 0.39 and 0.21 for all and {\it well-defined} outflows, respectively) is noted for $F_{\rm out}$ against core mass over 3 orders of magnitude. A linear fit to the data points shows a slope of 0.5 for all outflows and 0.4 for the {\it well-defined outflows}. 

To examine the overall clump-scale scenario, we plotted the total value of $F_{\rm out}$ for each region against the total mass of the clump \citep[obtained from][]{liu20a} along with the parameters for more than one hundred of ATLASGAL clumps studied by \citet{yang18} (see Figure~\ref{fig7}b). In the absence of bolometric luminosity of individual cores, we also plotted the total $F_{\rm out}$ of each target against the bolometric luminosity of the whole clump \citep[see][for detailed parameters]{urquhart18, liu20a} along with parameters from \citet{yang18} (see Figure~\ref{fig7}c). As can be seen in Figure~\ref{fig7}b,c, our sample typically follows a similar increasing trend that was seen by \citet{yang18}. The data points for our eleven targets roughly fall within the parametric distribution of \citet{yang18}. 

A hint of increasing trend for $F_{\rm out}$ with $M_{\rm core}$ that holds for 3 orders of magnitude of core mass indicates a possible similar mechanism for driving outflows over the whole range of low-mass to massive cores. However, a slightly different slope is obtained at core-scale (slope=0.5 and 0.4, Figure~\ref{fig7}a) compared to clump-scale (slope=0.8, Figure~\ref{fig7}b). A similar increasing trend was noted before for low-mass and massive clumps by \citet{maud15} and \citet{yang18} for $F_{\rm out}$ against bolometric luminosity. These authors, however, did not exclude the possibility of observational bias between the outflow force of low-luminosity and high-luminosity sources considering that they lie at different distances. The targets of this work are massive protoclusters where we primarily detected low-mass outflows along with a few high-mass outflows. We thus conclude that the increasing trend is possibly real as both low-mass and high-mass outflows in our study are identified in the same target region. 

Considering the overall trend of the outflow parameters with the parametrs of the cores, a general inclination might be to believe that the stars of all different masses produce ouflows by a similar mechanism. However, as also discussed by \citet{maud15}, the outflow parameters such as $\dot{M}_{\rm out}$ and $F_{\rm out}$ have an intrinsic dependency on the entrained mass by the outflows. Thus, these parameters can only infer the strength of the outflows, not the launching mechanism. The outflow launching mechanisms for low-mass and massive star formation could thus be substantially different. 

The mass accretion rate could provide us the information on the ongoing star formation activity in the host cores. In general, an accretion rate of $\lesssim$10$^{-5}$ M$_\odot$ yr$^{-1}$ is typical for low-mass star formation, whereas a much higher accretion rates of $\gtrsim$10$^{-4}$ M$_\odot$ yr$^{-1}$ is required for massive star formation \citep{mckee03}.
It is possible to get an estimate of the accretion rate from the derived outflow mechanical force if it assumed that (1) the observed outflows are driven by proto-stellar winds from accretion disks \citep{keto03}, (2) the momentum is conserved between the wind and the outflow, and (3) the wind and molecular gas interface is efficiently mixed \citep{richer00}. Then, the accretion rate can be formulated as $\dot{M}_{\rm acc}$ = $k \dot{M}_{\rm wind}$, where $\dot{M}_{\rm wind}$ is the mass-loss rate of the wind which can be inferred from the $F_{\rm out}$ (= $\dot{M}_{\rm wind} v_{\rm wind}$). Here, $v_{\rm wind}$ is the wind velocity which we adopted to be 500 km s$^{-1}$ \citep{bally16,li20} and $k$ is the ratio between the mass accretion rate and the mass ejection rate which we assumed to be 3 \citep{shu00}. Accordingly, the derived accretion rates range from 4.6$\times$10$^{-8}$ to 3.0$\times$10$^{-4}$ M$_\odot$ yr$^{-1}$, with an average value of 2.4$\times$10$^{-5}$ M$_\odot$ yr$^{-1}$. Although the mass accretion rates derived here are typically small compared to the expected accretion rate (10$^{-4}$--10$^{-3}$ M$_\odot$ yr$^{-1}$) for high-mass star formation \citep{mckee03, wang10}, a couple of sources also show comparable accretion rates that are seen toward high-mass star-forming regions \citep{zhang05, qiu09, liu16, liu17}. 

These typically low accretion rates of $\lesssim$10$^{-5}$ M$_\odot$ yr$^{-1}$ imply that the majority of the outflows in our studied regions are associated with low-mass star formation \citep[see e.g.,][]{machida13, frank14, lee20}. In the case of low-mass star formation, it is observed that the accretion rate falls off with time at an approximate $t^{-1}$ law \citep{caratti12}. Additionally, the mass accretion rates for low-mass stars depends on the mass (M$_\ast$) of the forming star \citep[see][and references therein]{alcala14,biazzo19}. While the determination of the stellar mass and the evolutionary stages of the observed cores is beyond the scope of this paper, we also note that uncertainties in mass accretion rates are large, resulting from the uncertainties of outflow parameters and adopted assumptions. Although this calculation prone to large uncertainty, it may still provide us the information on the general star formation scenario in these studied regions. We found that currently these regions are mostly going through an active phase of low-mass star formation.

\begin{figure}
\includegraphics[width=\columnwidth]{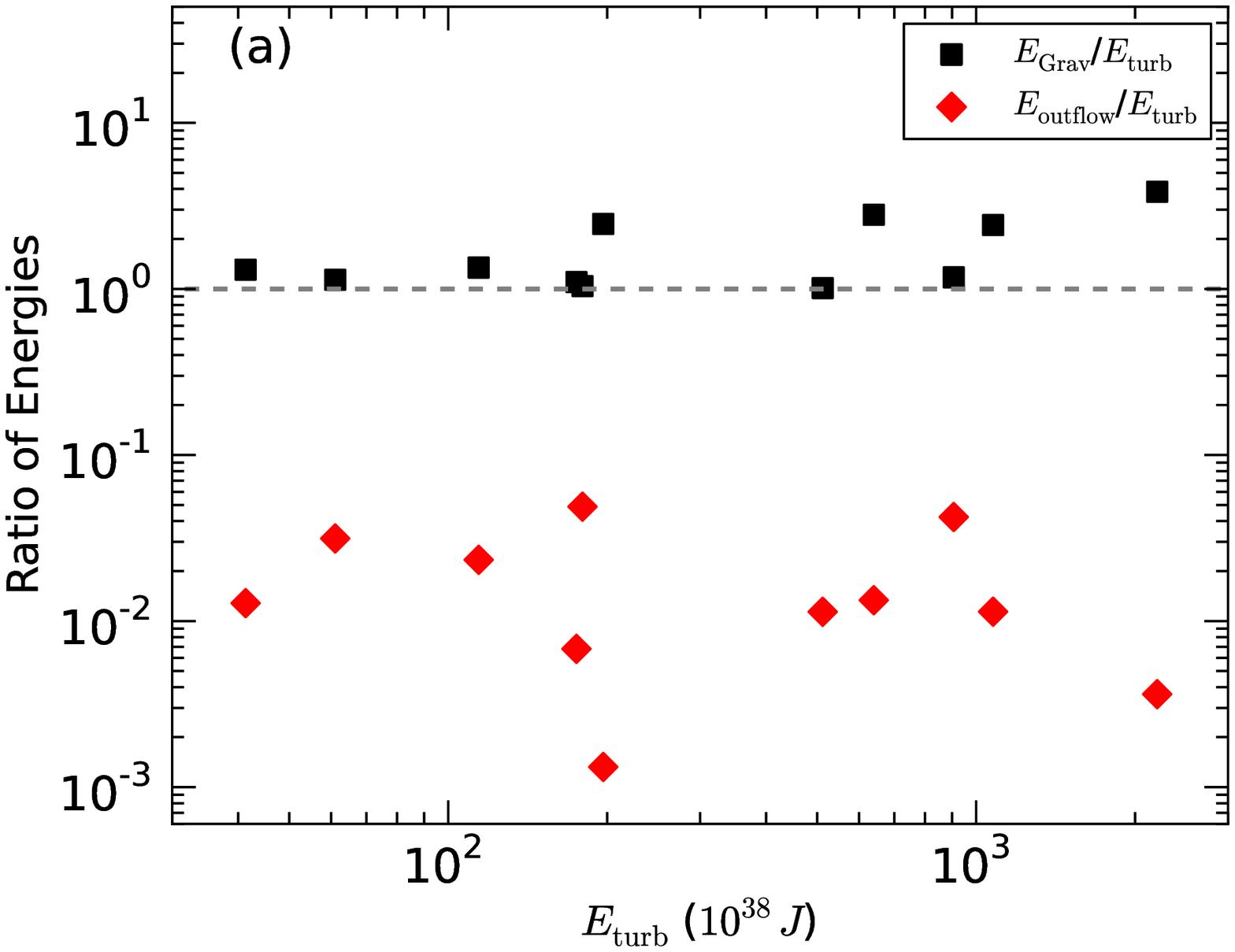}
\includegraphics[width=\columnwidth]{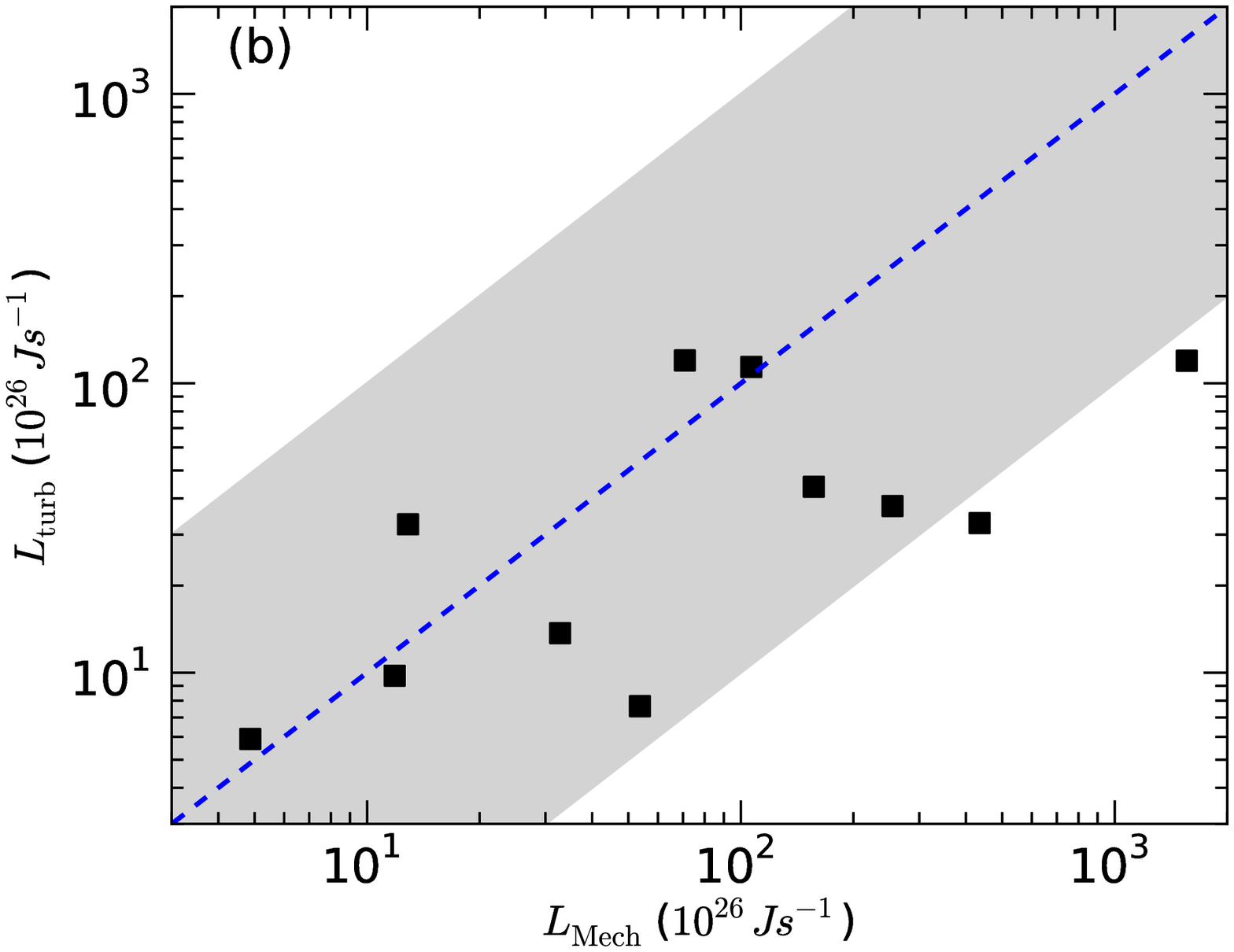}
\caption{(a) The ratios $E_{\rm grav}/E_{\rm turb}$ (black squares) and $E_{\rm outflow}/E_{\rm turb}$ (red diamonds) plotted as a function of the turbulent energy. The dashed black line represents the ratio of one. (b) The comparison between turbulent dissipation rate and the mechanical energy injection rate. The blue dashed line represents equal rates of turbulent dissipation and mechanical energy injection, and the grey-shaded area shows a variation of one order of magnitude around the blue dashed line. }
\label{fig7a}
\end{figure}

\begin{table*}
\centering
\caption{Energy budget of the observed protoclusters}
\label{table3}
\begin{tabular}{cccccrrccrrr} 
\hline
Source & Rad$^a$& Log($L_{\rm bol})^a$ & Log($M_{\rm clump})^a$ & $\Delta v ({\rm C^{17}O})^b$ &$E_{\rm grav}$ & $E_{\rm turb}$ & $E_{\rm outflow}$&  $t_{\rm diss}$     & $L_{\rm turb}$ & $L_{\rm Mech}$ \\
\cline{6-8} \cline{10-11}
	   &  (pc)  &   ($L_\odot$)        &    ($M_\odot$)         &  (km s$^{-1}$)         & \multicolumn{3}{c}{($\times$10$^{38}~J$)}         &($\times$10$^{5}$ yr)& \multicolumn{2}{c}{($\times$10$^{26}~J$ s$^{-1}$)} \\
\hline
 I14382-6017 &  1.68 & 5.2 & 3.6 &  3.03 &  484 &  197 &  1 & 6.4 &  10 &   17 \\
 I14498-5856 &  0.74 & 4.4 & 3.0 &  3.37 &   69 &   61 &  3 & 2.5 &   8 &   79 \\
 I15520-5234 &  0.67 & 5.1 & 3.2 &  4.53 &  192 &  175 &  2 & 1.7 &  33 &   19 \\
 I15596-5301 &  1.81 & 5.5 & 3.9 &  3.87 & 1789 &  640 & 13 & 5.4 &  38 &  373 \\
 I16060-5146 &  1.24 & 5.8 & 3.9 &  5.02 & 2612 & 1077 & 18 & 2.8 & 120 &  104 \\
 I16071-5142 &  1.21 & 4.8 & 3.7 &  5.80 & 1066 &  907 & 56 & 2.4 & 120 & 2286 \\
 I16076-5134 &  1.57 & 5.3 & 3.6 &  4.89 &  518 &  512 &  9 & 3.7 &  44 &  230 \\
 I16272-4837 &  0.84 & 4.3 & 3.2 &  3.66 &  154 &  114 &  4 & 2.6 &  14 &   48 \\
 I16351-4722 &  0.69 & 4.9 & 3.2 &  4.59 &  187 &  180 & 13 & 1.7 &  33 &  638 \\
 I17204-3636 &  0.60 & 4.2 & 2.9 &  3.11 &   54 &   41 &  1 & 2.2 &   6 &    7 \\
 I17220-3609 &  2.41 & 5.7 & 4.3 &  4.53 & 8479 & 2203 & 12 & 6.1 & 114 &  156 \\
\hline
\multicolumn{5}{l}{\footnotesize $^a$ Values from \citet{urquhart18}}\\
\multicolumn{5}{l}{\footnotesize $^b$ Values from Yue et al. (private communication)}
\end{tabular}
\end{table*}

\subsection{Energy budgets of the clumps and dissipation rate}
It is important to examine the cumulative impact of the outflows on the dynamics of their host clouds. We thus investigated whether the identified outflows have sufficient energy to drive the turbulence that counteracts the gravitational collapse of the host cloud. Estimation of the gravitational binding energy ($E_{\rm grav}$ = $3GM_{\rm cloud}^{2}/5R_{\rm cloud}$, where $G$ is gravitational constant) requires the total mass and the radius of the clouds. These values were obtained from \citet{urquhart18} (also see Table A1 of \citet{liu20a}). The mass, radius, and the calculated values of $E_{\rm grav}$ are listed in Table~\ref{table3}. The turbulent energy of a molecular cloud is given by

\begin{equation}
	E_{\rm turb} = \frac{1}{2} M_{\rm cloud} \sigma_{\rm 3d}^{2},
\end{equation}

where $\sigma_{\rm 3d}$ is the three-dimensional turbulent velocity dispersion, which can be calculated using

\begin{equation}
	\sigma_{\rm 3d} = \frac{\sqrt{3}}{2\sqrt{2~ln~2}} \Delta v_{\rm FWHM},
\end{equation}

where $\Delta v_{\rm FWHM}$ is the one-dimensional velocity width of an optically thin tracer of the molecular cloud. We obtained the $\Delta v_{\rm FWHM}$ values for all our regions from Yue et al. (private communication). They performed the C$^{17}$O (3--2) line observations of these sources using the Atacama Pathfinder Experiment (APEX) and obtained the $\Delta v_{\rm FWHM}$ values by performing a Gaussian fit to the observed line profile. The values of $\Delta v_{\rm FWHM}$ (i.e., $\Delta v ({\rm C^{17}O})$) and the calculated $E_{\rm turb}$ are listed in Table~\ref{table3}. The table also lists the total energy of all the outflows ($E_{\rm outflow}$) identified in each region. In Figure~\ref{fig7a}a we have shown the ratios of $E_{\rm grav}/E_{\rm turb}$ and $E_{\rm outflow}/E_{\rm turb}$ versus the turbulent energy. As can be seen in the figure, although $E_{\rm grav}$ is generally comparable to $E_{\rm turb}$, the total outflow energy, $E_{\rm outflow}$, is typically 1-3 orders of magnitude smaller than the turbulent energies in these regions, and thus, cannot drive the turbulence in these protoclusters. A similar conclusion was reached for outflows in the Taurus molecular cloud by \citet{li15}. However, the regions studied here are massive protoclusters and are mostly associated with the H {\sc ii} regions. Thus, a significant amount of turbulence might be contributed by the massive stars and the associated H {\sc ii} region. Note that the origin of such turbulence in the molecular clouds is unclear \citep[see][]{larson81, heyer04}. \citet{hennebelle12} suggested that such turbulence could be driven by external sources, such as, supernovae explosion, large-scale H{\sc i} streams \citep{ballesteros99}, shocks \citep{mckee07}, or galactic differential rotation \citep{klessen10, hennebelle12}.

We further estimated the total outflow energy injection rate (i.e., $L_{\rm Mech}$) and compare it with the energy rate needed to maintain the turbulence of the clump (i.e., $L_{\rm turb}$). The turbulent dissipation rate can be calculated as
\begin{equation}
	L_{\rm turb} = \frac{E_{\rm turb}}{t_{\rm diss}}, 
\end{equation}

where $t_{\rm diss}$ is the turbulent dissipation time. The turbulent dissipation time is calculated using the equation based on numerical simulations \citep{mckee07}.

\begin{equation}
	t_{\rm diss} \sim 0.5\frac{D_{\rm cloud}}{\sigma_{\rm 1d}},
\end{equation}

where $D_{\rm cloud}$ is the diameter of the clump and $\sigma_{\rm 1d}$ is the one-dimensional turbulent velocity dispersion along the line of sight (i.e., $\sigma_{\rm 1d} = \frac{\sigma_{\rm 3d}}{\sqrt{3}}$).

The calculated turbulence dissipation time, the turbulence energy dissipation rate, and the total outflow energy rate for all the targets in our sample are listed in Table~\ref{table3}. Note that these values should be treated only as order-of-magnitude estimates because there are several uncertainties involved in the estimated outflow parameters as already discussed in Section~\ref{SecDynProp}. In Figure~\ref{fig7a}b, the turbulent energy dissipation rate, $L_{\rm turb}$, is plotted with respect to the total outflow energy injection rate, $L_{\rm Mech}$, for our eleven protoclusters. It can be seen in the figure that the $L_{\rm turb}$ is typically comparable within an order of magnitude of the $L_{\rm Mech}$ values for all the clumps. Thus, it seems that the outflow observed here cannot account for the generation of turbulence, but can sustain the turbulence at the current epoch as it is similar to the estimated dissipation rate, similar to what was found in low mass star forming regions \citep[e.g.,][]{li15}.
\section{Conclusions}
\label{sec:conclusions}
We studied the properties of the outflows toward 11 Galactic massive protoclusters. Outflows are identified using three different tracers, the CO (3--2), HCN (4--3), and HCO$^{+}$ (4--3) lines. A total of 106 outflow lobes have been detected in these regions as already reported in Paper I. About 40\% of these outflows are identified with a single lobe, possibly because of the dense clustering nature of the massive protoclusters. All the physical parameters studied here are derived from CO. The main findings of this study are the following.

$\bullet$ Although the position angles of outflow lobes do not differ in the three tracers, HCN and HCO$^{+}$ generally detect outflows at lower terminal velocities compared to CO. This is possibly because CO is more sensitive to detect the low density and high-velocity outflowing material compared to the HCN and HCO$^{+}$.
 
$\bullet$ Identified outflows are young (typical $t_{\rm dyn}~ \sim$ 10$^2$--10$^{4}$ yr) and are composed of low-mass to (a few) massive outflows (e.g., $\dot{M}_{\rm out}\sim$ 10$^{-7}$--10$^{-4} M_\odot$ yr$^{-1}$; $P_{\rm out}\sim$ 10$^{-3}$--10$^{+1} M_\odot$ km s$^{-1}$). 

$\bullet$ An anti-correlation trend is noted for $\dot{M}_{\rm out}$ versus $t_{\rm dyn}$. This particular result may indicate that the outflow rate decreases with time. Although not significant, increase of $t_{\rm dyn}$ with the mass of the associated core probably indicates that the massive cores might have longer accretion history than the low mass cores in these protoclusters.

$\bullet$ The outflow mass for our sample shows a trend that increase with core mass with a power-law index of $\sim$0.4-0.5. This index is in agreement with clump-scale massive outflows reported by \citet{yang18}, but shallower (power index $\sim$0.9) than found in a few previous findings \citep{beuther02, lopez09, devilliers14, li18}. A similar rising trendency is seen for $F_{\rm out}$ with respect to core mass which could be a result of increasing entrained mass with increasing mass of the driving source, rather than the outflow launching mechanism.

$\bullet$ Calculation of the energy budget reveals that the kinetic energy of outflows alone cannot balance the gravitational binding energy, and is also unable to generate the observed turbulence in the host molecular clouds of our target regions. However, the energy injection rates of outflows in most of the regions can sustain the turbulence at the current epoch.

\section*{Acknowledgements}
We thank an anonymous referee for the thorough review and critical comments that helped us to improve the scientific content and the presentation of the manuscript. The work is supported by National Science Foundation of China (11988101, 11721303, 11973013, U1631102, 11373010). TB and KW are supported by the National Key Research and Development Program of China (2017YFA0402702, 2019YFA0405100). KW also acknowledges the support from a starting grant at the Kavli Institute for Astronomy and  Astrophysics, Peking University (7101502016). TB acknowledges funding from the China Postdoctoral Science Foundation through grant 2018M631241 and the PKU-Tokyo Partner fund. TB also acknowledge the support from S. N. Bose National Centre for Basic Sciences under the Department of Science and Technology (DST), Govt. of India. This work was carried out in part at  the Jet Propulsion Laboratory, operated for NASA by the California Institute of Technology. L.B. acknowledges support from CONICYT project Basal AFB-170002. C.W.L. is supported by the Basic Science Research Program through the National Research Foundation of Korea (NRF) funded by the Ministry of Education, Science and Technology (NRF-2019R1A2C1010851). This research made use of Astropy, a community-developed core Python package for astronomy \citep{astropy18}. This paper makes use of the following ALMA data: ADS/JAO.ALMA\#2017.1.00545.S. ALMA is a partnership of ESO (representing its member states), NSF (USA) and NINS (Japan), together with NRC (Canada), MOST and ASIAA (Taiwan), and KASI (Republic of Korea), in cooperation with the Republic of Chile. The Joint ALMA Observatory is operated by ESO, AUI/NRAO, and NAOJ. 

\section*{Data Availability}
The raw data that support  this study are publicly available for download from the ALMA archive. The reduced data are available from the corresponding author upon reasonable request.



\bibliographystyle{mnras}
\bibliography{ProtoclusterPaper2_TB_Rev2} 




\appendix

\section{A. Outflow lobes for all the regions}
\label{appendix}

\begin{figure*}
\includegraphics[width=8.6cm,height=7.3cm]{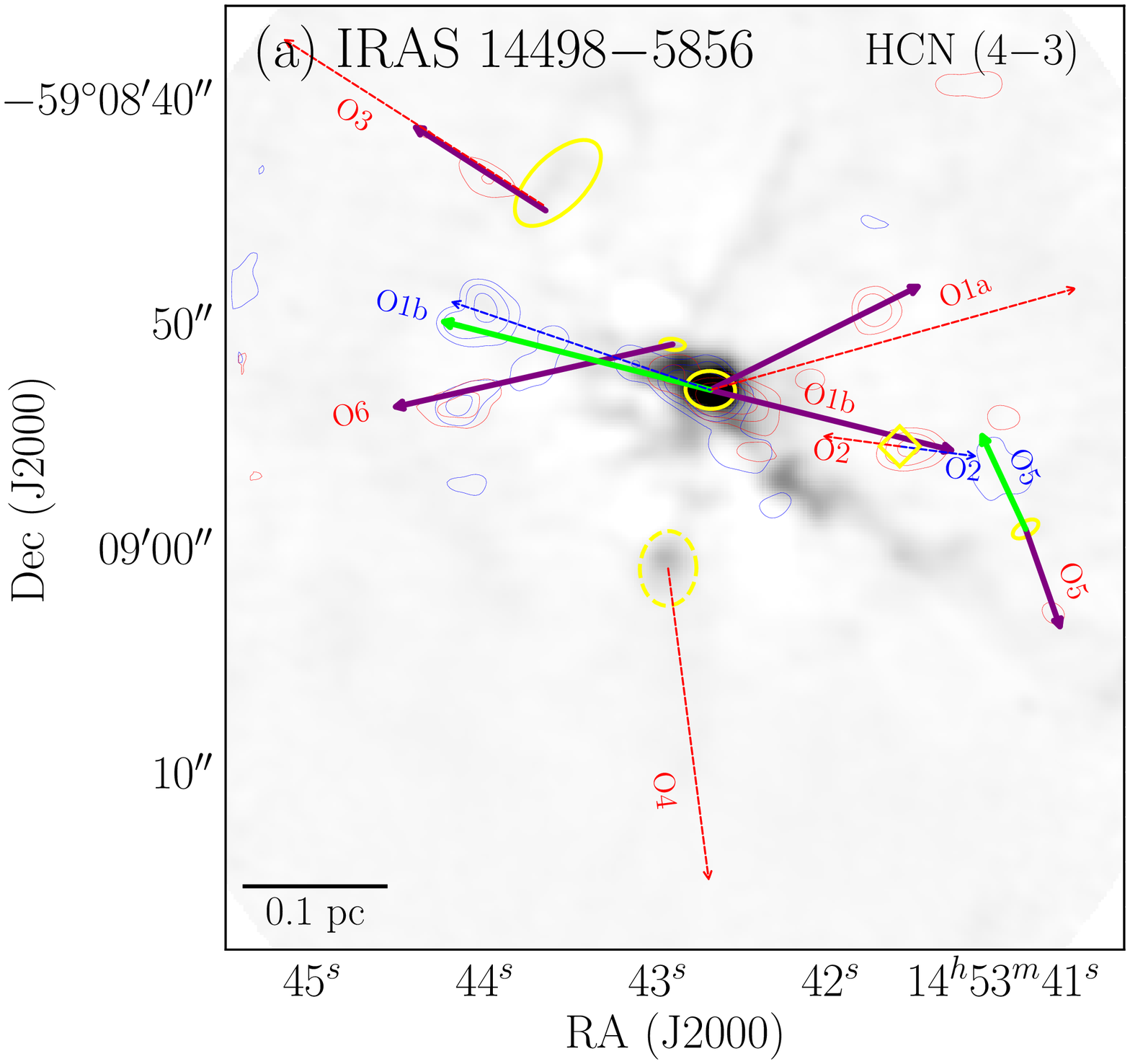}
\includegraphics[width=8.6cm,height=7.3cm]{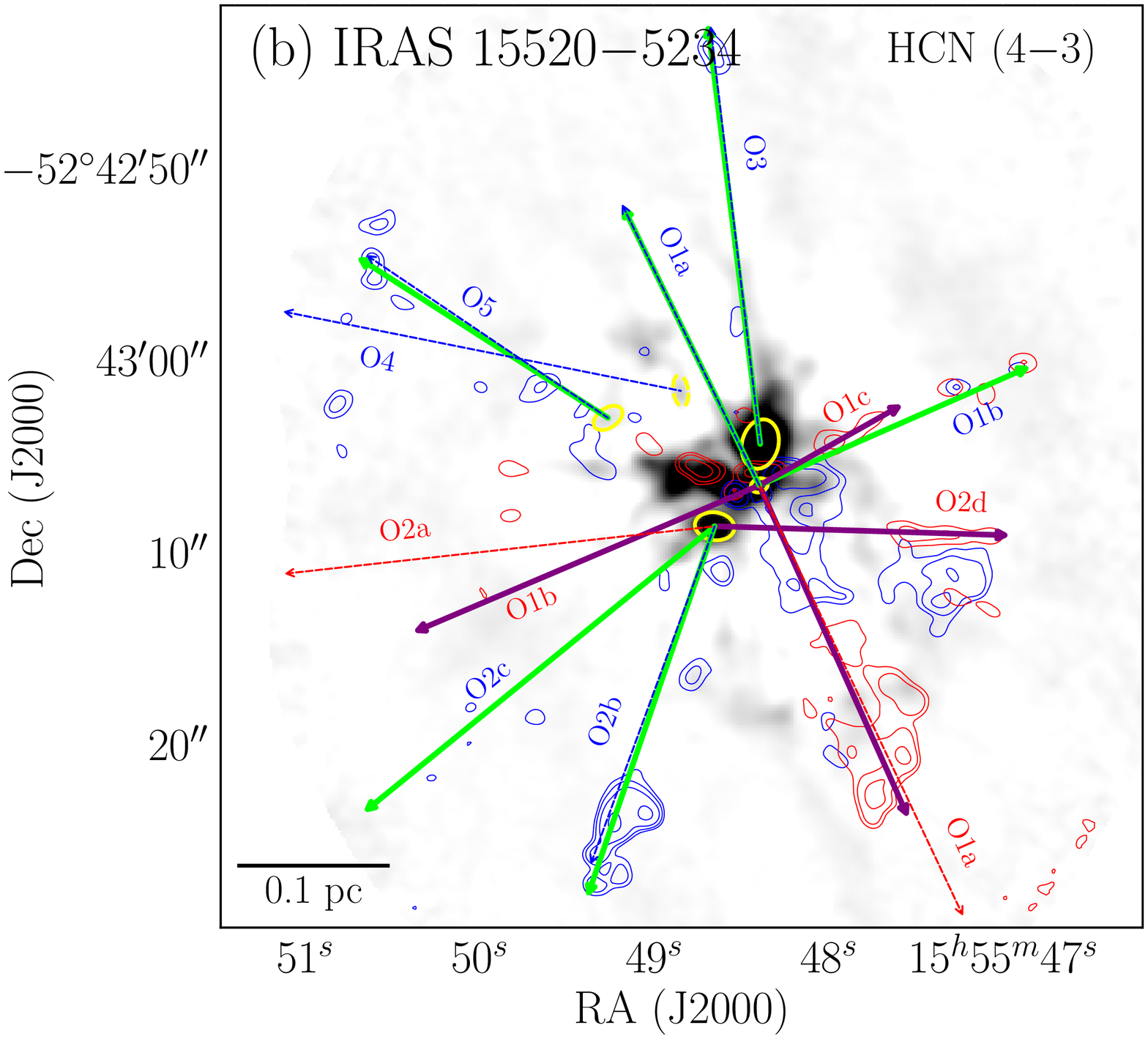}
\includegraphics[width=8.6cm,height=7.3cm]{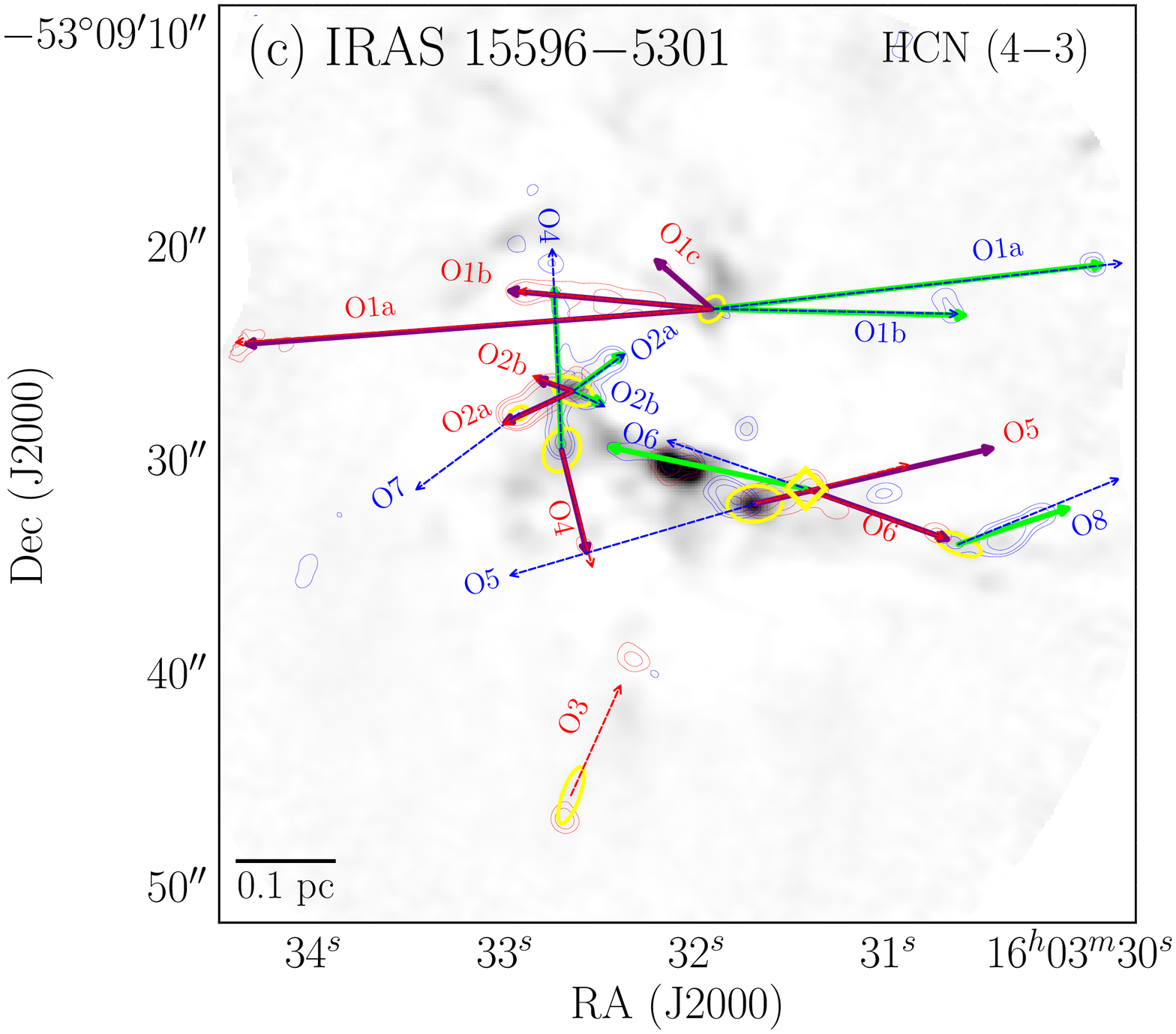}
\includegraphics[width=8.6cm,height=7.3cm]{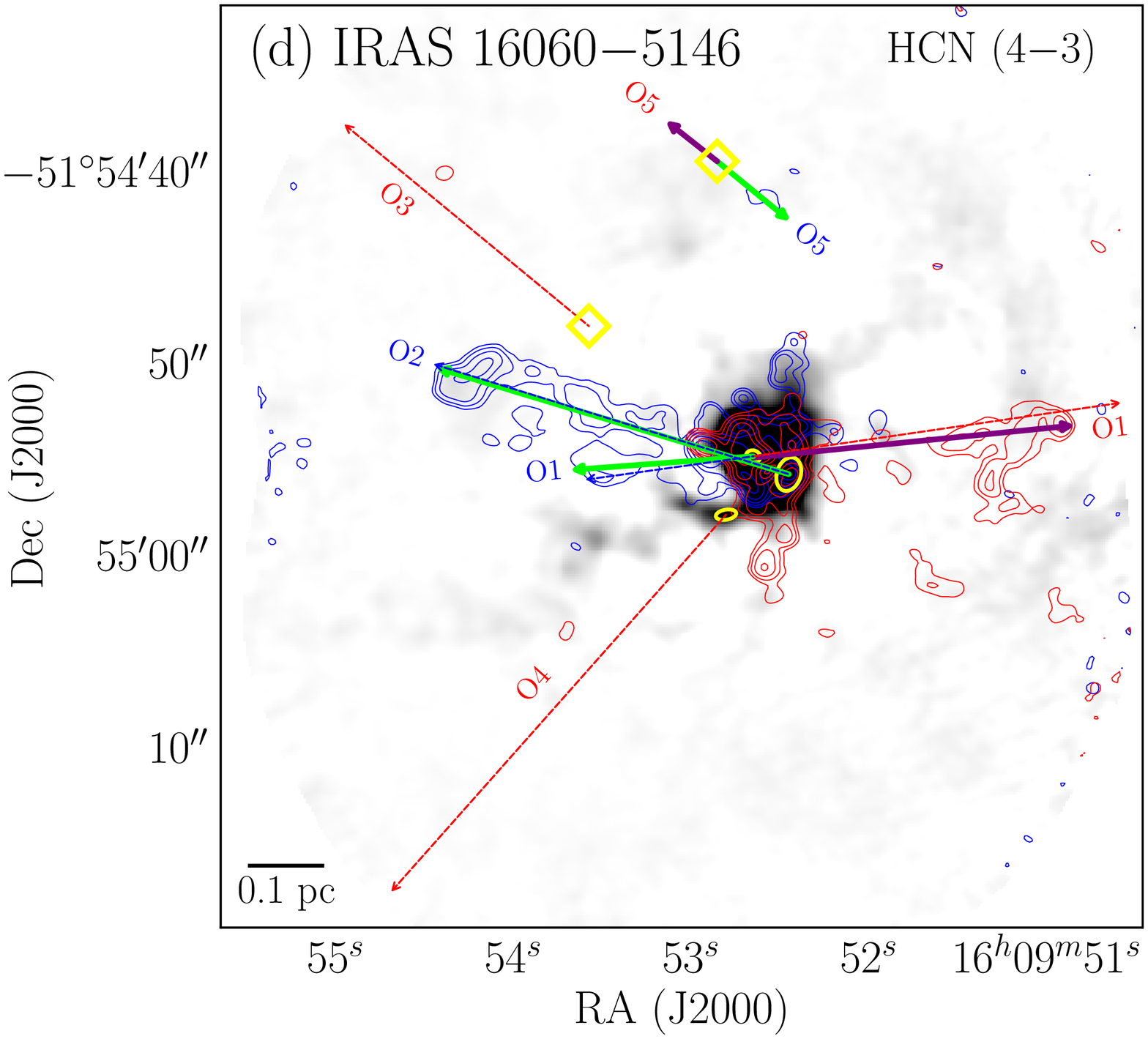}
\includegraphics[width=8.6cm,height=7.3cm]{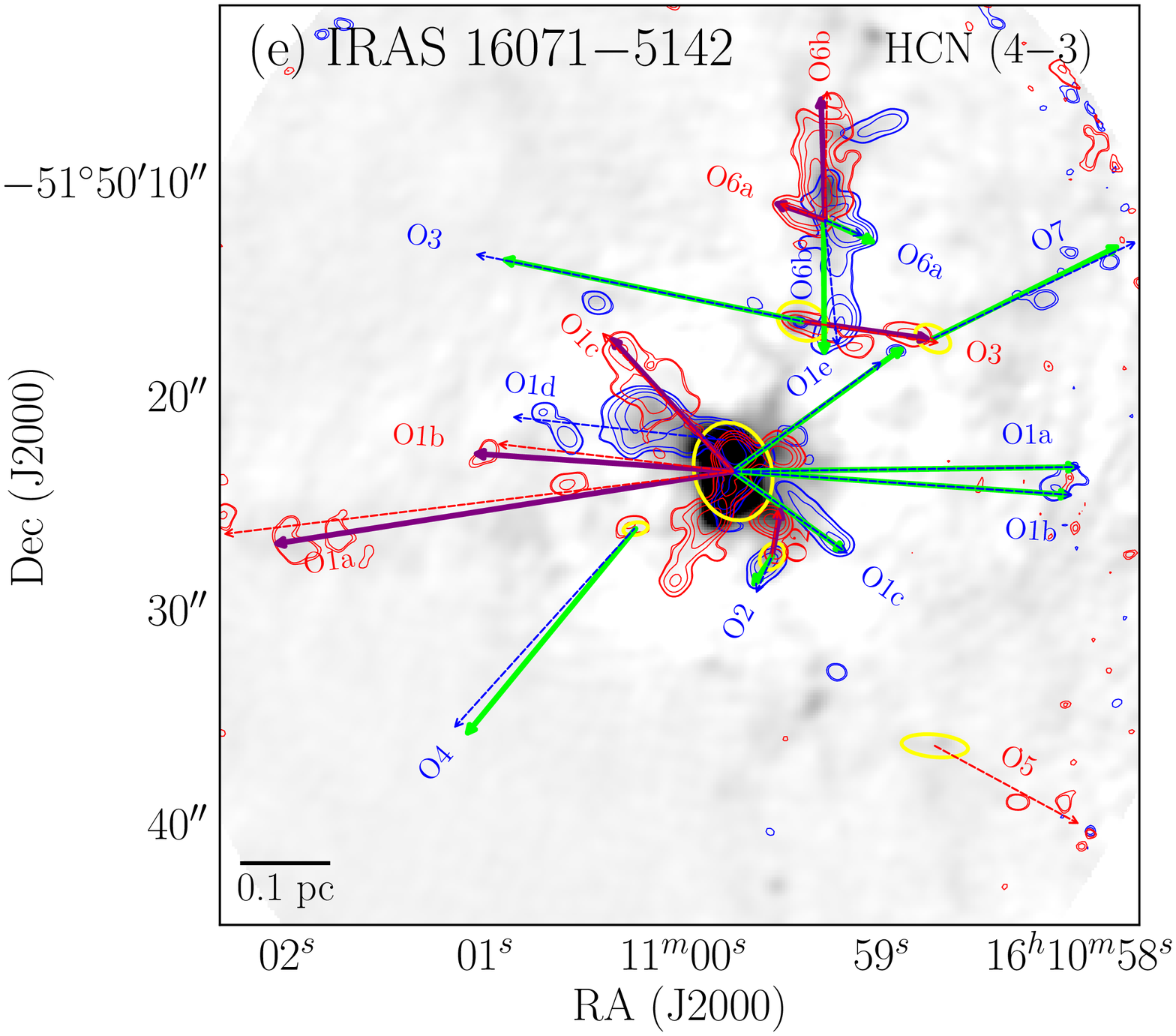}
\includegraphics[width=9.0cm,height=7.4cm]{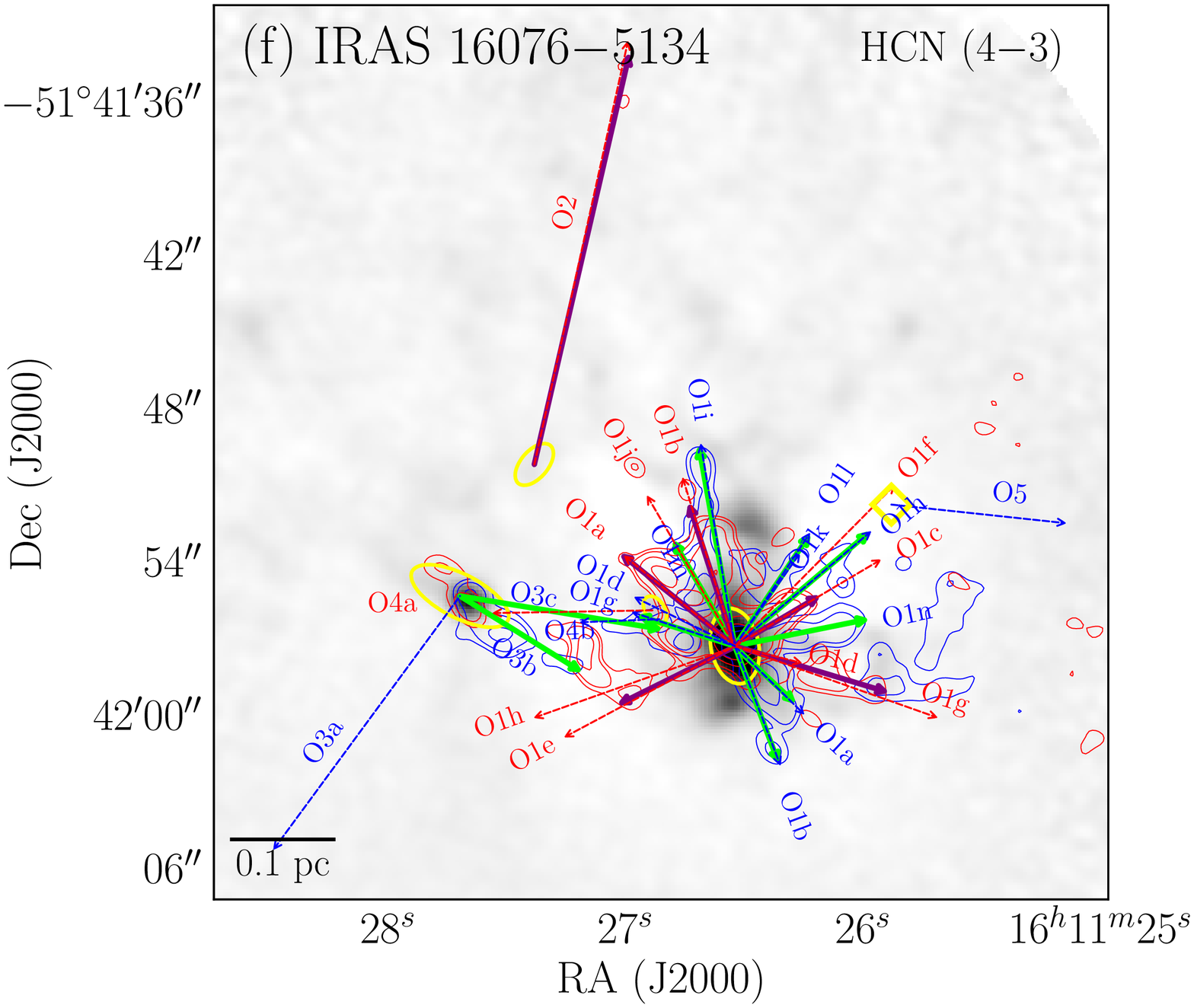}
\caption{Images of HCN outflows in the six target fields. Background grey-scale images are the ALMA 0.9 mm continuum maps. The red and blue contours correspond to redshifted and blueshifted HCN gas integrated over carefully selected velocity ranges to depict the outflow lobes. The blueshifted and redshifted outflow lobes of HCN are marked by solid green and purple arrows, respectively, while the outflow lobes identified using CO are also shown by blue and red dashed arrows, respectively. The driving sources detected in the continuum map are marked in yellow ellipses. Outflow lobes with yellow diamonds are those for which no continuum sources are identified (see caption of Figure~\ref{fig1}).}
\label{figA1}
\end{figure*}

\begin{figure*}
\includegraphics[width=8.6cm,height=7.3cm]{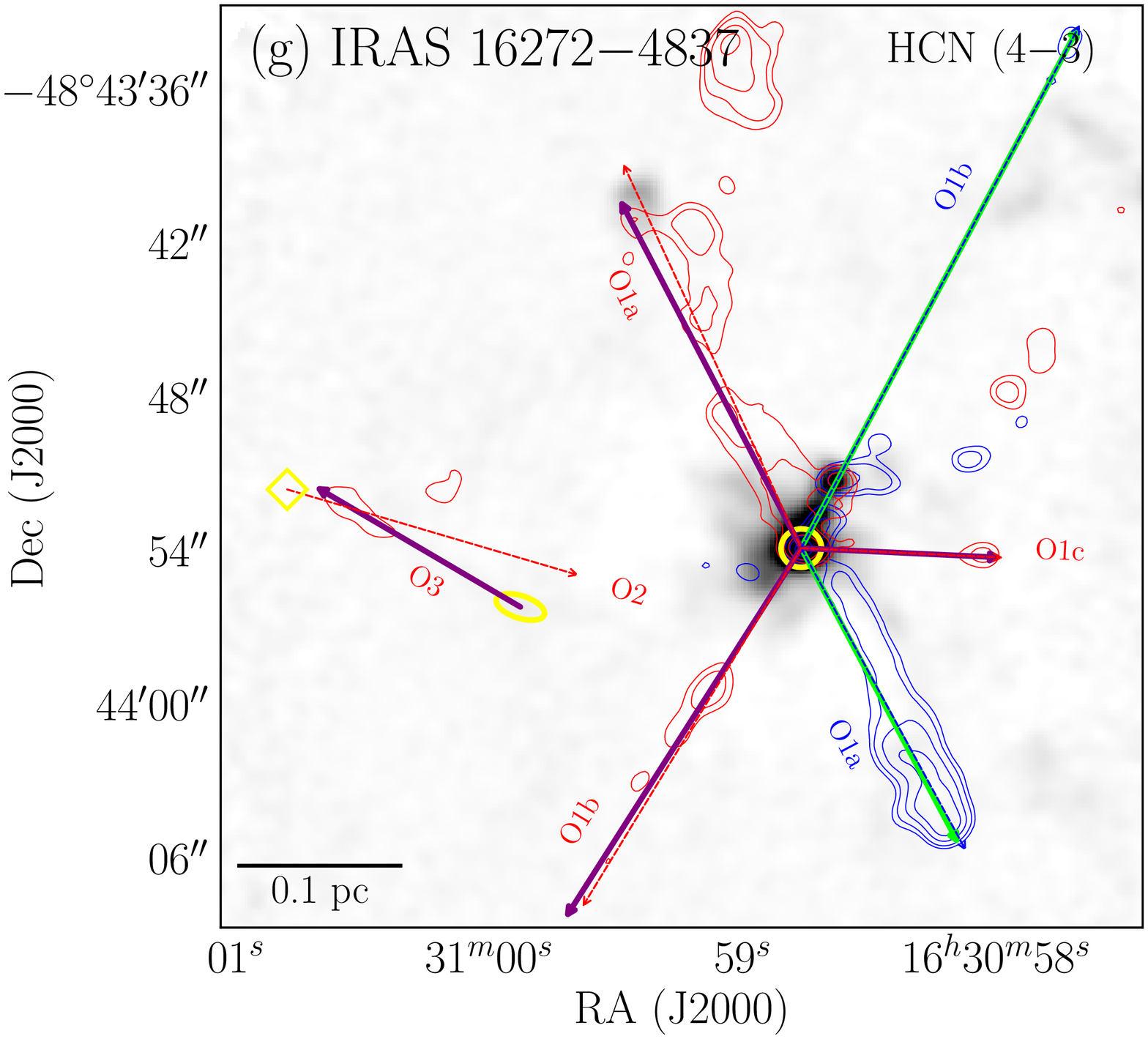}
\includegraphics[width=9.0cm,height=7.4cm]{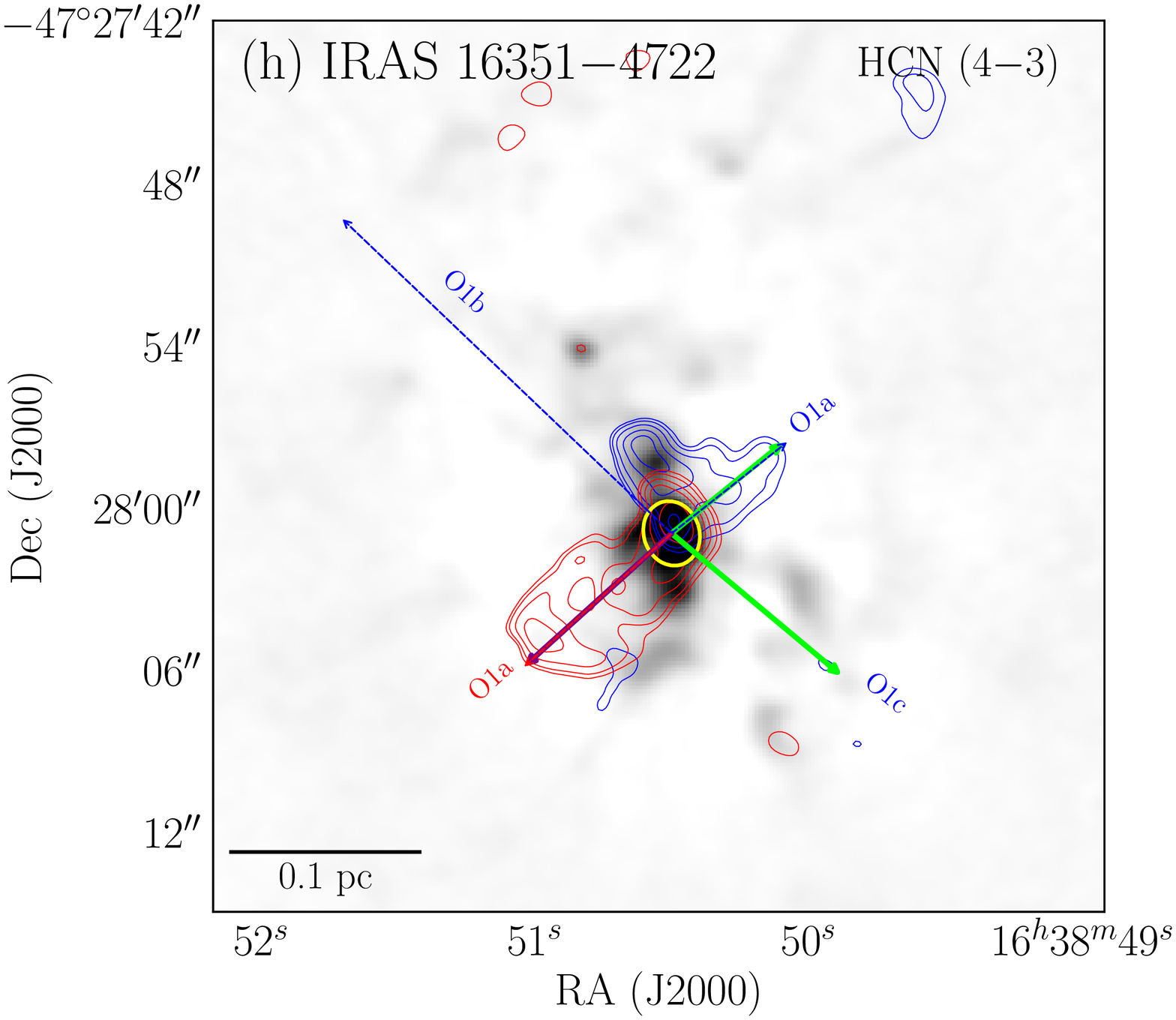}
\includegraphics[width=8.6cm,height=7.3cm]{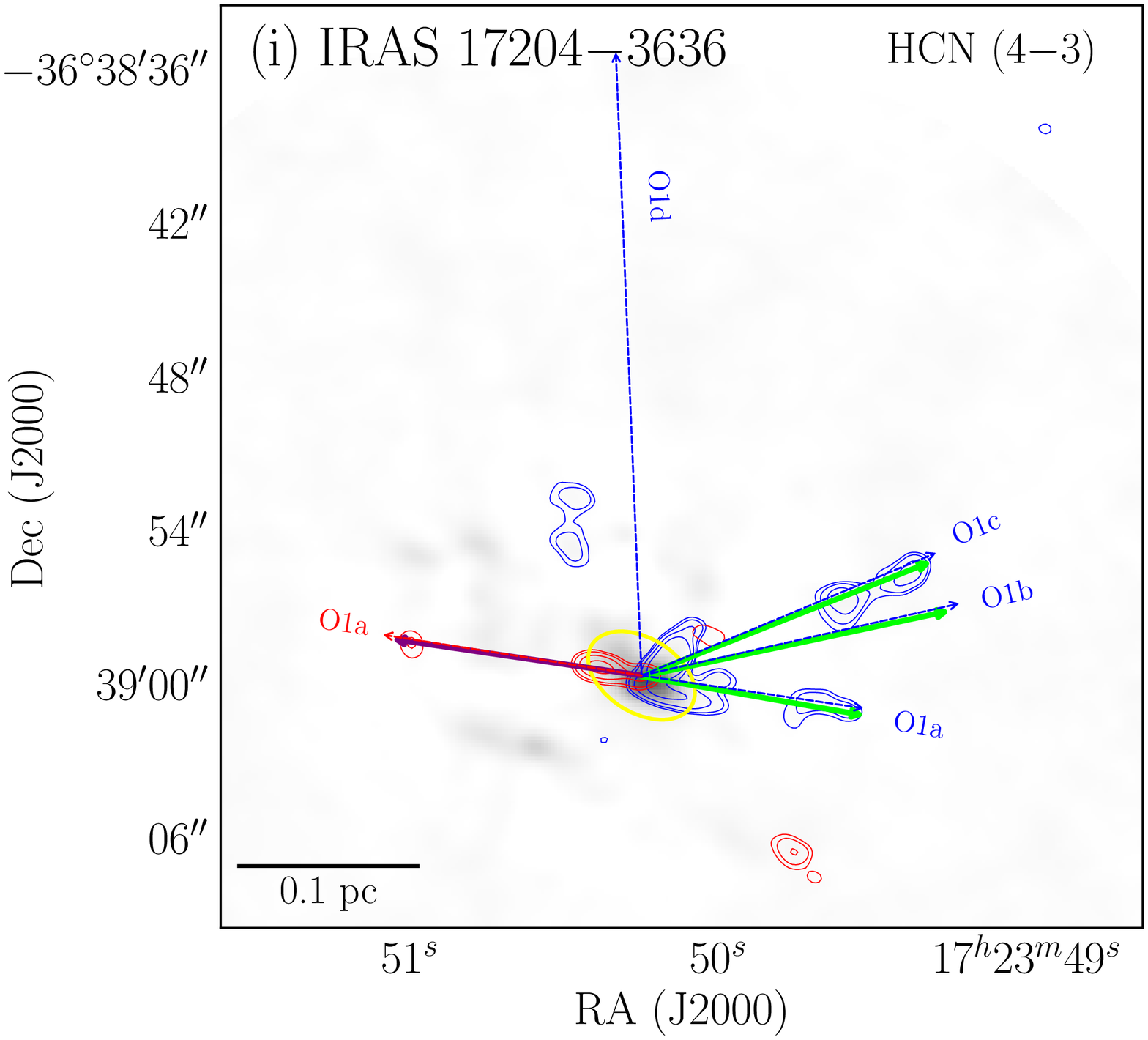}
\includegraphics[width=8.6cm,height=7.3cm]{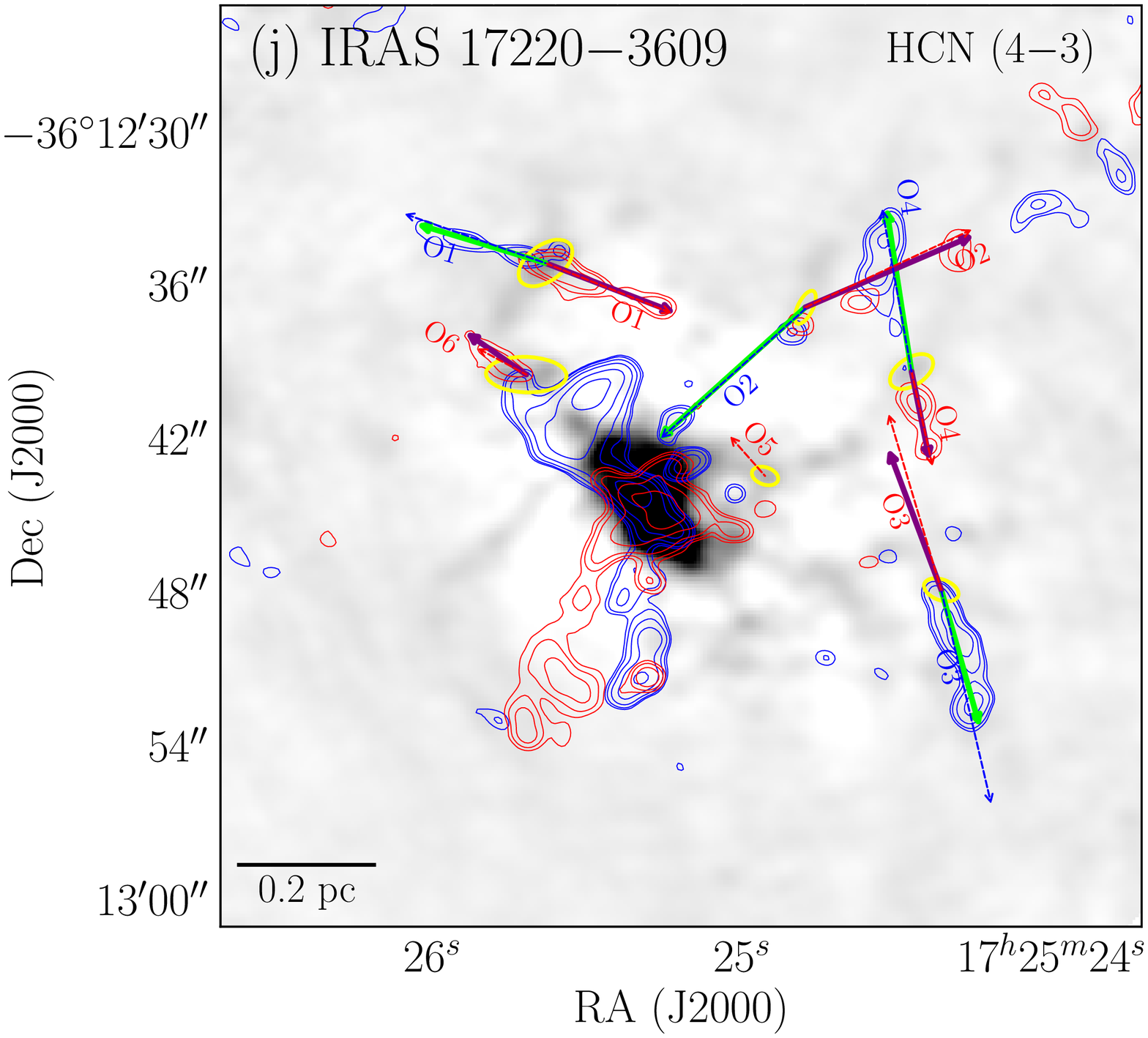}
\caption{Images of HCN outflows for the remaining four target fields (see caption of Figure~\ref{figA1}).}
\label{figA2}
\end{figure*}

\begin{figure*}
\includegraphics[width=8.6cm,height=7.3cm]{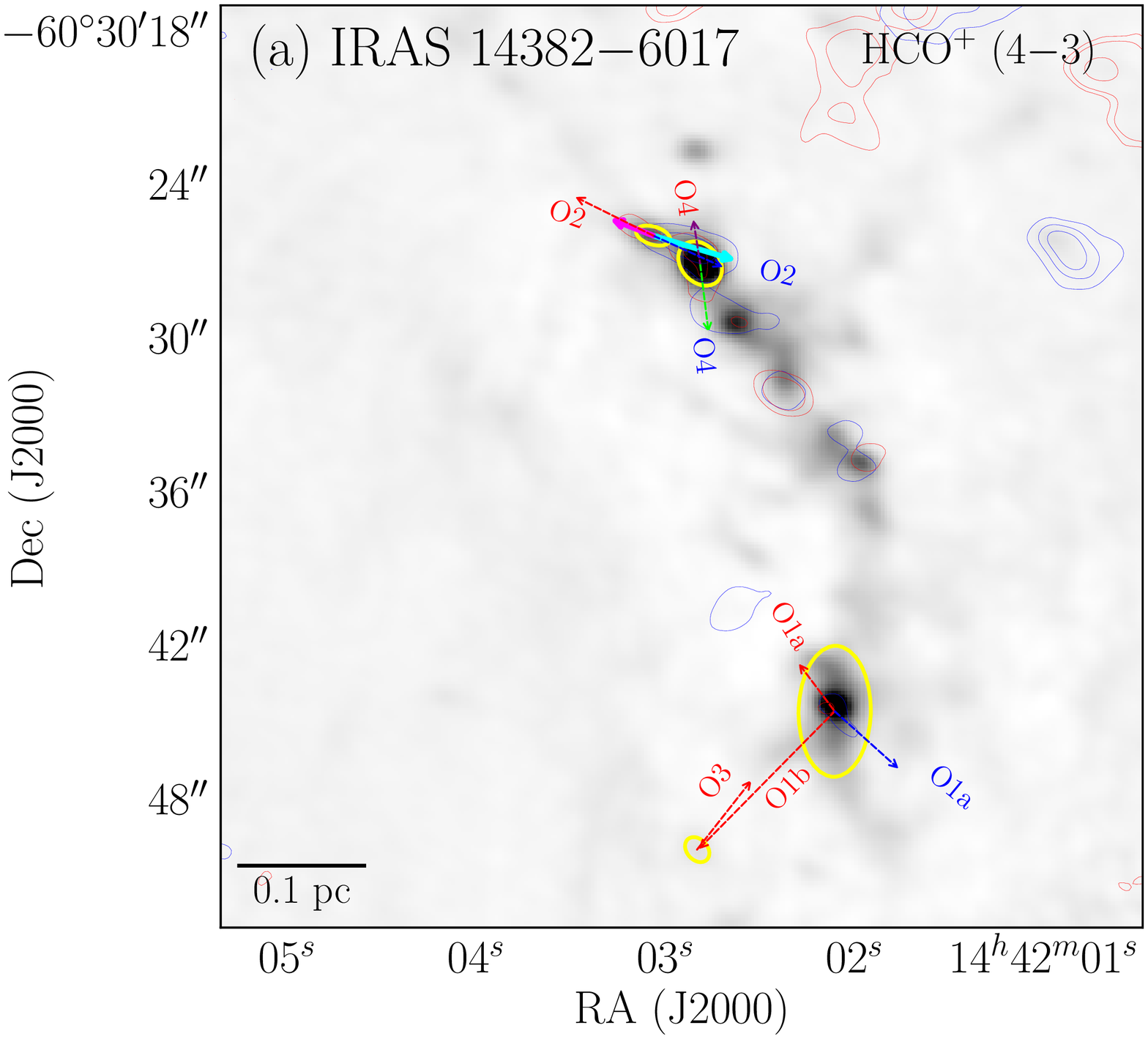}
\includegraphics[width=8.6cm,height=7.3cm]{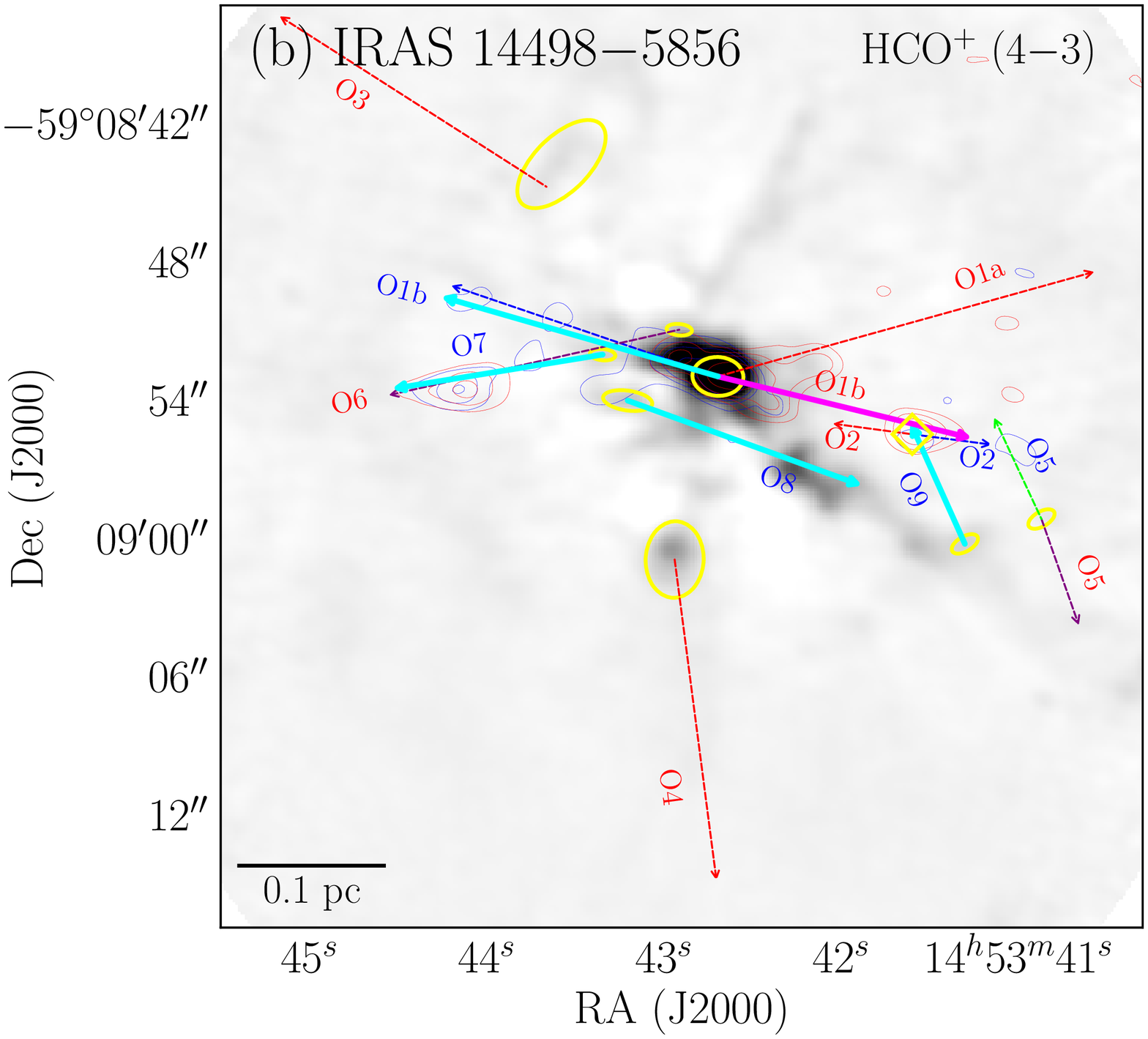}
\includegraphics[width=8.6cm,height=7.3cm]{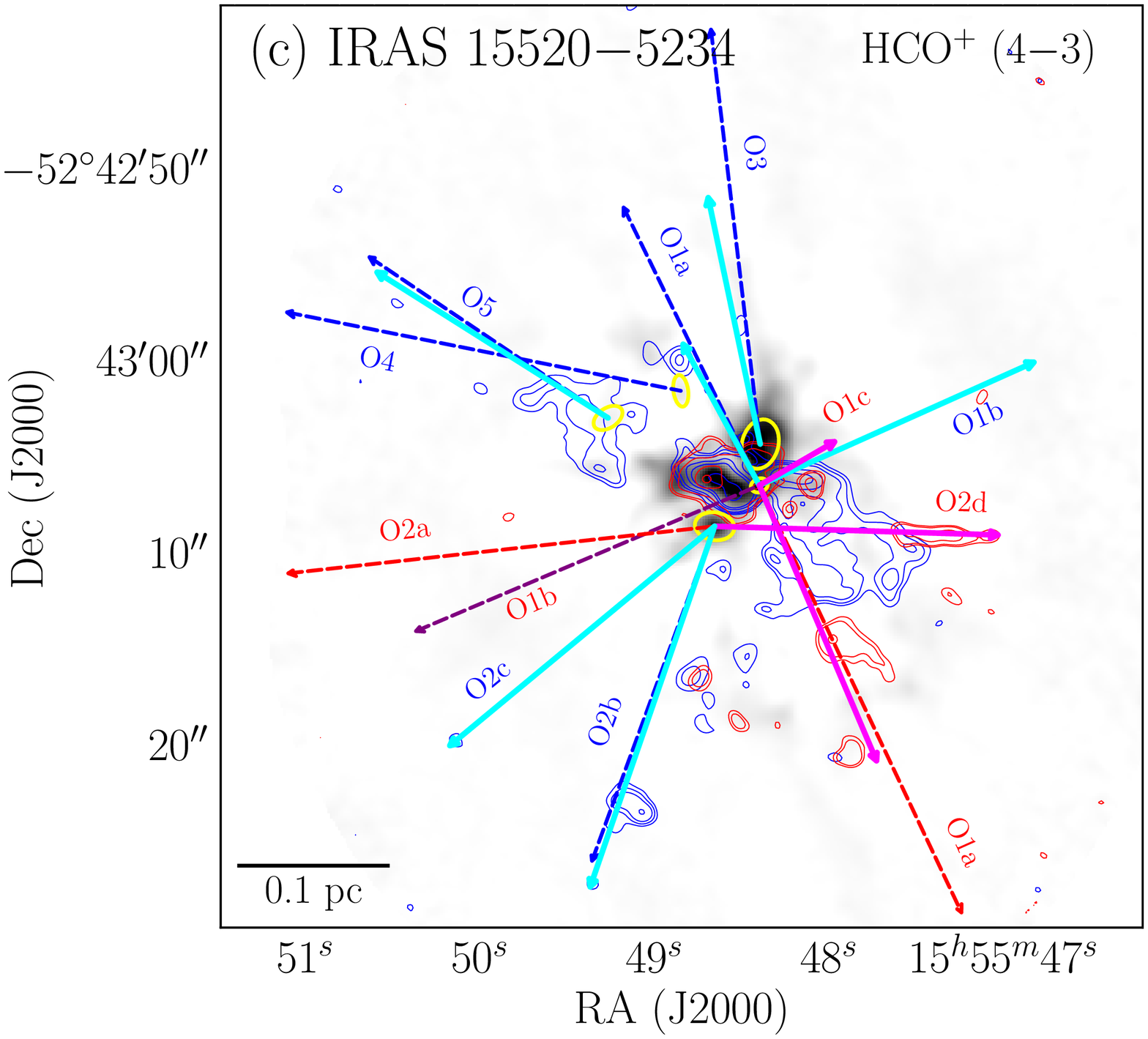}
\includegraphics[width=8.6cm,height=7.3cm]{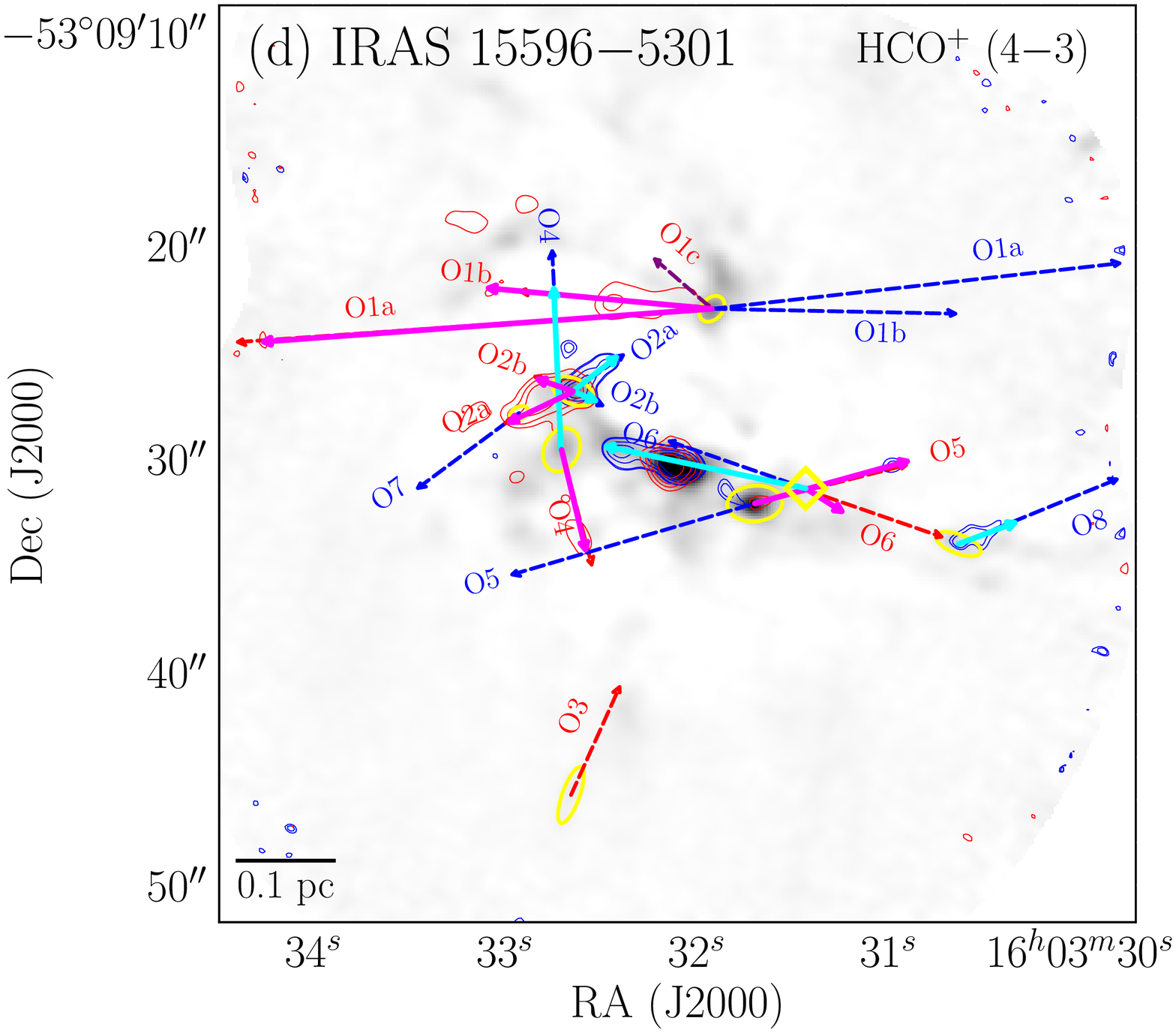}
\includegraphics[width=8.6cm,height=7.3cm]{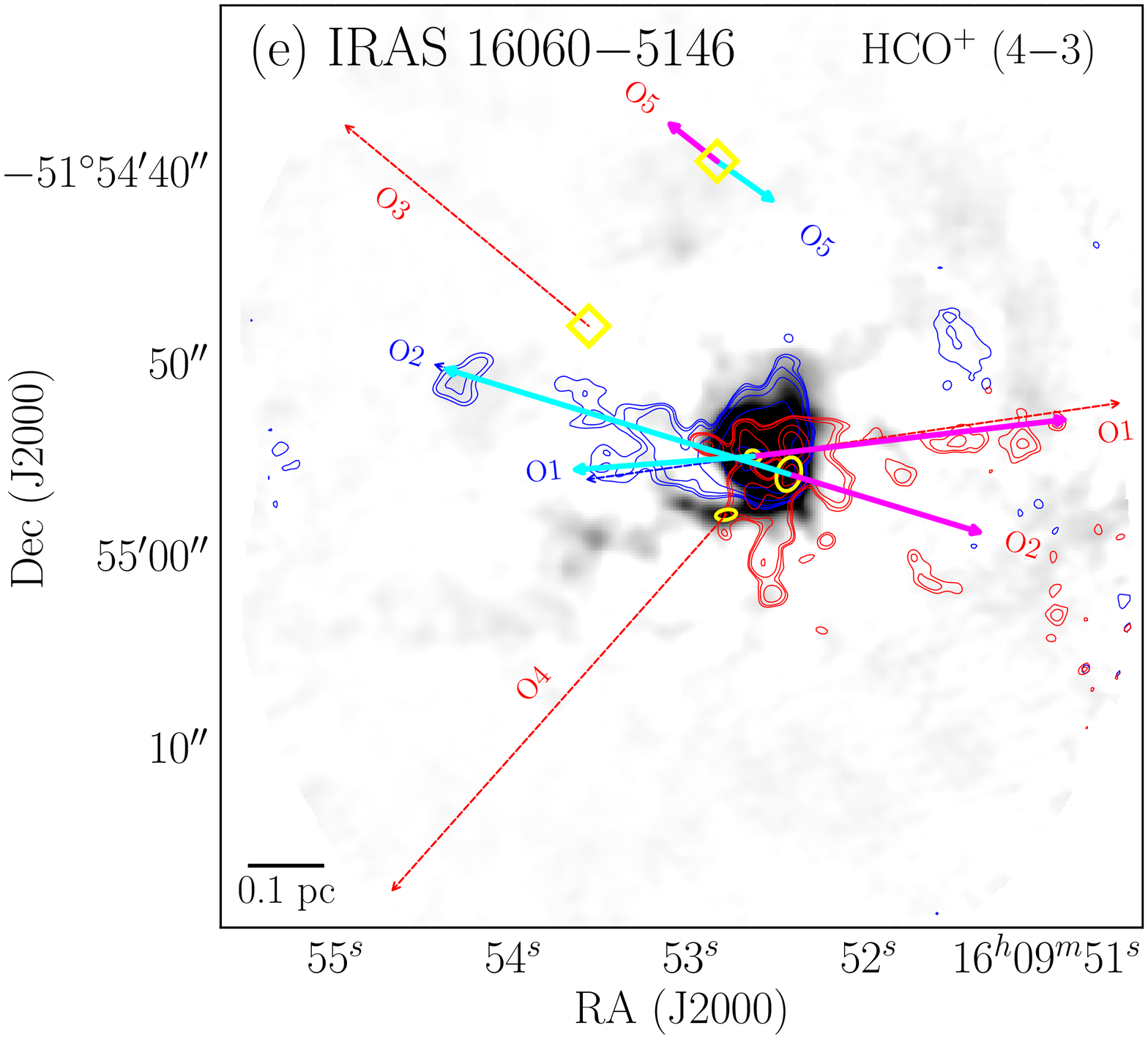}
\includegraphics[width=8.6cm,height=7.3cm]{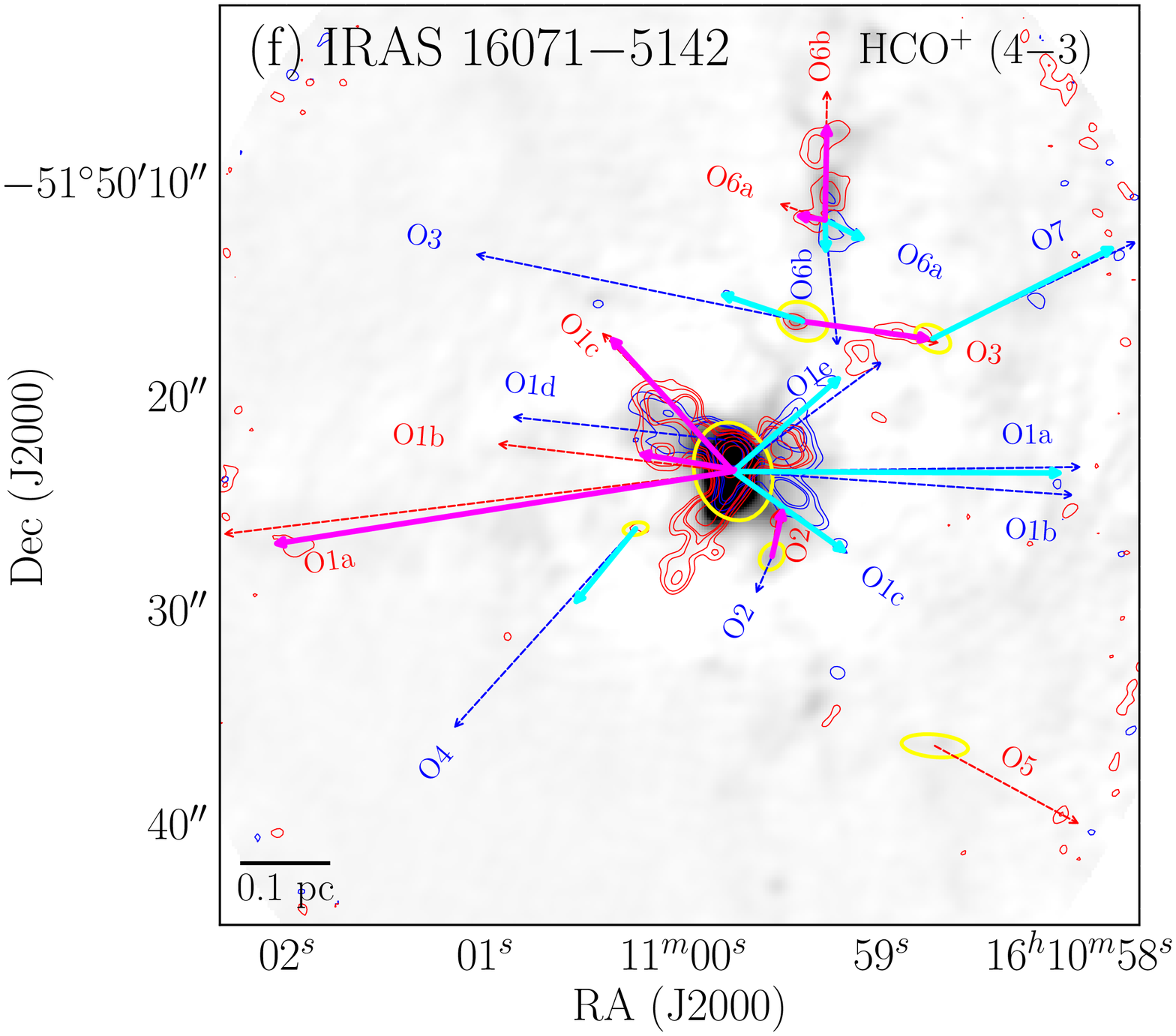}
\caption{Images of HCO$^{+}$ outflows in six target fields. The red and blue contours correspond to redshifted and blueshifted HCO$^{+}$ gas integrated over carefully selected velocity ranges to depict the outflow lobes. The blueshifted and redshifted outflow lobes of HCO$^{+}$ are marked by solid cyan and magenta arrows, respectively. The blueshifted and redshifted outflow lobes identified using CO are shown by blue and red dashed arrows, while a few additional blueshifted and redshifted lobes that are identified only in HCN are marked by green and purple arrows, respectively. The remaining symbols are the same as shown in Figure~\ref{fig1} and Figure~\ref{figA1}.}
\label{figA3}
\end{figure*}

\begin{figure*}
\includegraphics[width=9.0cm,height=7.4cm]{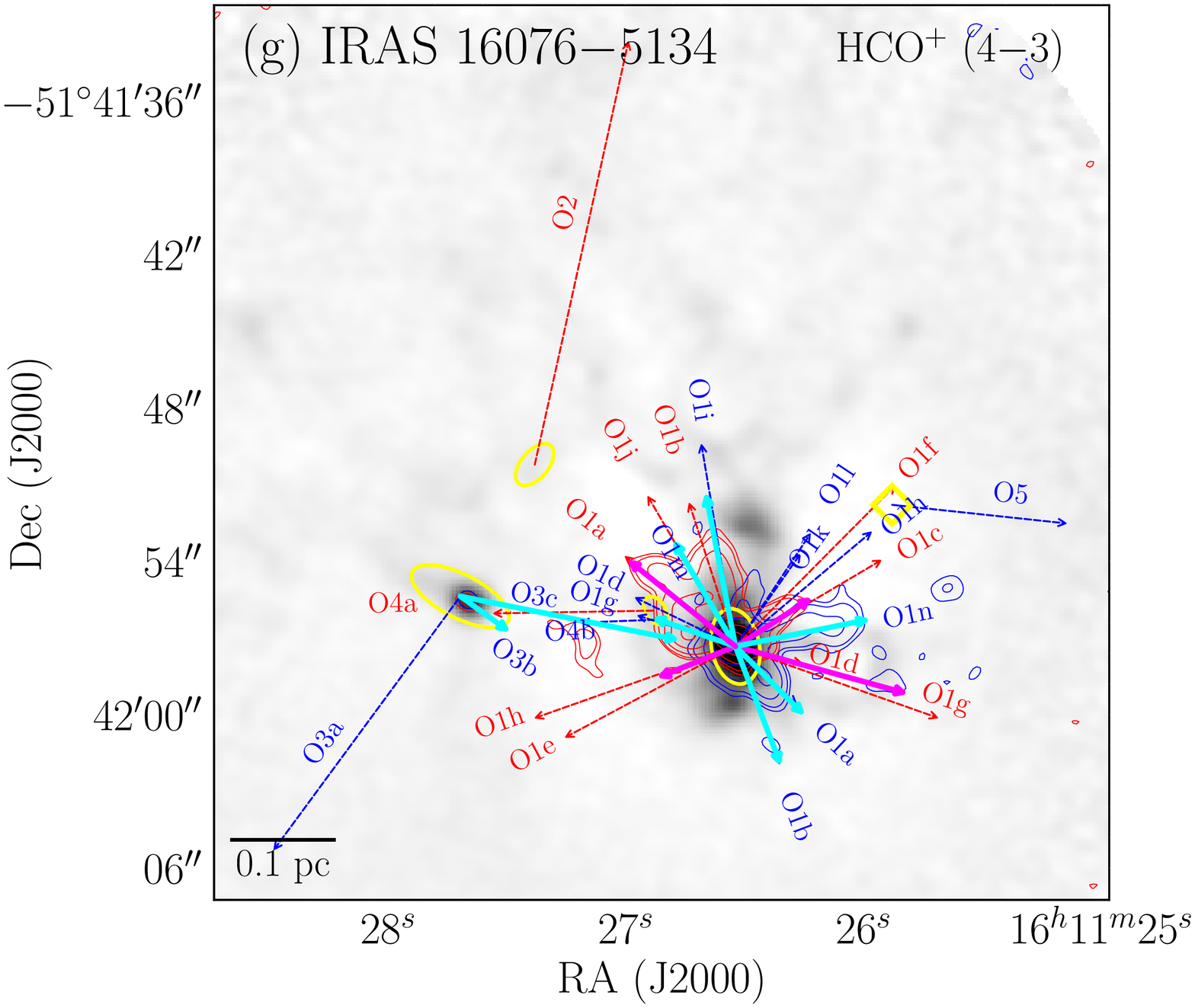}
\includegraphics[width=8.6cm,height=7.3cm]{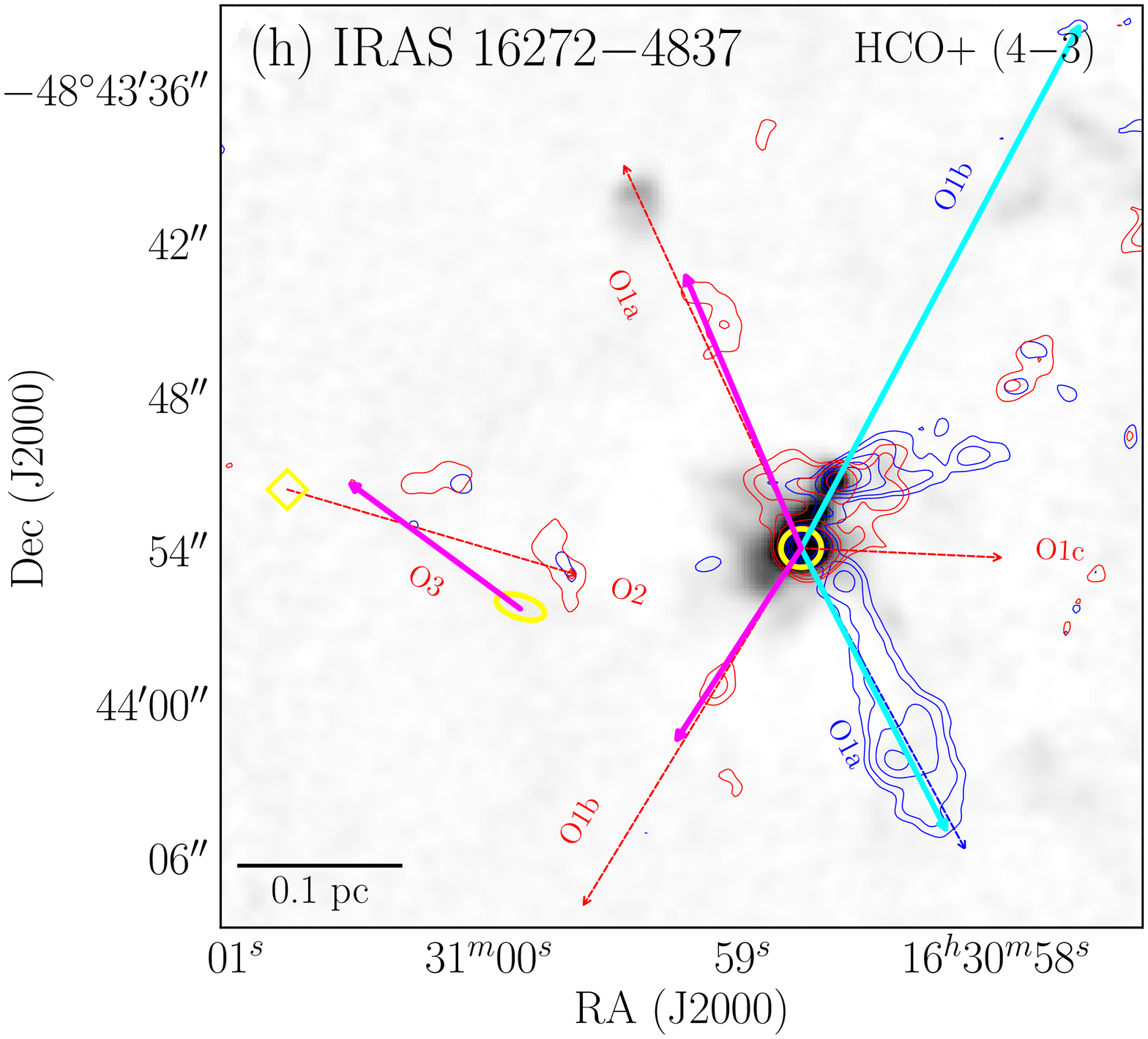}
\includegraphics[width=9.0cm,height=7.4cm]{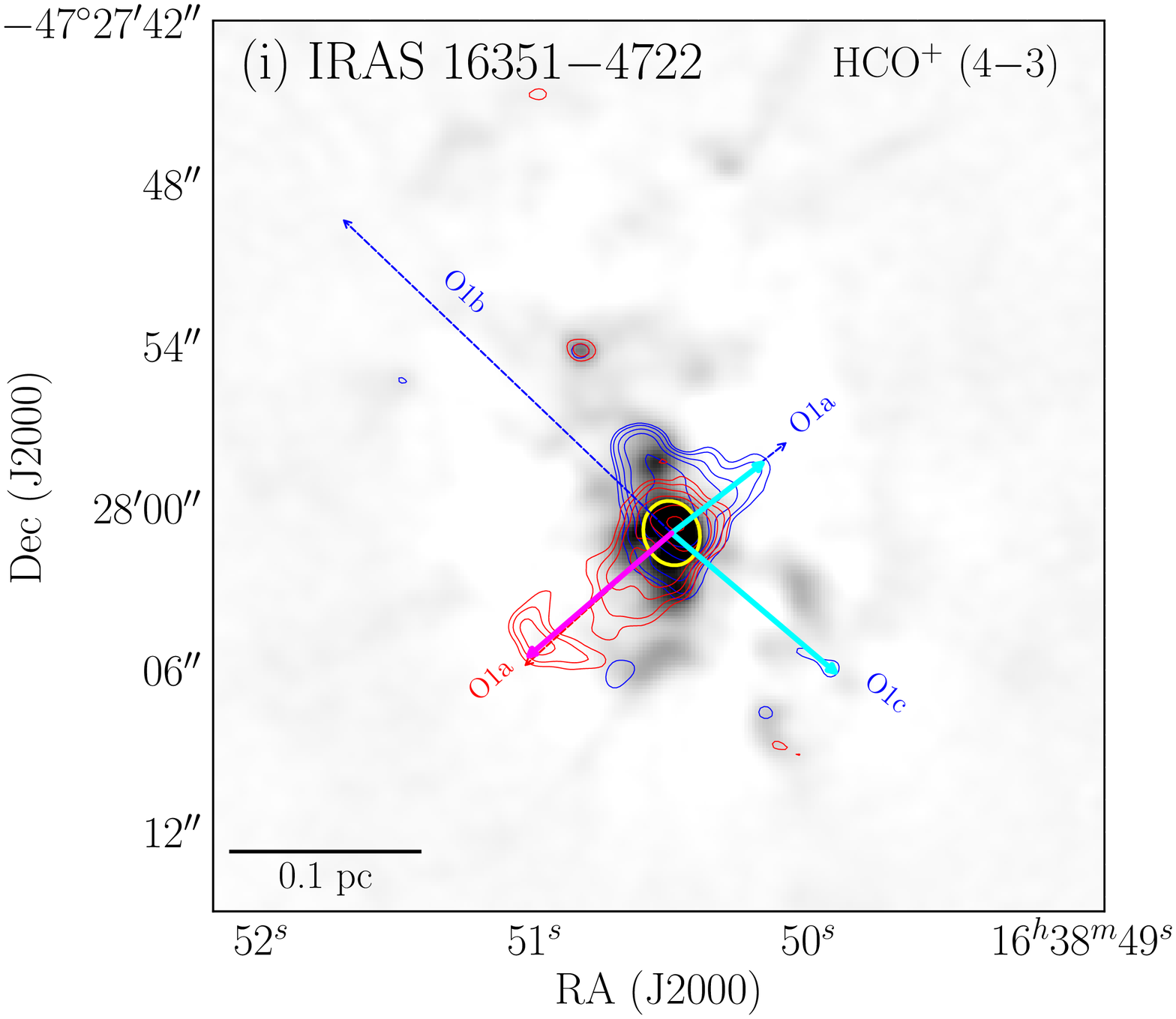}
\includegraphics[width=8.6cm,height=7.3cm]{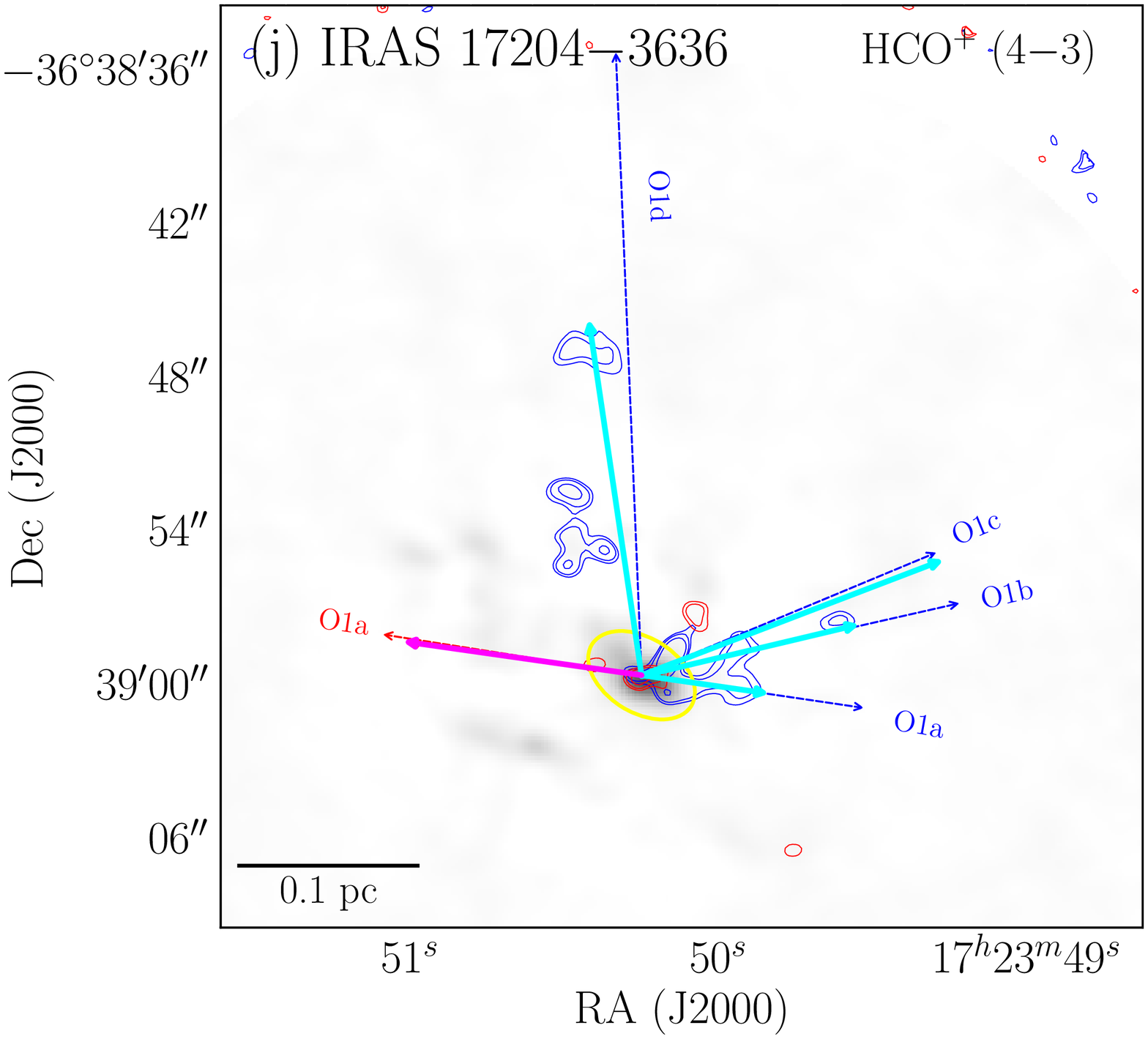}
\includegraphics[width=8.6cm,height=7.3cm]{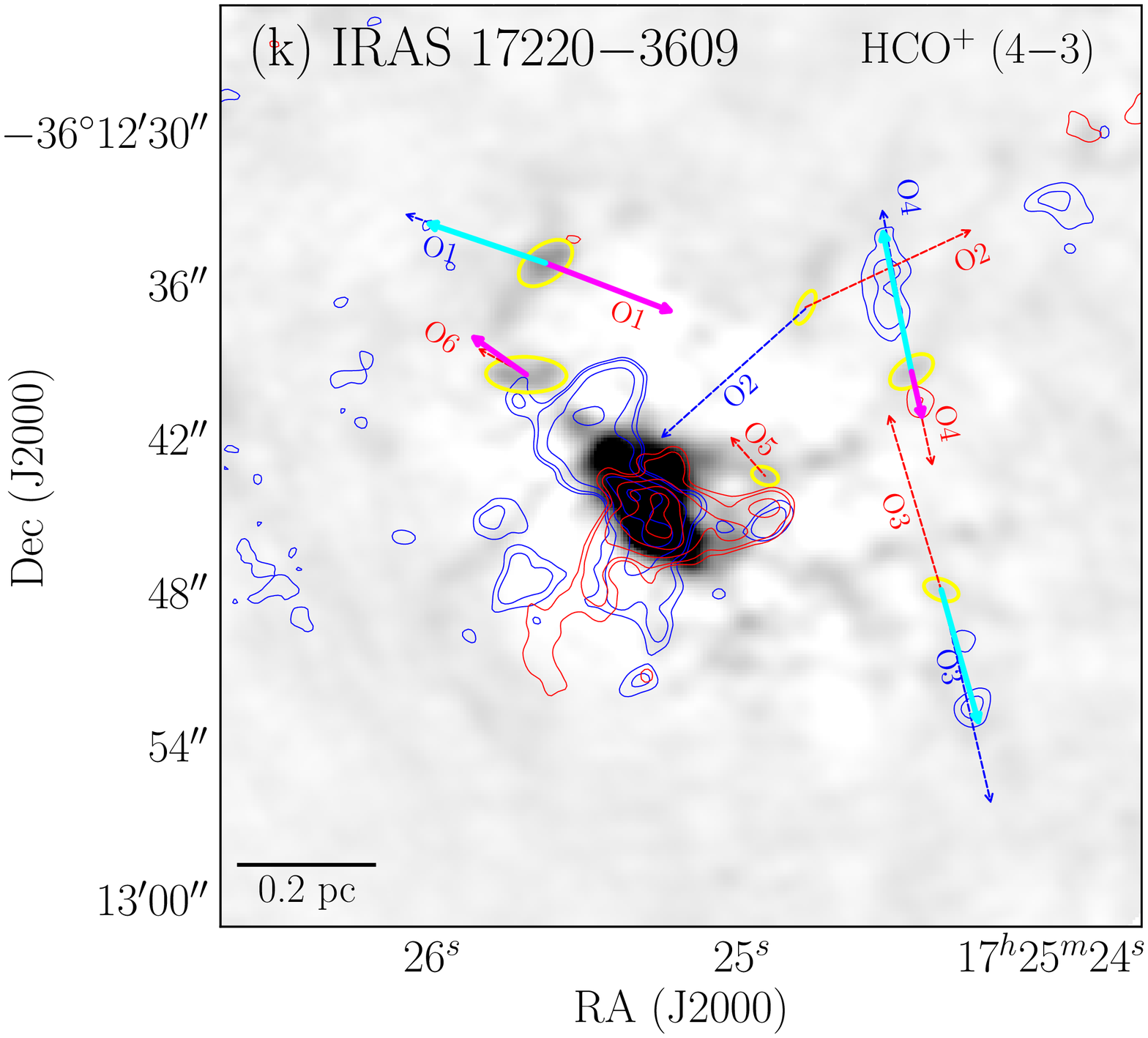}
\caption{Images of HCO$^{+}$ outflows for the remaining five target fields. For details of the markers and symbols see caption of Figure~\ref{figA3}.}
\label{figA4}
\end{figure*}

As presented in Section~\ref{sec:IndetifyOutflows}, we identified the outflow lobes using three different tracers (i.e., CO, HCN, and HCO$^{+}$). An example figure for CO and HCN lobes is presented in Figure~\ref{fig1}. Figures corresponding to the rest of the regions with HCN lobes are presented in Figures~\ref{figA1} and \ref{figA2}. The figures with HCO$^{+}$ outflow lobes are also presented in Figures~\ref{figA3}~and~\ref{figA4}. 
 
\bsp	
\label{lastpage}
\end{document}